\documentclass[11pt,a4paper]{article}
\usepackage{a4wide}
\usepackage{graphicx}
\usepackage{latexsym}
\usepackage{amssymb}
\usepackage{amsthm}
\usepackage[fleqn]{amsmath}
\usepackage{microtype}
\allowdisplaybreaks
\usepackage{url}
\newtheoremstyle{myplain}
  {1.5ex} % spaceabove
  {1.5ex} % spacebelow
  {\addtolength{\leftskip}{2\parindent}\itshape%
    \renewcommand{\emph}[1]{\textbf{#1}}%
  }       % bodyfont
  {-2\parindent}  % indent
  {}      % headfont
  {.}     % headpunct
  { }     % headspace
  {\textsc{\thmname{#1}\thmnumber{ #2}} \thmnote{#3}%
  }       % headspec
\newtheoremstyle{myplainshort}
  {1.5ex} % spaceabove
  {1.5ex} % spacebelow
  {\addtolength{\leftskip}{2\parindent}\itshape%
    \renewcommand{\emph}[1]{\textbf{#1}}%
  }       % bodyfont
  {-2\parindent}  % indent
  {}      % headfont
  {.}     % headpunct
  { }     % headspace
  {\textsc{\thmname{#1}\thmnumber{ #2}}%
  }       % headspec
\newtheoremstyle{mydef}
  {1.5ex} % spaceabove
  {1.5ex} % spacebelow
  {\addtolength{\leftskip}{2\parindent}%
  }       % bodyfont
  {-2\parindent}  % indent
  {}      % headfont
  {.}     % headpunct
  { }     % headspace
  {\textsc{\thmname{#1}\thmnumber{ #2}} (\thmnote{#3})%
  }       % headspec
\theoremstyle{myplain}
\newtheorem{theorem}{Theorem}
\newtheorem{proposition}[theorem]{Proposition}
\newtheorem{lemma}[theorem]{Lemma}

\theoremstyle{myplainshort}
\theoremstyle{mydef}
\newtheorem{definition}{Definition}
\theoremstyle{remark}

\newcommand{\runinhead}[1]{\medskip\par\noindent\textbf{#1}}
\begin{document}
\title{Multivariate Medians for Image and Shape Analysis}
\author{Martin Welk\\
Institute of Biomedical Image Analysis\\
UMIT TIROL -- Private University for Health Sciences,\\Biomedical Informatics 
and Technology\\
Eduard-Walln{\"o}fer-Zentrum 1, 6060 Hall/Tyrol, Austria\\
\url{martin.welk@umit-tirol.at}
}
\date{June 25, 2021}
\maketitle
\begin{abstract}
Having been studied since long by statisticians, multivariate median
concepts found their way into the image processing literature in the
course of the last decades, being used to construct robust and efficient
denoising filters for multivariate images such as colour images but also
matrix-valued images. 
Based on the similarities between image and geometric data as results of
the sampling of continuous physical quantities, it can be expected that the
understanding of multivariate median filters for images provides a starting
point for the development of shape processing techniques.
This paper presents an overview of multivariate median concepts relevant
for image and shape processing. It focusses on their mathematical principles
and discusses important properties especially in the context of image 
processing.
\end{abstract}

\section{Introduction}

For almost half a century, the median filter has established itself as
an efficient tool for filtering of univariate signals and images
\cite{Tukey-Book71}.
Despite its simplicity, it has favourable properties as a denoising
method. Whereas average filters and their variants such as Gaussian
smoothing are useful for denoising e.g.\ additive Gaussian noise, they
break down in the presence of noise featuring more outliers such as
impulsive noise. In contrast, the median filter performs surprisingly well
in removing such noise; at the same time, it is capable of preserving sharp
signal or image boundaries. Moreover, an interesting connection to
partial differential equations (PDEs) could be proven \cite{Guichard-sana97}.

The success of univariate median filtering has inspired researchers to
investigate similar procedures for multivariate images. In several works
on this subject \cite{Astola-PIEEE90,Spence-icip07,Welk-dagm03}, 
the minimisation of an objective function composed of distances of the
median point to the input values was considered, building on or partly
rediscovering concepts that had been around long before in the statistical
community \cite{Gini-Met29} and are nowadays often referred to as
$L^1$ medians. In the statistical literature, however, shortcomings of this
concept have been discussed since the 1940s \cite{Haldane-Biomet48}, and
various alternative ideas have been developed since then \cite{Small-ISR90}.
Recently, attempts have been made to introduce some of these alternative
concepts to image processing \cite{Welk-JMIV16,Welk-ssvm19}. 

The ability of the median to reduce a data set to a single value representing
its position, and doing so in a way that is hardly sensitive to outliers
in the data, makes multivariate medians also an interesting candidate for
the analysis of geometric data such as point clouds in the real
affine space, or more generally on manifolds. Resulting from the sampling
of continuous-scale physical quantities, such geometric data are not far
from image data, and image analysis techniques have been adapted
successfully for their processing, see \cite{Clarenz-vis00} for an example,
or \cite{Clarenz-TIP04} for combined processing of shape and image data
in the case of textured surfaces. 
Although few references to median filtering of geometric data can be
found in the literature so far -- see e.g.\ \cite[Sec.~7]{Small-ISR90}
for a short discussion of medians for points on a sphere --, 
it can therefore be expected that a
thorough understanding of multivariate median filtering of images can serve
as a starting point for robust shape processing techniques.

This paper aims at giving an overview of multivariate median concepts
that can be useful for processing images and shapes. In the course of the
paper, the underlying mathematical structures and principles are 
presented, and linked to resulting properties of the filters that are
particularly relevant for the processing of image and shape data. 
Results on connections between multivariate median filters and PDE-based
image filters are reviewed. Results from existing work, including some
by the author of this paper, are reported. In order to present the
overarching principles and ideas, proofs of individual results are
generally omitted, although hints at important ideas behind these proofs
are given where appropriate.

The remainder of this paper is organised as follows. 
In Section~\ref{sec-med}, different ways to define univariate medians
for discrete sets and continuous distributions are juxtaposed, and some
equivariance properties and algorithmic aspects are reviewed. 
Median filtering of univariate (grey-scale) images is discussed in
Section~\ref{sec-medf}, covering also a space-continuous model and its
relation to PDEs as well as an extension to
adaptive neighbourhoods (so-called morphological amoebas).
In Section~\ref{sec-mvmed}, a selection of multivariate median concepts
are presented, namely, the $L^1$ median, Oja's simplex median, a 
transformation--retransformation $L^1$ median, the half-space median and
the convex-hull-stripping median. 
It will be shown how these concepts are derived from
generalising different variants of univariate median definitions.
Properties of multivariate median concepts that are relevant to image
and shape processing are pointed out, with emphasis on the relation between
discrete and continuous models, and equivariance properties. Remarks on
algorithmic aspects are included. 
The short Section~\ref{sec-mvmedf} points out some specific aspects of
median filtering of multivariate images, focussing on the proper handling
of different dimensionality of image domains and data ranges.
Section~\ref{sec-mvmedpde} gives an overview of the results on approximations
of PDEs by multivariate median filters that have been achieved so far,
including one result the proof of which is still awaiting 
(and which is stated as conjecture therefore).
The paper ends with a short summary in Section~\ref{sec-sum}.

\runinhead{Notations.} Let us briefly introduce a few notations that will
be used throughout this paper. By $\mathbb{R}$ we denote the set of 
real numbers. The symbol $\mathbb{R}^+$ stands for positive real numbers
whereas $\mathbb{R}^+_0$ names nonnegative real numbers. As usual,
$\mathbb{R}^n$ is the $n$-dimensional real vector or affine space which
will sometimes be equipped with the standard Euclidean structure.
In this case, $\lVert\boldsymbol{v}\rVert$ is the Euclidean norm of a vector 
$\boldsymbol{v}\in\mathbb{R}^n$, and 
$\langle \boldsymbol{v},\boldsymbol{w}\rangle$ the scalar product of
vectors $\boldsymbol{v},\boldsymbol{w}\in\mathbb{R}^n$.
By $\mathbb{Z}$ we denote the set of integers.

A \emph{multiset} is a set with (integer) multiplicities; standard 
notations for sets, including union and intersection, are used for 
multisets in an intuitive way. The cardinality of a set or multiset $S$
will be written as $\#S$.
The convex hull of a set or multiset $S\subset\mathbb{R}^n$
will be denoted by $[S]$, where $[x_1,\ldots,x_n]$ abbreviates
$[\{x_1,\ldots,x_n\}]$ for finite multisets. The volume ($n$-dimensional
measure) of a measurable point set $S\subset\mathbb{R}^n$ will be 
written as $\lvert S\rvert$. In particular, $\bigl\lvert[S]\bigr\rvert$
is the volume of the convex hull of $S$. 

The concatenation of functions $g$ and $f$ will be denoted by 
$f\circ g$, i.e.\ $(f\circ g)(x) := f (g (x))$.
At some points, reference will be made to distributions over $\mathbb{R}$
or $\mathbb{R}^n$; their spaces will be denoted by $\mathcal{D}'(\mathbb{R})$
or $\mathcal{D}'(\mathbb{R}^n)$, respectively, compare 
\cite{Vladimirov-Book67en,Walter-Book94}.

For error estimates as well as for algorithmic complexity statements,
we will use the common $\mathcal{O}$ Landau notation. Just recall that for
error estimates, $\mathcal{O}(f(\varepsilon))$ denotes the set of functions 
$g(\varepsilon)$ which for $\varepsilon\to0$ are bounded by a constant 
multiple of $f(\varepsilon)$, whereas for complexity statements
$\mathcal{O}(f(N))$ denotes the set of functions $g(N)$ which are bounded
by constant multiples of $f(N)$ for $N\to\infty$.

\section{Univariate Median}
\label{sec-med}

In this section, we collect basic facts about the median of real-valued
data in various settings, ranging from finite multisets via countable
weighted sets up to continuous densities over $\mathbb{R}$. 

Whereas practical computations always act on finite, thus discrete, data,
the continuous case is important for a proper theoretical foundation of the
intended application of medians in image and shape analysis since 
we always understand discrete images and shapes as samples of
space-continuous objects.

We put special emphasis on characterisations
of the median by minimisation properties as these play a central role
for the later generalisation to the multivariate case.

\subsection{Median for Unweighted Discrete Data}
\label{ssec-med}

\runinhead{Rank-order definition.}
The most basic concept of the median relies on rank order, taking
the middle element of an ordered sequence.

\begin{definition}[Median by rank-order]
Let a finite multiset 
$\mathcal{X}=\{x_1,\ldots,x_N\}$ of real numbers
$x_1,\ldots,x_N\in\mathbb{R}$ be given. 
Assume that the elements of $\mathcal{X}$ are 
arranged in ascending order, $x_1\le \ldots\le x_N$.
If $N$ is odd, the \emph{median} of $\mathcal{X}$ is given by $x_{(N+1)/2}$. 
If $N$ is even, the median is given by the interval $[x_{N/2},x_{N/2+1}]$.
\end{definition}

\noindent
In the case of even $N$, there is an ambiguity; one could define either
of the two values in the middle or e.g.\ their arithmetic mean as the median
of $\mathcal{X}$. 
As this ambiguity does not play a central role in the context of this 
work, we will prefer the set-valued interpretation as stated in the Definition
and not further elaborate on disambiguation strategies. We will still 
colloquially speak of ``the'' median even in this case.

\runinhead{An extrema-stripping procedure to determine the median.}
An (albeit inefficient) way to determine the median of a multiset can be
based directly on this rank-order definition: 

\begin{definition}[Median by extrema-stripping]
\label{def-med-extr}
Let a finite multiset $\mathcal{X}$ of real numbers be given.
Delete its smallest and greatest element, 
then again the remaining smallest and greatest element etc., until an
empty set is obtained. The one or two points that were deleted in the
very last step of this sequence are medians, with the entire median
set given as their convex hull.
\end{definition}

\runinhead{The median as $1/2$-quantile.}
The median is also the $1/2$-quantile of the given set, i.e.\ the
smallest number $\mu$ for which at least half of the given data are
less or equal $\mu$.
In the following definition, 
$\#S$ denotes the cardinality of a finite multiset $S$.

\begin{definition}[Median as 1/2-quantile]
Let $\mathcal{X}$ be a finite multiset of real numbers. The median of
$\mathcal{X}$ is defined as
\begin{align}
\operatorname{med}(\mathcal{X}) &:=
\min \left\{\mu\in\mathbb{R}~|~
\#(\mathcal{X}\cap(-\infty,\mu])\ge\tfrac12\#\mathcal{X}\right\} \;.
\label{med-lhalf}
\end{align}
\end{definition}

\noindent
Note that the ambiguity in the case of even $\#\mathcal{X}=N$ is
resolved here by taking the minimum. A more symmetric definition
again implies a set-valued median.

\begin{definition}[Median as 1/2-quantile, symmetric]
Let $\mathcal{X}$ be a finite multiset of real numbers. The median of
$\mathcal{X}$ is defined as
\begin{align}
\operatorname{med}(\mathcal{X}) &:= \phantom{{}\cap{}}
\left\{\mu\in\mathbb{R}~|~
\#(\mathcal{X}\cap(-\infty,\mu])\ge\tfrac12\#\mathcal{X}
\right\} 
\notag\\*&\phantom{{}:={}}
{{}\cap{}}
\left\{\mu\in\mathbb{R}~|~
\#(\mathcal{X}\cap[\mu,+\infty))\ge\tfrac12\#\mathcal{X}
\right\} \;.
\label{med-shalf}
\end{align}
\end{definition}

\runinhead{The median as point of maximal half-line depth.}
The latter characterisation can be rephrased by defining the median
as the set of points which split the data multiset as symmetric as
possible.
This characterisation of the median goes back to the work
by Hotelling \cite{Hotelling-EJ29} and has been picked up by
Tukey \cite{Tukey-icm74}. 

\begin{definition}[Median as point of maximal half-line depth]
For a finite multiset $\mathcal{X}$ of real numbers, define the
median of $\mathcal{X}$ as
\begin{align}
\operatorname{med}(\mathcal{X}) &:= 
\mathop{\operatorname{argmax}}\limits_{\mu\in\mathbb{R}}
\min\{\#(\mathcal{X}\cap(-\infty,\mu]),\#(\mathcal{X}\cap[\mu,+\infty)\}
\;.
\label{med-hs}
\end{align}
The count 
$\min\{\#(\mathcal{X}\cap(-\infty,\mu]),\#(\mathcal{X}\cap[\mu,+\infty)\}$
that indicates for any $\mu$ 
the number of data points that are contained at least in a
half-line starting at $\mu$, is called \emph{half-line depth}
of $\mu$ w.r.t.\ $\mathcal{X}$. 
\end{definition}

\noindent
Note that in a finite set (without multiplicities)
the half-line depth of a data point is simply its rank or reverse
rank. The extrema-stripping procedure from Definition~\ref{def-med-extr}
then actually
enumerates the data points in the order of increasing half-line depth;
the half-line depth of each data point is the ordinal number of the step
in which it is deleted.

\runinhead{The median as minimiser of a convex function.}
As has been observed first by Jackson \cite{Jackson-BAMS21},
the median of real numbers
can also be defined via an optimisation property.

\begin{definition}[Median as minimiser]
Let $\mathcal{X}=\{x_1,\ldots,x_N\}$ be a multiset of real numbers
$x_1,\ldots,x_N\in\mathbb{R}$. 
The median of $\mathcal{X}$ is
\begin{align}
\operatorname{med}(\mathcal{X})&:=
\mathop{\operatorname{argmin}}\limits_{\mu\in\mathbb{R}} 
E_{\mathcal{X}}(\mu)\;, &
E_{\mathcal{X}}(\mu)&:=
\sum\limits_{i=1}^N
\lvert \mu-x_i\rvert
\;.
\label{med-E}
\end{align}
\end{definition}

\noindent
In this formulation, the order of real numbers is not explicitly
invoked. The objective function $E_{\mathcal{X}}(\mu)$ is a continuous,
piecewise linear function with a kink at every data value $\mu=x_i$.
For even $N$, there will be a constant segment over the interval formed
by the two middle values in the rank order.

\subsection{Median for Weighted Discrete Data}
\label{ssec-medw}

The median concept can easily be extended to weighted data.
We assume that $\mathcal{X}$ is a countable set of real numbers,
on which a positive weight function $w:\mathcal{X}\to\mathbb{R}^+$ is given.
We will require $w$ to have unit total weight, i.e.\
$\sum\nolimits_{x\in\mathcal{X}}w(x)=1$.

\runinhead{Quantile-based definition.}
We can define the median of weighted data 
analogously to \eqref{med-lhalf} or \eqref{med-shalf}.

\begin{definition}[Weighted median as $1/2$-quantile]
\label{def-medw-lhalf}
Let $\mathcal{X}$ be a countable set of real numbers with a positive
weight function $w:\mathcal{X}\to\mathbb{R}^+$ of unit total weight.
The median of the weighted set $\mathcal{X}$ is
\begin{align}
\operatorname{med}(\mathcal{X},w) &:=
\min \left\{\mu\in\mathbb{R}~\left|~
\sum\nolimits_{x\in\mathcal{X},x\le\mu}w(x)\ge\tfrac12\right.\right\} \;.
\label{medw-lhalf}
\end{align}
\end{definition}

\begin{definition}[Weighted median as $1/2$-quantile, symmetric]
Let $\mathcal{X}$, $w$ be as in Definition~\ref{def-medw-lhalf}.
The median of the weighted set $\mathcal{X}$ is
\begin{align}
\operatorname{med}(\mathcal{X},w) &:= \phantom{{}\cap{}}
\left\{\mu\in\mathbb{R}~\left|~
\sum\nolimits_{x\in\mathcal{X},x\le\mu}w(x)\ge\tfrac12\right.\right\} 
\notag\\*&\phantom{{}:={}}
{{}\cap{}}
\left\{\mu\in\mathbb{R}~\left|~
\sum\nolimits_{x\in\mathcal{X},x\ge\mu}w(x)\ge\tfrac12\right.\right\} \;.
\label{medw-shalf}
\end{align}
\end{definition}

\runinhead{Weighted half-line depth.}
The half-line depth definition of the median can be translated as follows.
\begin{definition}[Weighted median as point of maximal half-line depth]
Let $\mathcal{X}$, $w$ be as in Definition~\ref{def-medw-lhalf}.
The median of the weighted set $\mathcal{X}$ is
\begin{align}
\operatorname{med}(\mathcal{X},w) &:= 
\mathop{\operatorname{argmax}}\limits_{\mu\in\mathbb{R}}
\min\left\{
\sum\nolimits_{x\in\mathcal{X},x\le\mu}w(x),
\sum\nolimits_{x\in\mathcal{X},x\ge\mu}w(x)
\right\} \;.
\label{medw-hs}
\end{align}
For $\mu\in\mathbb{R}$, we call
$\min\left\{
\sum\nolimits_{x\in\mathcal{X},x\le\mu}w(x),
\sum\nolimits_{x\in\mathcal{X},x\ge\mu}w(x)
\right\}$
\emph{half-line depth} of $\mu$ w.r.t.\ $(\mathcal{X},w)$.
\end{definition}

\noindent
If $\mathcal{X}$ is finite,
adapting the extrema-stripping procedure to weighted medians is possible
with proper bookkeeping of the weighted half-line depth of points during
the deletion process; in each step, one deletes the extremum point (minimum
or maximum) with lesser half-line depth, or both simultaneously if their
half-line depths are equal.

\runinhead{Minimisation property.}
The minimisation definition \eqref{med-E}, too, 
can easily be adapted to weighted data.

\begin{definition}[Weighted median as minimiser]
Let $\mathcal{X}$, $w$ be as in Definition~\ref{def-medw-lhalf}.
The median of the weighted set $\mathcal{X}$ is
\begin{align}
\operatorname{med}(\mathcal{X},w)&:=
\mathop{\operatorname{argmin}}\limits_{\mu\in\mathbb{R}} 
E_{\mathcal{X},w}(\mu)\;, &
E_{\mathcal{X},w}(\mu)&:=
\sum\limits_{x\in\mathcal{X}}
w(x)\lvert \mu-x\rvert
\;.
\label{medw-E}
\end{align}
\end{definition}

\subsection{Median of Continuous Data}
\label{ssec-medd}

The various definitions of the median for weighted sets can easily be 
generalised to continuous data described by densities. 

\begin{definition}[Normalised regular density on $\mathbb{R}$]
Let $\gamma:\mathbb{R}\to\mathbb{R}^+_0$ be a nonnegative
integrable function with unit total weight
$\int\nolimits_\mathbb{R}\gamma(x)\,\mathrm{d}x=1$. Then $\gamma$
is called \emph{normalised regular density} on $\mathbb{R}$.
\end{definition}

\runinhead{Quantile-based definition.}
We start again by definitions of the median as a $1/2$-quantile, see
\cite{Haldane-Biomet48} where the so-defined median is called
\emph{arithmetic median.}

\begin{definition}[Continuous median as $1/2$-quantile]
The median of a normalised regular density $\gamma$ is defined as
\begin{align}
\operatorname{med}(\gamma) &:=
\min \left\{\mu\in\mathbb{R}~\left|~
\int\nolimits_{-\infty}^\mu\gamma(x)\,\mathrm{d}x
\ge\frac12\right.\right\} \;,
\label{medd-lhalf}
\end{align}
\end{definition}

\begin{definition}[Continuous median as $1/2$-quantile, symmetric]
The median of a normalised regular density $\gamma$ is defined as
\begin{align}
\operatorname{med}(\gamma) &:= \phantom{{}\cap{}}
\left\{\mu\in\mathbb{R}~\left|~
\int\nolimits_{-\infty}^\mu\gamma(x)\,\mathrm{d}x
\ge\frac12\right.\right\} 
\notag\\*&\phantom{{}:={}}
{{}\cap{}}
\left\{\mu\in\mathbb{R}~\left|~
\int\nolimits_{\mu}^{+\infty}\gamma(x)\,\mathrm{d}x
\ge\frac12\right.\right\} \;.
\label{medd-shalf}
\end{align}
\end{definition}

\noindent
For regular densities as specified so far, the inequalities
$\int\gamma\,\mathrm{d}x\ge1/2$ will even be fulfilled with equality.
However, \eqref{medd-lhalf} and \eqref{medd-shalf} can even be used
verbatim for a more general class of distribution-valued densities
$\gamma\in\mathcal{D}'(\mathbb{R})$ that
can be decomposed into a regular (real-valued) density $\{\gamma\}$ with a 
countable sum of weighted delta (Dirac) peaks, thus representing the
combination of a continuous distribution of values (in which each particular
number has negligible weight) with a weighted discrete set as in
\eqref{medw-lhalf}, \eqref{medw-shalf}. In this case the inequalities
$\int\gamma\,\mathrm{d}x\ge1/2$ may not be satisfied with equality if
the median happens to be at a delta peak of $\gamma$.

\runinhead{Continuous half-line depth.}
The characterisation \eqref{medd-shalf} can again be rewritten in a 
half-line depth form, where the continuous half-line depth is determined
by integrals.

\begin{definition}[Continuous median as point of maximal half-line depth]
For a normalised regular density $\gamma$, its median is defined as
\begin{align}
\operatorname{med}(\gamma) &:=
\mathop{\operatorname{argmax}}\limits_{\mu\in\mathbb{R}}
\min\left\{
\int\nolimits_{-\infty}^{\mu}\gamma(x)\,\mathrm{d}x,
\int\nolimits_{\mu}^{+\infty}\gamma(x)\,\mathrm{d}x
\right\}
\;.
\label{medd-hs}
\end{align}
\end{definition}

\runinhead{Continuous extrema stripping.}
Even an extrema-stripping formulation of the continuous median is possible.
To this end, the stripping procedure must be stated as a time-continuous
process. We will give this definition for simplicity in the case of a 
density with a connected compact support set.

\begin{definition}[Continuous median by extrema-stripping]
Let $\gamma$ be a normalised regular density which is zero outside some
interval $[a_0,b_0]$ and nonzero in its interior $(a_0,b_0)$. Consider then
the ordinary differential equations
\begin{align}
a'(t)&=\bigl(\gamma(a(t))\bigr)^{-1}\;,&
b'(t)&=-\bigl(\gamma(b(t))\bigr)^{-1}
\label{medd-exstrip-ode}
\end{align}
with the initial conditions
\begin{align}
a(0)&=a_0\;, & b(0)&=b_0\;.
\end{align}
This initial boundary value problem yields a monotonically increasing 
$a:[0,T]\to\mathbb{R}$ and monotonically decreasing $b:[0,T]\to\mathbb{R}$
on an interval $[0,T]$ where $T$ is chosen such that $a(T)=b(T)$, and
the median of $\gamma$ is defined as
\begin{align}
\operatorname{med}(\gamma) &:= a(T)\;.
\end{align}
For each $t\in[0,T]$, we call $t$ the continuous half-line depth of $a(t)$ 
and $b(t)$.
\end{definition}

\runinhead{Minimisation property.}
Provided that $\gamma$ is compactly supported, a reformulation using a 
minimisation property is again possible.
This median is called \emph{geometric median} in \cite{Haldane-Biomet48}.

\begin{definition}[Continuous median as minimiser]
Let $\gamma$ be a normalised regular density with compact support.
The median of $\gamma$ is
\begin{align}
\operatorname{med}(\gamma)&:=
\mathop{\operatorname{argmin}}\limits_{\mu\in\mathbb{R}} 
E_{\gamma}(\mu)\;, &
E_{\gamma}(\mu)&:=
\int\nolimits_{\mathbb{R}}
\gamma(x)\lvert \mu-x\rvert
\,\mathrm{d}x
\;.
\label{medd-E}
\end{align}
\end{definition}

\noindent
The requirement of compact support of $\gamma$ can be replaced by a
weaker condition that ensures that the integral in $E_\gamma$ converges.

Non-uniqueness of the median is much less of a problem here than in the
case of discrete data as it occurs only if the support of $\gamma$ is
contained in two separate intervals each of which contains exactly half
the total weight of $\gamma$. In particular, if the support of $\gamma$ is 
an interval, and $\gamma$ has only isolated zeros within this interval, the 
median is unique.

\subsection{Equivariance Properties}

As the median in the sense of \eqref{med-lhalf}, \eqref{medw-lhalf} 
or \eqref{med-shalf}, \eqref{medw-shalf}
depends on nothing but the total order of the data, it can be generalised
to data in any totally ordered set $R$ instead of $\mathbb{R}$. 
Moreover, the median is equivariant w.r.t.\ any strictly monotonically
increasing map $T:R\to R$, i.e.\
\begin{align}
T\left(\operatorname{med}(\mathcal{X})\right) &= 
\operatorname{med}\bigl(T(\mathcal{X})\bigr)
\end{align}
where 
\begin{align}
T(\mathcal{X}) &:= \{ T(x) ~|~ x\in\mathcal{X} \}
\end{align}
denotes element-wise application of $T$ to all members of $\mathcal{X}$
(retaining their cardinalities), and similarly
\begin{align}
T\left(\operatorname{med}(\mathcal{X},w)\right) &=
\operatorname{med}\bigl(T(\mathcal{X},w\circ T^{-1})\bigr) \;.
\end{align}
Note that the weight function $w$ needs to be adapted using the
inverse map $T^{-1}$ which obviously exists.

Specifically for $R=\mathbb{R}$ we notice that the symmetric definitions
\eqref{med-shalf}, \eqref{medw-shalf} are also equivariant under
strictly monotonically decreasing maps $T$.

For the median of continuous densities \eqref{medd-lhalf} or \eqref{medd-shalf}
over $\mathbb{R}$ the equivariance w.r.t.\ strictly monotone transformations 
$T$ is given by
\begin{align}
T\left(\operatorname{med}(\gamma)\right) &=
\operatorname{med}\bigl(\gamma^T)\bigr) \;.
\label{medd-equv}
\end{align}
where 
$\gamma^T$ is given as
\begin{align}
\gamma^T(x) &:= \frac{\mathrm{d}}{\mathrm{d}y}
\left(\int\nolimits_{-\infty}^{T^{-1}(y)}\gamma(z)\,\mathrm{d}z\right)(x)\;,
\end{align}
i.e.\ the uniquely determined density for which
\begin{align}
\int\nolimits_{-\infty}^{T(x)}\gamma^T(y)\,\mathrm{d}y
&= \int\nolimits_{-\infty}^x\gamma(y)\,\mathrm{d}y\quad
\text{for all $x\in\mathbb{R}$.}
\end{align}
In the case of \eqref{medd-shalf}, the equivariance property \eqref{medd-equv}
again holds for strictly monotonically decreasing maps $T$, too.

\subsection{Algorithmic Aspects}

In the processing of images by median filtering, see Section~\ref{sec-medf},
median computations will have to be carried out multiple times. Efficient
algorithms are therefore relevant for the practicability of such filters.

\runinhead{Unweighted median.}
Finding the median of a finite multiset $\mathcal{X}$ of size $N$ 
is what is known in algorithmics as a \emph{selection problem.} 
Whereas in principle one could sort the data in $\mathcal{X}$ based on
their total order (for which several algorithms with
$\mathcal{O}(N\log N)$ computational complexity exist), the selection
problem does actually not require the complete order of the data, and
can be solved even by $\mathcal{O}(N)$ operations. A popular selection
algorithm is Quick-Select which has $\mathcal{O}(N)$ average complexity
but $\mathcal{O}(N^2)$ worst-case complexity. 
By using the \emph{median of medians} procedure \cite{Blum-JCSS73}
for selecting the so-called \emph{pivot} element in Quick-Select, its
worst-case complexity can be improved to $\mathcal{O}(N)$.

\runinhead{Weighted median.}
The median of a finite set $\mathcal{X}$ with weights 
$w:\mathcal{X}\to\mathbb{R}^+$ can be computed by sorting $\mathcal{X}$,
which takes $\mathcal{O}(N\log N)$ operations, 
followed by cumulating the weights
for the members of $\mathcal{X}$ in ascending order, which requires
$\mathcal{O}(N)$ effort, yielding an overall $\mathcal{O}(N\log N)$
complexity, both worst-case and average. In \cite{Rauh-icip10} an
algorithm with $\mathcal{O}(N)$ average complexity is proposed.

\section{Univariate Median Filtering of Images}
\label{sec-medf}

The definition of median filters for grey-value images and
some important properties of these are discussed in this section.

\subsection{Standard Discrete Median Filtering}

A discrete grey-value image $u$ is an array of intensities $u_{i,j}$
where $i=1,\ldots,N_x$ and $j=1,\ldots,N_y$ enumerate the locations 
(pixels) of a regular grid in horizontal and vertical direction, 
respectively.

The classical median filter \cite{Tukey-Book71}
processes a discrete grey-value image $u$ 
using a \emph{sliding window}. A sliding window can be described by
a set $S=\{(\Delta i_1,\Delta j_1),\ldots,(\Delta i_r,\Delta j_r)\}$
of integer pairs $(\Delta i_k,\Delta j_k)$; typically $(0,0)$ is one of
these pairs, and the others are centered around $(0,0)$, often in a 
symmetric way. For example, a $(2\ell+1)\times(2\ell+1)$ window is
given by $S=\{-\ell,\ldots,+\ell\}\times\{-\ell,\ldots,+\ell\}$, or an 
approximately disc-shaped window of radius $r$ by 
$S=\{(\Delta i,\Delta j)~|~i,j\in\mathbb{Z},i^2+j^2\le r^2\}$.
{\sloppy\par}

For each pixel $(i,j)$, the sliding window is shifted to this pixel
to select a set of pixels around $(i,j)$ and generate new (filtered)
pixel values from the selected pixels.
Using the notation 
$(i,j)+S:=\{(i+\Delta i,j+\Delta j)~|~(\Delta i,\Delta j)\in S\}$,
the median filter assigns the median of all $u_{i',j'}$ for
$(i',j')\in(i,j)+S$ to pixel $(i,j)$ in the median-filtered image $v$,
\begin{align}
v_{i,j} &:= \operatorname{med}\{u_{i',j'}~|~(i',j')\in(i,j)+S\}
\;.
\end{align}
For pixels near the boundary where $(i,j)+S$ contains locations outside
the image domain $\{1,\ldots,N_x\}\times\{1,\ldots,N_y\}$, a suitable
boundary treatment needs to be defined, which we do not discuss further
here.

It is important to notice that the filter consists of two steps: the
\emph{selection step} that extracts values from the input image
using the sliding window, and the \emph{aggregation step} that uses the
median to generate the new value from the extracted values. This is
the universal structure of local image filters; the specification of
selection and aggregation step defines the particular filter.

The entire process can be iterated by applying the median filter to the
median-filtered image, and so on. This is called \emph{iterative median
filtering.}

\runinhead{Properties.}
The median is known as a robust position measure \cite{Tukey-Book71}
in statistics, and bequeathes its robustness to the median filter for
images. In particular, the median filter is known for its denoising
capabilities even in the presence of some sorts of non-Gaussian
(heavy-tailed) noise like salt-and-pepper or uniform impulse noise.
At the same time, it can preserve sharp edges, since for pixels near
an edge, the sliding window contains some pixels from either side of
the edge, and the median will be on the side of the majority, thus
yielding an intensity clearly belonging to one side, and not creating
intermediate values.

When iterating median filtering, several phenomena are observed. On one
hand, small image structures are eliminated. Whereas filtered images
still show sharp edges, corners are progressively rounded.

On the other hand, non-trivial stationary patterns can appear
during the iterated filtering process, most prominent 
with small sliding windows such as $3\times3$ or $5\times5$, but not limited
to these.
These so-called \emph{root signals}
have been studied e.g.\ by Eckhardt \cite{Eckhardt-JMIV03}.
The existence of such patterns is inherent to the discrete median filtering
process, whereas their actual occurrence depends on the interaction
between small-scale image structures and sliding windows, and can be
considered as filtering artifacts related to the resolution limit of the
discrete image. Worst-case
examples are a checkerboard pattern (say, all pixels $(i,j)$ with $i+j$ 
even are black, and all others white) which is stationary under
whatever $(2\ell+1)\times(2\ell+1)$ median filter, and a stripe pattern
(say, all pixels $(i,j)$ with $i$ even are black, and all others white)
which is inverted in each iteration.

\subsection{Space-Continuous Median Filtering}
\label{ssec-meddf}

Whereas practical image processing always has to deal with discrete images, 
a continuous viewpoint is helpful for the theoretical understanding
of image processing. Image acquisition creates discrete images by sampling
spatially continuous physical quantities, and involves the choice of a 
sampling grid, which is, in essence, arbitrary. For image processing methods
to be meaningful, they need to be consistent with a continuous reality,
and depend as little as possible on sampling.

To support theoretical analysis of median filtering, it is therefore
of interest to consider a median filtering process for space-continuous
images. Thus, we interpret a (planar) grey-value image as a 
bounded function from a suitable function space over a region of 
$\mathbb{R}^2$, which should be compact and connected.
For a deeper analysis, several function spaces have been proposed,
such as functions of bounded variation (BV) or sums of the BV space
(to represent piecewise smooth images) and Sobolev spaces (to represent
noise and/or texture components), 
see e.g.\ \cite{Aujol-IJCV05,Aujol-JMIV05,Osher-MMS03,Yin-vlsm05} and the 
numerous references therein.
In this work, we will for simplicity assume that images are smooth
functions.

\begin{definition}[Smooth image]
\label{def-img}
Let $\varOmega\subset\mathbb{R}^2$ be compact and equal to the closure of
its interior.
Let $u:\varOmega\to\mathbb{R}$ be a bounded smooth function on
$\varOmega$. We call $u$ 
\emph{smooth image with domain $\varOmega$}.

We denote by $\boldsymbol{\nabla}u$ the gradient of $u$,
and by $\mathrm{D}^2u$ its Hessian, i.e.\ the symmetric
$2\times2$-tensor of its second-order derivatives.
We call $\boldsymbol{x}\in\varOmega$ \emph{regular point of $u$} 
if $\boldsymbol{\nabla}u(\boldsymbol{x})\ne\boldsymbol{0}$.
\end{definition}

\noindent
We will now give some definitions to formalise the two steps of a local
filter in a space-continuous setting. We start with the selection step.

\begin{definition}[Selector]
\label{def-sel}
Let $u$ be a smooth image with domain 
$\varOmega\subset\mathbb{R}^2$.
Let $S$ be a map that assigns to each location $\boldsymbol{x}\in\varOmega$
a closed neighbourhood $S(\boldsymbol{x})\subset\mathbb{R}^2$.
Then $S$ is called \emph{selector} for $u$.
\end{definition}

\noindent
Whereas Definition~\ref{def-sel} actually allows some more generality
(with regard to the subsequent section), 
the continuous median filter discussed in this section will strictly
follow the sliding-window idea by deriving
all $S(\boldsymbol{x})$ from just one set $S_0$ by shifting, 
$S(\boldsymbol{x})=\boldsymbol{x}+S_0:=
\{\boldsymbol{x}+\boldsymbol{y}\in\mathbb{R}^2~|~\boldsymbol{y}\in S_0\}$.

Varying the size of the neighbourhood for filtering will be an important
step in the theoretical analysis, thus we specify such a procedure by
a further definition.

\begin{definition}[Scaled selector]
\label{def-scalsel}
Let $\varSigma$ be a map that assigns to any smooth image $u$
and any positive $\varrho$ 
a selector $\varSigma_{u,\varrho}$ for $u$. 
Assume that for each $u$ and each location
$\boldsymbol{x}$ and for all $\varrho<\sigma$ the inclusion 
$\varSigma_{u,\varrho}(\boldsymbol{x})\subset 
\varSigma_{u,\sigma}(\boldsymbol{x})$
holds, and 
$\bigcap\nolimits_{\varrho>0}
\varSigma_{u,\varrho(\boldsymbol{x})}
=\{\boldsymbol{x}\}$. 
Then $\varSigma$ is called \emph{scaled selector for $u$.}
\end{definition}

\noindent
Our main example of a scaled selector in this section will be the family 
$\mathcal{D}=\{D_\varrho~|~\varrho>0\}$ of
discs with radius $\varrho$ given by 
$D_\varrho(\boldsymbol{x})=\{\boldsymbol{y}\in\mathbb{R}^2~|~
\lVert\boldsymbol{y-x}\rVert\le\varrho\}$.

Let us now turn to the second step of a local filter, aggregation.

\begin{definition}[Aggregator]
Let $\varGamma$ be a family of integrable densities
$\gamma:\mathbb{R}\to\mathbb{R}^+_0$. 
Let $A$ be a functional that maps densities $\gamma\in\varGamma$
to values $A(\gamma)\in\mathbb{R}$.
Then $A$ is called a \emph{$\varGamma$-aggregator on $\mathbb{R}$.}
\end{definition}

\noindent
In particular, the space-continuous median is an aggregator, where
$\varGamma$ can be either the regular densities mainly considered in
Section~\ref{ssec-medd} or the distributions $\mathcal{D}'(\mathbb{R})$.
Both steps are combined in the following definition.

\begin{definition}[Local filters]
\label{def-localfilter-2-1}
Let $u$ be a smooth image with domain $\varOmega$,
and $S$ a selector for $u$.
For each $\boldsymbol{x}\in\varOmega$, 
let $u{\upharpoonright} S(\boldsymbol{x})$
be the restriction of $u$ to $\varOmega\cap S(\boldsymbol{x})$,
i.e.\ $u{\upharpoonright} S(\boldsymbol{x}):
\varOmega\cap S(\boldsymbol{x})\to\mathbb{R}$,
$(u{\upharpoonright} S(\boldsymbol{x}))(\boldsymbol{y})
=u(\boldsymbol{y})$ for $\boldsymbol{y}\in\varOmega\cap S(\boldsymbol{x})$.

Let $A$ be a $\varGamma$-aggregator on $\mathbb{R}$.
If for each $\boldsymbol{x}\in\varOmega$, the density 
$\gamma(u{\upharpoonright}S(\boldsymbol{x})):
\mathbb{R}\to\mathbb{R}^+_0$ of the values of 
$u{\upharpoonright}S(\boldsymbol{x})$ belongs to $\varGamma$,
we define $A(u,S):\varOmega\to\mathbb{R}$ by
$A(u,S)(\boldsymbol{x}):=
A\bigl(\gamma(u{\upharpoonright} S(\boldsymbol{x}))\bigr)$ 
for all $\boldsymbol{x}\in\varOmega$
and call it \emph{$(A,S)$-filtering of $u$.}

If $\varSigma$ is a scaled selector for $u$, and the 
$(A,\varSigma_{u,\varrho})$-filtering
of $u$ exists for all $\varrho>0$, we call the family
of $(A,\varSigma_{u,\varrho})$-filterings the 
\emph{$(A,\varSigma)$-filtering of $u$.}
\end{definition}

\noindent
Continuous median filtering can then be described as 
$(\operatorname{med},S)$-filtering with a suitable selector.
For example, one step of $(\operatorname{med},D_\varrho)$-filtering assigns
to each location $\boldsymbol{x}$ as filtered value the median of the
grey-value density within a disc of radius $\varrho$ around this location.

In the following we will be interested in the family of such filters,
i.e.\ the (scaled) $(\operatorname{med},\mathcal{D})$-filtering.

\subsection{Infinitesimal Analysis of Median Filtering}
\label{ssec-meddf-pde}

When the neighbourhood radius $\varrho$ of a 
$(\operatorname{med},D_\varrho)$-filtering is reduced, the filtering result
$\operatorname{med}(u,D_\varrho)$ will differ less and less from $u$ itself.
However, in iterated filtering the reduced filtering effect of a single
step can be compensated by increasing the number of iterations by a suitable
scale. In the limit $\varrho\to0$, the number of filtering steps goes to
infinity whereas the effect of each step goes to zero. Instead of a sequence
of progressively filtered images, an image evolution with a continuous
time parameter results.

\begin{definition}[Image evolution]
Let $\varOmega\subset\mathbb{R}^2$ be as in Definition~\ref{def-img}, 
and $[0,T]\subset\mathbb{R}$ an interval.
The coordinates of $\Omega$ will be called \emph{spatial coordinates,}
the additional coordinate from $[0,T]$ \emph{time coordinate.}
Let $u:\varOmega\times[0,T]\to\mathbb{R}$ be a bounded smooth 
function. We call $u$ an 
\emph{image evolution with spatial domain $\varOmega$.}

We denote by $\boldsymbol{\nabla}u$ the spatial gradient of 
$u$, i.e.\ the vector of its first-order derivatives
w.r.t.\ the spatial coordinates,
and by $\mathrm{D}^2u$ its spatial Hessian, i.e.\ the
$2\times2$-tensor of its second-order derivatives w.r.t.\ the
spatial coordinates.

By $u(t^*)$ for $t^*\in[0,T]$ we denote the smooth image
with $u(t^*)(\boldsymbol{x}):=u(\boldsymbol{x},t^*)$
for all $\boldsymbol{x}\in\Omega$.
\end{definition}

\noindent
Time-continuous image evolutions can be described by partial differential
equations (PDEs). 
To capture the limit transition $\varrho\to0$ of an
iterated local filter, including the appropriate scaling of the iteration
count, we use the following definition.

\begin{definition}[Regular PDE limit]
\label{def-regpdelimit}
Let $\varSigma$ be a scaled selector in $\mathbb{R}^2$, 
and $A$ a $\varGamma$-aggregator on $\mathbb{R}$.

The second-order PDE 
$u_t=F(\boldsymbol{\nabla}u,\mathrm{D}^2u)$
is the \emph{regular PDE limit of $(A,\varSigma)$-filtering} with 
\emph{time scale $\tau(\varrho)$}
if for each smooth image $u$ the $(A,\varSigma)$-filtering
exists, and 
if there exists an $\varepsilon>0$ such that
for each regular point $\boldsymbol{x}$ of $u$
one has for $\varrho\to0$
\begin{align}
\frac{A(u,\varSigma_{u,\varrho})
(\boldsymbol{x})
-u(\boldsymbol{x})}{\tau(\varrho)}
- F\bigl(\boldsymbol{\nabla}u(\boldsymbol{x}),
\mathrm{D}^2u(\boldsymbol{x})\bigr) =
\mathcal{O}(\varrho^{\varepsilon}) \;.
\label{regpdelimit}
\end{align}
\end{definition}
 
\runinhead{PDE limit of median filtering.}
Guichard and Morel \cite{Guichard-sana97} proved that
space-continuous median filtering approximates for $\varrho\to0$
a well-known image filtering PDE evolution, namely 
\emph{(mean) curvature flow} \cite{Alvarez-SINUM92}.

With our definitions above, we can state their result as follows.

\begin{proposition}[\cite{Guichard-sana97}]
\label{prop-gm}
For a smooth image,
$(\operatorname{med},\mathcal{D})$-filtering with the median as aggregator
and the family of discs $\mathcal{D}$ as scaled selector
has the regular PDE limit with time scale $\tau(\varrho)=\varrho^2/6$
given by
\begin{align}
u_t &= \lVert\boldsymbol{\nabla}u\rVert\,\operatorname{div}\,
\left(\frac{\boldsymbol{\nabla}u}{\lVert\boldsymbol{\nabla}u\rVert}\right)
\;.
\label{mcm}
\end{align}
\end{proposition}

\noindent
{\sloppy
The PDE \eqref{mcm} can be written in various forms. We mention that
$\operatorname{div}\bigl(\boldsymbol{\nabla}u/\lVert\boldsymbol{\nabla}u\rVert
\bigr)$ gives at any regular image location $\boldsymbol{x}$ the curvature 
$\kappa(\boldsymbol{x})$ of the level line passing through $\boldsymbol{x}$.
Therefore \eqref{mcm} can be written as \cite[(2.25)]{Sapiro-Book01}
\begin{align}
u_t &= \kappa\,\lVert\boldsymbol{\nabla}u\rVert\;.
\end{align}
Moreover, if $\boldsymbol{\xi}$ denotes at each image location a unit vector 
perpendicular to the gradient, thus, in the direction of the local level line,
one can write the same equation as \cite[(2.26)]{Sapiro-Book01}
\begin{align}
u_t &= u_{\boldsymbol{\xi\xi}}\;,
\end{align}
i.e.\ a diffusion process in level line direction, which gives reason to the
names \emph{geometric diffusion} or \emph{geometric heat flow}
that are also often used.
\par}

As an application of the result in Proposition~\ref{prop-gm}, it has been
proposed that median filtering may be used as a
discretisation of mean curvature flow, see e.g.\ \cite{Merriman-tr92} which
anticipated the relation between the two processes, and \cite{Jalba-ismm09}.

\subsection{Combination with Adaptive Structuring Elements}

An interesting modification of median filtering is obtained by altering the
selection step to use image-adaptive neighbourhoods. Instead of a sliding
window that is shifted to each pixel, the window shape is determined
at each pixel individually. One approach to do this are the
\emph{morphological amoebas} introduced by Lerallut et al.\
\cite{Lerallut-ismm05,Lerallut-IVC07}. Here, the distance between neighbouring
pixels is measured by an \emph{amoeba distance} combining standard Euclidean
distance with contrast information, i.e.\
\begin{align}
d_\beta\bigl((i,j),(i',j')\bigr)&:=1+\beta\,\lvert u_{i,j}-u_{i',j'}\rvert
\label{damoeba-lerallut}
\end{align}
for two pixels $(i,j)$ and $(i',j')$ in the image $u$ which are either
horizontal or vertical neighbours, i.e.\ 
$\lvert i-i'\rvert+\lvert j-j'\rvert\le1$. 
This distance measure is extended to
non-adjacent pixels via shortest paths, and for each pixel $(i,j)$ the
amoeba $A_{\beta,\varrho}(i,j)$ of radius $\varrho$ is defined as the set of 
all pixels the amoeba distance of which to $(i,j)$ does not exceed $\varrho$.

An amoeba median filter is then obtained by 
\begin{align}
v_{i,j} &:= \operatorname{med}\{u_{i',j'}~|~(i',j')\in A_{\beta,\varrho}(i,j)
\} \;.
\end{align}

Retaining this basic procedure, the amoeba distance \eqref{damoeba-lerallut}
can be modified in several ways. Following \cite{Welk-JMIV11}, the 
$4$-neighbourhoods in which the amoeba distance is defined first can be 
extended to $8$-neighbourhoods for a better approximation of rotational
invariance. Moreover, the $L^1$ summation of spatial Euclidean distance and
contrast can be replaced with a Euclidean summation. Combining these
changes, one has \cite{Welk-JMIV11}
\begin{align}
d_\beta\bigl((i,j),(i',j')\bigr)&:=
\sqrt{(i-i')^2+(j-j')^2+\beta^2(u_{i,j}-u_{i',j'})^2}
\label{damoeba-welk}
\end{align}
for $\lvert i-i'\rvert\le1$, $\lvert j-j'\rvert\le1$, which is again extended
to non-adjacent pixels via shortest paths.

\runinhead{Continuous amoeba median filtering.}
In \cite{Welk-JMIV11}, a space-continuous formulation of amoeba median
filtering of planar grey-value images was proposed. 
Modelling the space-continuous selectors after
\eqref{damoeba-welk}, they are defined as
\begin{align}
A_{\beta,\varrho}(\boldsymbol{x})&:=
\{\boldsymbol{x}'\in\mathbb{R}^2~|~d_\beta(\boldsymbol{x},\boldsymbol{x}')
\le\varrho\}
\end{align}
where the space-continuous amoeba metric $d_\beta$ depends on the
smooth image $u:\mathbb{R}^2\supset\varOmega\to\mathbb{R}$ and is given by
\begin{align}
d_\beta(\boldsymbol{x},\boldsymbol{x}') &:=
\min\limits_{\boldsymbol{c}:[0,1]\to\mathbb{R}^2}\int\nolimits_0^1
\sqrt{
\lVert\boldsymbol{c}'(s)\rVert^2
+\beta^2\lvert(u\circ \boldsymbol{c})'(s)\rvert^2
}\,\mathrm{d}s \;,
\label{damoeba-welk-cont}
\end{align}
minimising over regular curves $\boldsymbol{c}:[0,1]\to\mathbb{R}^2$ with
$\boldsymbol{c}(0)=\boldsymbol{x}$, $\boldsymbol{c}(1)=\boldsymbol{x}'$.
The metric $d_\beta$ can immediately be interpreted as the distance on the
Riemannian surface 
$\{(\boldsymbol{x},\beta\,u(\boldsymbol{x}))~|~\boldsymbol{x}\in\varOmega\}
\subset\mathbb{R}^3$, i.e.\ the (vertically rescaled) \emph{graph} of the 
function $u$.

If $L^1$ summation between the spatial distance and contrast is preferred,
$d_\beta$ may be defined instead as
\begin{align}
d_\beta(\boldsymbol{x},\boldsymbol{x}') &:=
\min\limits_{\boldsymbol{c}:[0,1]\to\mathbb{R}^2}\int\nolimits_0^1
\lVert\boldsymbol{c}'(s)\rVert+\beta\,\lvert(u\circ \boldsymbol{c})'(s)\rvert
\,\mathrm{d}s
\label{damoeba-welk-cont1}
\end{align}
which corresponds to a different, Finslerian \cite{Bao-Book00}, metric on 
the image graph.

Obviously, the family $\mathcal{A}_{\beta,\varrho}:=
\{A_{\beta,\varrho}(\boldsymbol{x})~|~\boldsymbol{x}\in\varOmega\}$
is a selector, and 
$\mathcal{A}_\beta:=\{\mathcal{A}_{\beta,\varrho}~|~\varrho>0\}$ is a 
scaled selector for $u$ in the sense of our definitions from 
Section~\ref{ssec-meddf}.

\runinhead{PDE limit of continuous amoeba median filtering.}
In \cite{Welk-JMIV11}, the following result was proved which establishes
a relation between amoebas as a space-adaptive modification of median 
filtering, and a space-adaptive modification which
is fairly standard in the area of image filtering PDEs.

\begin{proposition}[{\cite[Section~3.2]{Welk-JMIV11}}]
\label{prop-meddfa-pde}
For a smooth image,
$(\operatorname{med},\mathcal{A}_\beta)$-filtering with the median as 
aggregator and the family of amoebas $\mathcal{A}_\beta$ derived from
the amoeba metric \eqref{damoeba-welk-cont} as scaled selector
has the regular PDE limit with time scale $\tau(\varrho)=\varrho^2/6$
given by
\begin{align}
u_t &= \lVert\boldsymbol{\nabla}u\rVert\,\operatorname{div}\,
\left(g\bigl(\lVert\boldsymbol{\nabla}u\rVert\bigr)\,
\frac{\boldsymbol{\nabla}u}{\lVert\boldsymbol{\nabla}u\rVert}\right)
\;.
\label{selfsnakes}
\end{align}
with the edge-detector function $g:\mathbb{R}^+_0\to\mathbb{R}^+_0$ 
defined by
\begin{align}
g(s)&=\frac1{1+\beta^2s^2} \;.
\label{g-pm-beta}
\end{align}
\end{proposition}

\noindent
The PDE \eqref{selfsnakes} describes the \emph{self-snakes} 
proposed in \cite{Sapiro-icip96a} as an image-shar\-pe\-ning evolution. 
It differs from (mean) curvature motion
\eqref{mcm} just by inserting a decreasing edge-detector function $g$ in the
divergence expression. 
The specific function $g$ from \eqref{g-pm-beta}
coincides with one (the most popular one)
of the diffusivity functions proposed for nonlinear
isotropic diffusion by Perona and Malik \cite{Perona-PAMI90}, with the
diffusivity parameter set to $1/\beta$. These functions were also proposed
for use in the self-snakes equation in \cite{Sapiro-icip96a}.

As shown further in \cite{Welk-JMIV11}, amoeba median filtering with
the amoeba metric \eqref{damoeba-welk-cont1} instead of 
\eqref{damoeba-welk-cont} yields the same PDE limit \eqref{selfsnakes}
but with a different edge-detector function $g$.

\section{Multivariate Median Concepts}
\label{sec-mvmed}

The simplicity of the univariate median and its properties as a
robust position measure in data sets make it attractive to define
similar concepts for multivariate ($\mathbb{R}^n$-valued) data.
Research interest in this question can be traced back to Hayford's
1902 work \cite{Hayford-JASA1902} where componentwise medians of
geographical coordinates were used to find an approximate ``center''
of the population of the United States. However, componentwise 
application of the univariate median was recognised soon as 
dependent on the choice of coordinate systems, inducing the quest for
proper multivariate concepts without such a dependency.

The first of these concepts was introduced in 1909 by Weber 
\cite{Weber-Book1909} in location theory; it was identified as a 
multivariate median in the 1920's and introduced to the statistics
literature \cite{Gini-Met29} where it was followed by numerous
papers 
over the subsequent decades 
(some of which rediscovered the concept)
that considered
theoretical properties 
\cite{Galvani-Met33,Gower-AS74,Haldane-Biomet48,Milasevic-AS87},
efficient numerical concepts
\cite{Austin-Met59,Seymour-Met70,Vardi-MP01,Weiszfeld-TMJ37}, 
and applications \cite{Brown-JRSSB83}. Deficiencies of this
concept, which is now known as $L^1$ median, caused the introduction
of alternative multivariate median concepts since the 1970s
\cite{Barnett-JRSSA76,Brown-JRSSB87,Chakraborty-PAMS96,
Hettmansperger-Biomet02,Liu-AS90,Nevalainen-CJS07,Oja-StPL83,Oja-JRSSB85,
Tukey-icm74}. 
For an overview see \cite{Small-ISR90}.

In the following we describe the $L^1$ median
and some of the more recent concepts that will be used in the further
course of this paper. With focus on their application for image and
shape analysis we will pay specific attention to continuous formulations
and equivariance properties.

\subsection{$L^1$ Median}

The $L^1$ median \cite{Weber-Book1909,Gini-Met29}
results from a straightforward generalisation of the
minimisation property \eqref{med-E} of the univariate median in
which absolute differences $\lvert \mu-x_i\rvert$ in $\mathbb{R}$ are
replaced by Euclidean distances 
$\lVert\boldsymbol{\mu}-\boldsymbol{x}_i\rVert$
in $\mathbb{R}^n$.

\runinhead{Discrete data.}
We start again by the case of finite multisets describing discrete data.

\begin{definition}[Discrete $L^1$ median]
Let a
finite multiset $\mathcal{X}=\{\boldsymbol{x}_1,\ldots,\boldsymbol{x}_N\}$
of data points $\boldsymbol{x}_i\in\mathbb{R}^n$ ($n\ge2$) be given.
The \emph{$L^1$ median} $\operatorname{med}_{L^1}(\mathcal{X})$ of
$\mathcal{X}$ is given by
\begin{align}
\operatorname{med}_{L^1}(\mathcal{X})&:=
\mathop{\operatorname{argmin}}\limits_{\boldsymbol{\mu}\in\mathbb{R}^n} 
E^{L^1}_{\mathcal{X}}(\boldsymbol{\mu})\;, &
E^{L^1}_{\mathcal{X}}(\boldsymbol{\mu})&:=
\sum\limits_{i=1}^N
\lVert \boldsymbol{\mu}-\boldsymbol{x}_i\rVert
\;.
\label{L1med-E}
\end{align}
\end{definition}

\noindent
The name $L^1$ median refers to the $L^1$ summation of the distances
$\lVert\boldsymbol{\mu}-\boldsymbol{x}_i\rVert$ in 
$E^{L^1}_{\mathcal{X}}(\boldsymbol{\mu})$.

Each of the individual distance functions
$d_i(\boldsymbol{\mu}):=\lVert\boldsymbol{\mu}-\boldsymbol{x}_i\rVert$
is convex in $\mathbb{R}^n$ but not strictly
convex because it is linear along each ray starting from $\boldsymbol{x}_i$.
As a consequence, if all data points $\boldsymbol{x}_i\in\mathcal{X}$ are 
situated on one straight line in $\mathbb{R}^n$ (an embedding of the affine 
space $\mathbb{R}$ into $\mathbb{R}^n$), the $L^1$ median is equivalent to
the univariate median on this line (via the embedding), and is therefore
non-unique (set-valued) if $n$ is even.
In all other cases, the objective function 
$E^{L^1}_{\mathcal{X}}(\boldsymbol{\mu})$
is strictly convex since on each given straight line at least one of the
$d_i$ is strictly convex, and the $L^1$ median therefore unique.
If all data points in $\mathcal{X}$ are located on a hyperplane, their
$L^1$ median obviously coincides with its $(n-1)$-dimensional counterpart.

\begin{figure}[t!]
\unitlength0.01\textwidth
\begin{picture}(100,73)
\put(0,41){
\put(0,0){
\put(0,33){\rotatebox{270}{\includegraphics[height=32\unitlength]
  {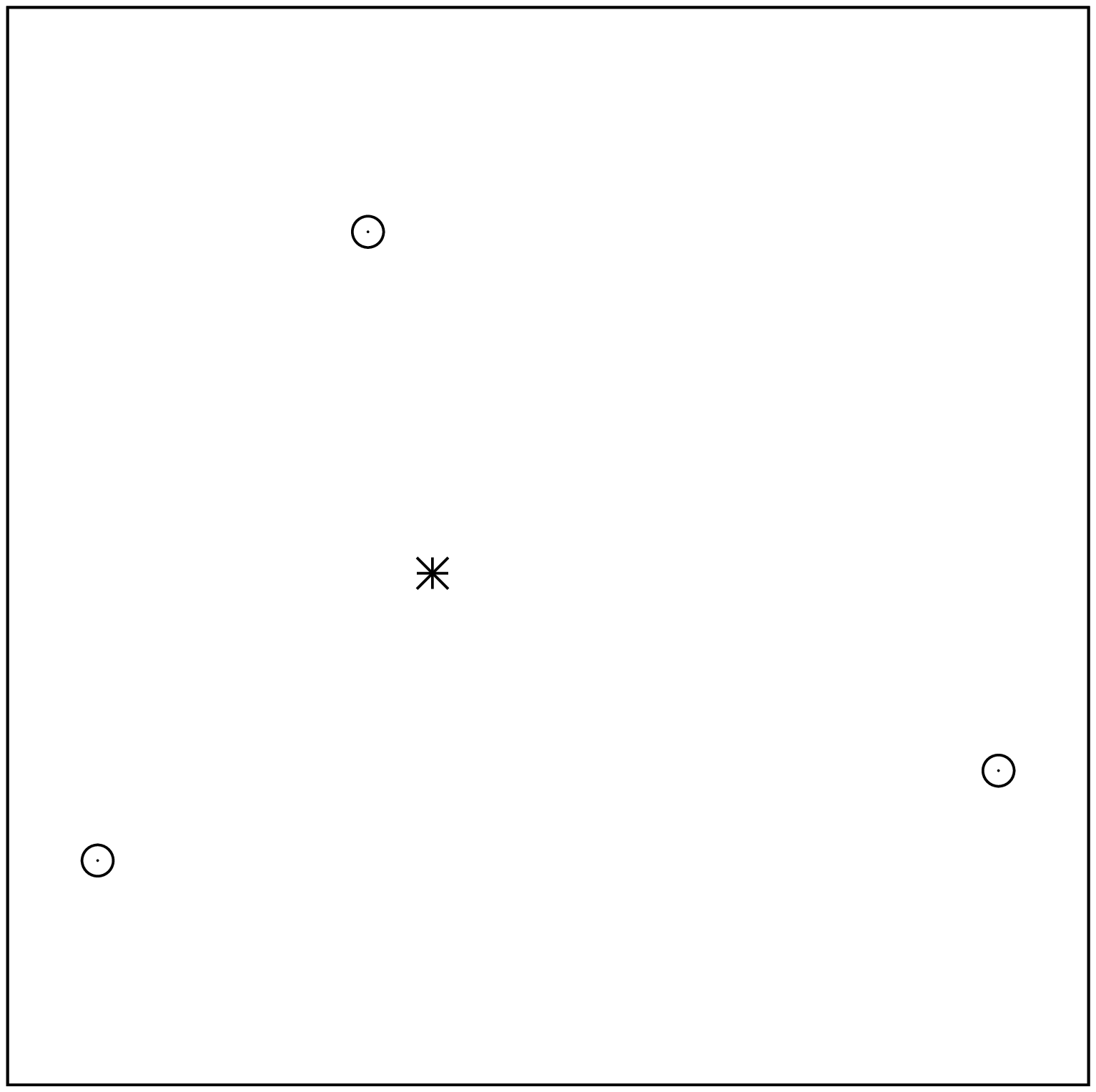}}}
\put(14,0){(a)}
}
\put(34,0){
\put(0,33){\rotatebox{270}{\includegraphics[height=32\unitlength]
  {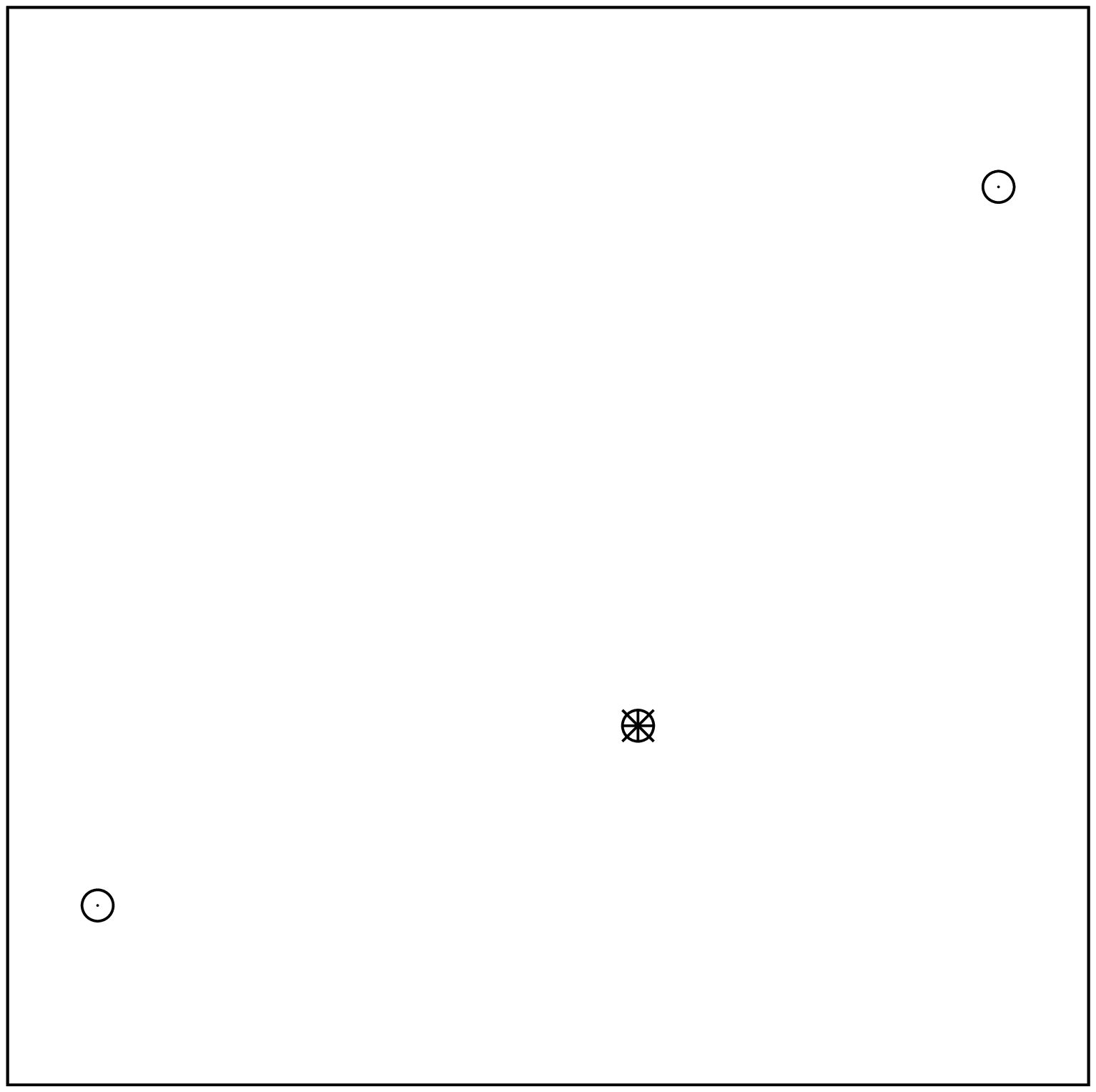}}}
\put(14,0){(b)}
}
\put(68,0){
\put(0,33){\rotatebox{270}{\includegraphics[height=32\unitlength]
  {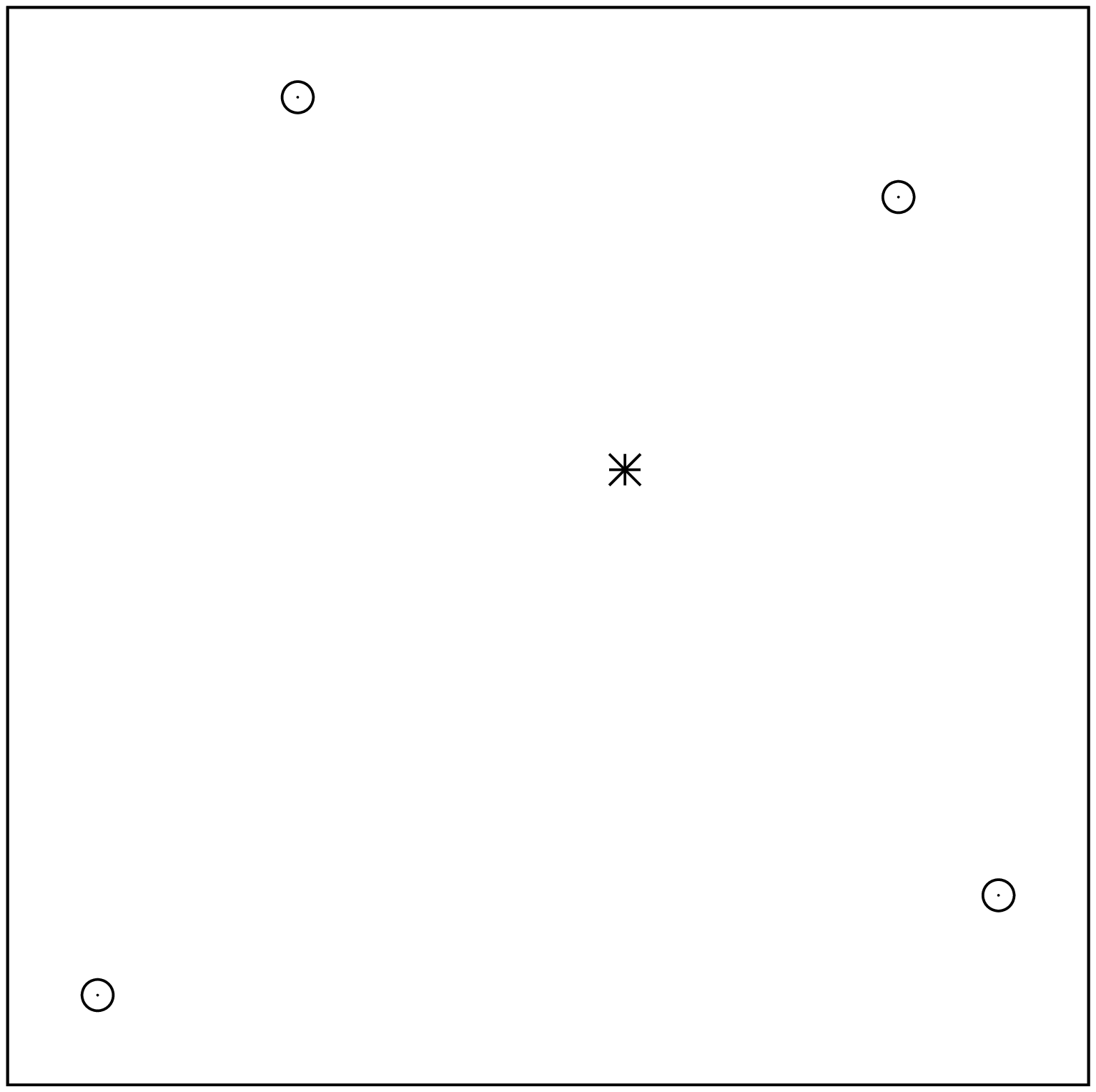}}}
\put(14,0){(c)}
}
}
\put(0,6){
\put(0,0){
\put(0,33){\rotatebox{270}{\includegraphics[height=32\unitlength]
  {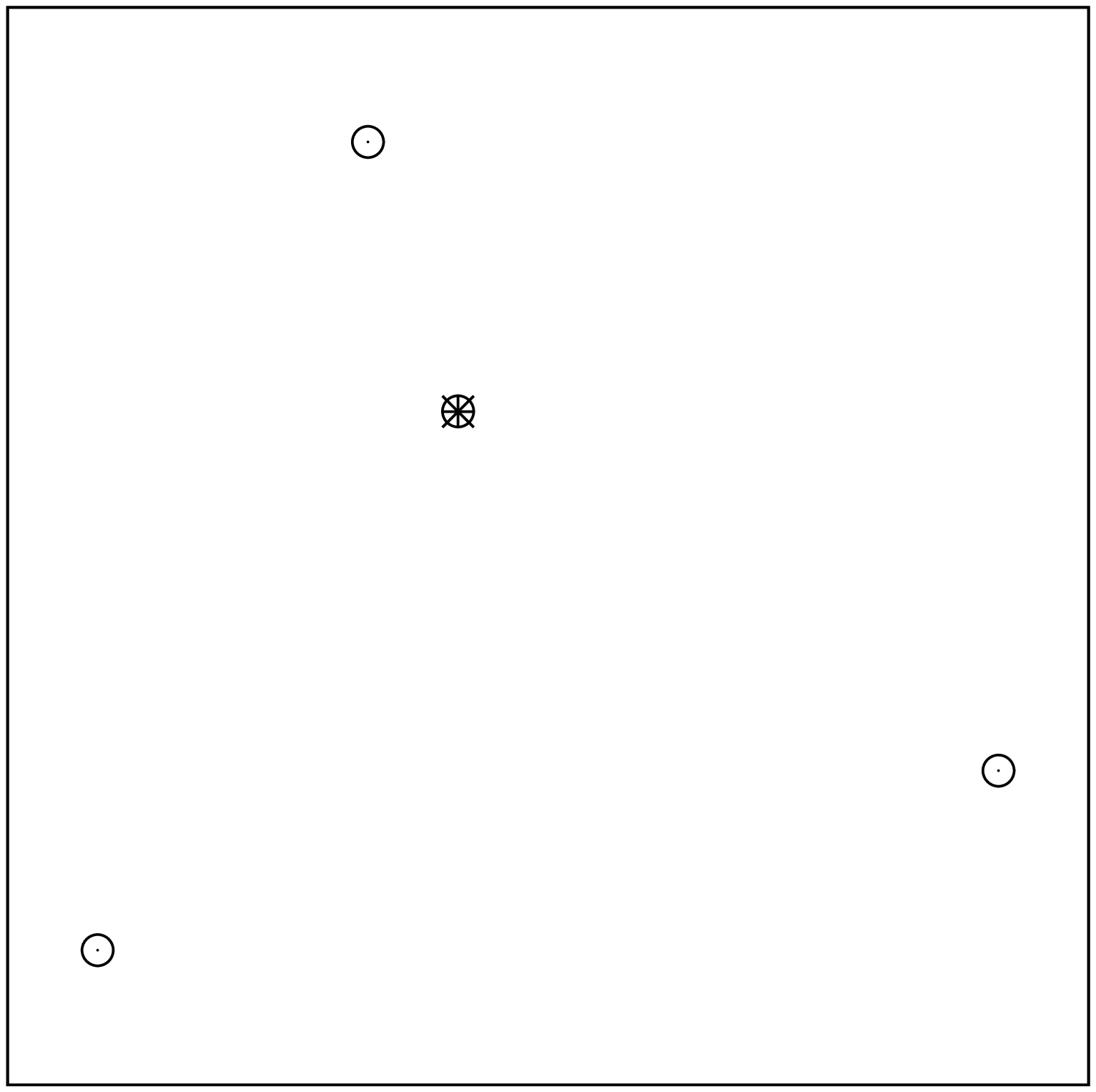}}}
\put(14,0){(d)}
}
\put(34,0){
\put(0,33){\rotatebox{270}{\includegraphics[height=32\unitlength]
  {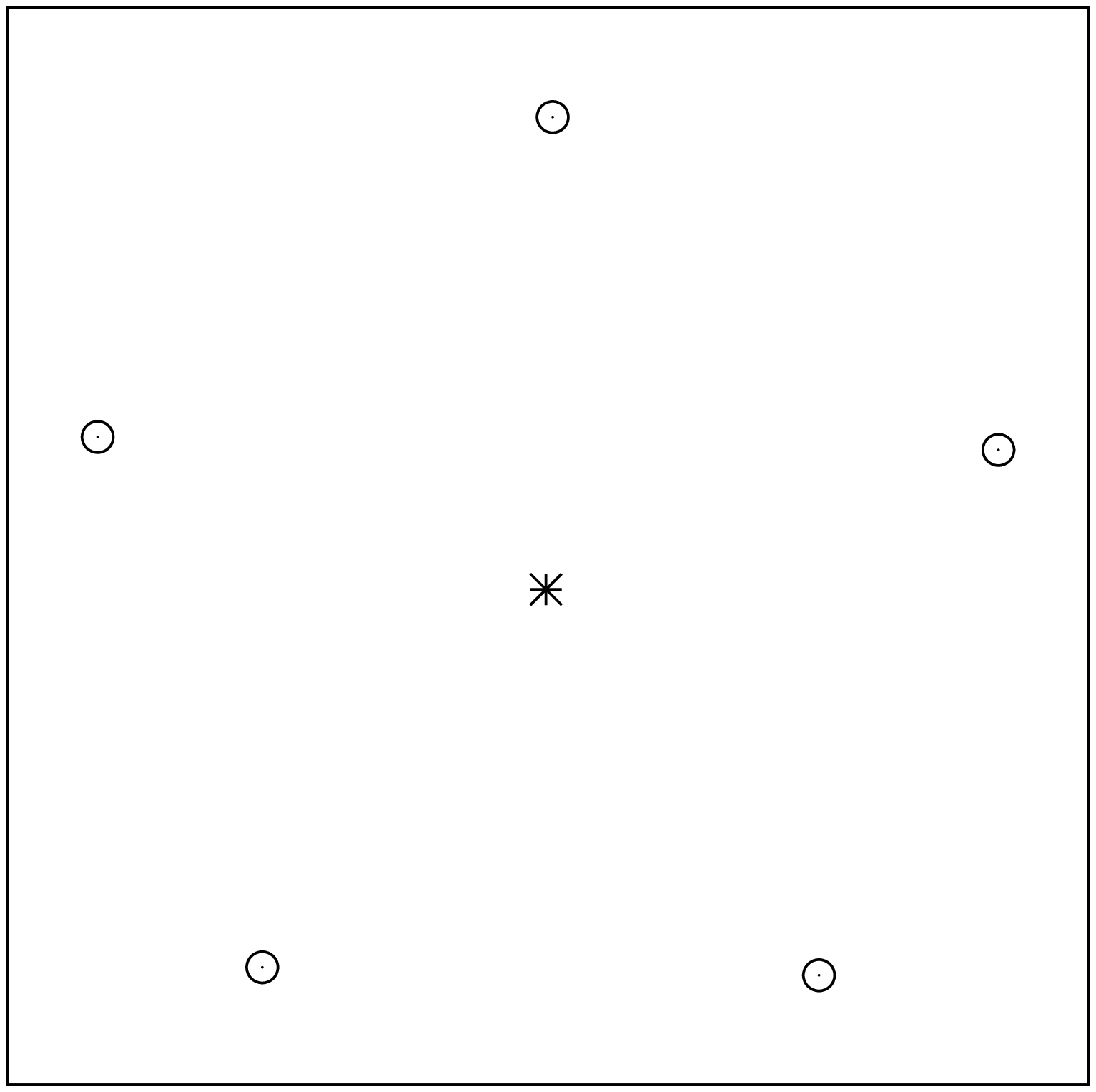}}}
\put(14,0){(e)}
}
\put(68,0){
\put(0,33){\rotatebox{270}{\includegraphics[height=32\unitlength]
  {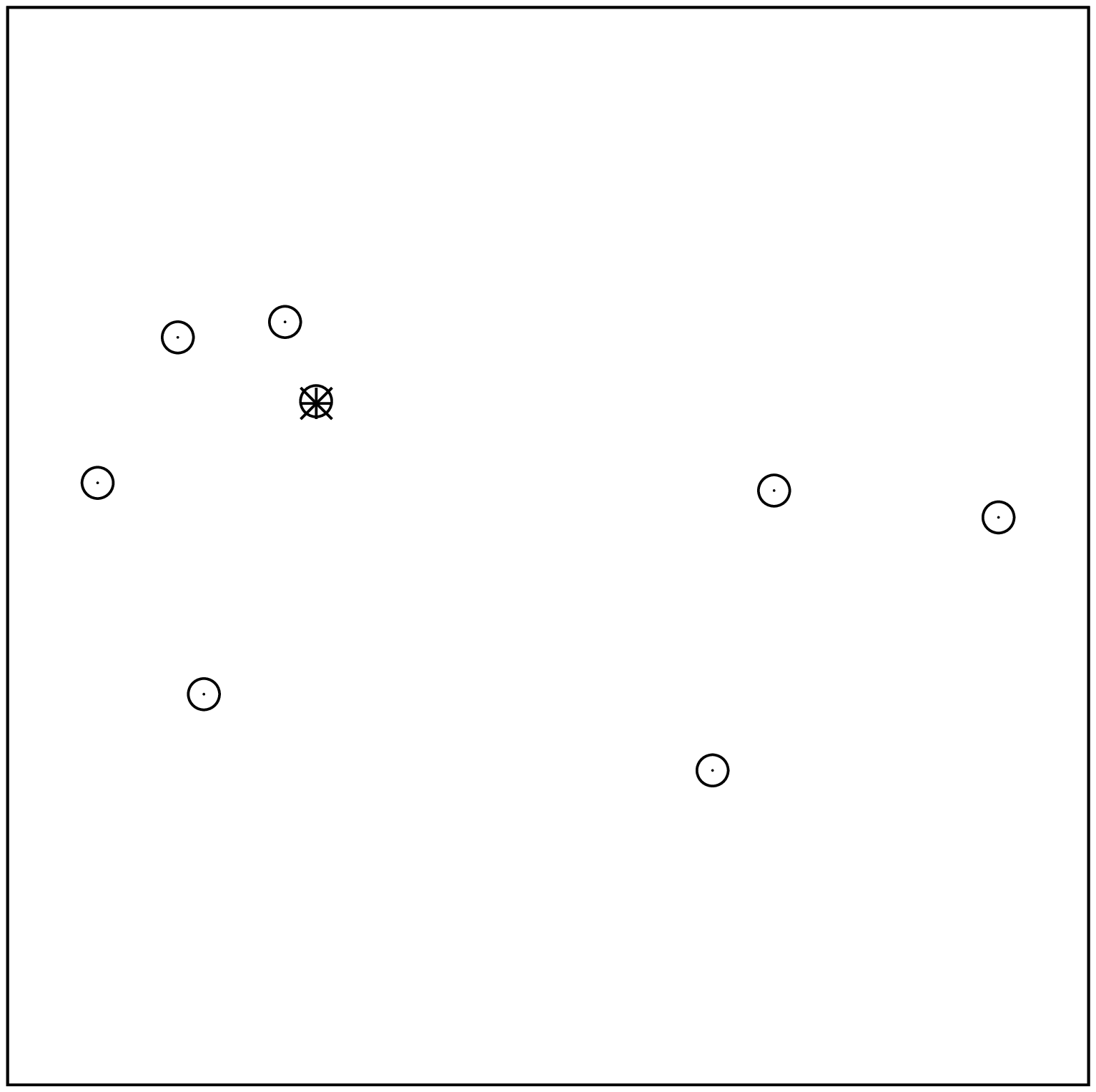}}}
\put(14,0){(f)}
}
}
\put(0,2){
\parbox[t]{100\unitlength}{\small\raggedright Legend:
\raisebox{2\unitlength}{\rotatebox{270}{\includegraphics[width=2\unitlength]{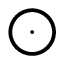}}}
Data points \quad
\raisebox{2\unitlength}{\rotatebox{270}{\includegraphics[width=2\unitlength]{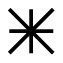}}}
$L^1$ median 
}
}
\end{picture}
\caption{Simple point sets with their $L^1$ medians. 
\textbf{(a)} In a triangle with no angle of $120$ degrees or greater, 
the $L^1$ median is the Fermat-Torricelli point from 
which all sides of the triangle are seen under $120$ degree angles. --
\textbf{(b)} In a triangle with one angle greater or equal $120$ degrees, the
obtuse corner is the $L^1$ median. -- 
\textbf{(c)} In a convex quadrilateral, the
intersection point of the diagonals is the $L^1$ median. --
\textbf{(d)} From four points with a triangle as convex hull, the inner data
point is the $L^1$ median. -- 
\textbf{(e)} Regular pentagon; the $L^1$ median is in the 
symmetry centre. -- 
\textbf{(f)} A configuration of $8$ points the $L^1$ median of
which is one of the data points.
}
\label{fig-l1med-examples}
\end{figure}

Unlike in the univariate case where $E_{\mathcal{X}}(\mu)$ is always 
minimised by at least one of the given data points, the $L^1$ median
will often be located somewhere inbetween, although 
in some generic cases the $L^1$ median still happens to be one of
the given data points. For example, for four points in the plane that
are corners of a convex quadrangle, their $L^1$ median is the intersection
point of its diagonals; but if from four points in the plane one is located
within the triangle defined by the others, this data point will be the
$L^1$ median.
Some simple point sets with their $L^1$ medians are depicted in 
Figure~\ref{fig-l1med-examples}.

Variants of the $L^1$ median concept are obtained if the Euclidean norm
$\lVert\,\cdot\,\rVert$ in the above definition
is replaced by other norms in $\mathbb{R}^n$. 
With the $L^1$ (city-block) metric, it boils down to the componentwise
median.
Another possibility would be the $L^\infty$ (maximum) norm. Dependent
on the choice of the norm, such an $L^1$ median may be non-unique in more
cases than with the Euclidean norm. We do not further discuss non-Euclidean
$L^1$ medians here. 

The generalisation of \eqref{L1med-E} to weighted countable sets is 
straightforward.

\runinhead{Continuous data.}
Similarly as in the univariate case, we formulate space-continuous versions
of multivariate median definitions for normalised regular densities.

\begin{definition}[Normalised regular density on $\mathbb{R}^n$]
Let $\gamma:\mathbb{R}^n\to\mathbb{R}^+_0$ be a compactly supported integrable
function with unit total weight,
$\int\nolimits_{\mathbb{R}^n}\gamma(\boldsymbol{x})\,\mathrm{d}
\boldsymbol{x}=1$. Then $\gamma$ is called \emph{normalised regular density}
on $\mathbb{R}^n$.
\end{definition}

\noindent
Generalising \eqref{medd-E}, we can define the $L^1$ median of such a density.

\begin{definition}[Continuous $L^1$ median]
Let $\gamma$ be a normalised regular density on $\mathbb{R}^n$. The 
\emph{$L^1$ median} of $\gamma$ is given by
\begin{align}
\operatorname{med}_{L^1}(\gamma)&:=
\mathop{\operatorname{argmin}}\limits_{\boldsymbol{\mu}\in\mathbb{R}^n} 
E^{L^1}_{\gamma}(\boldsymbol{\mu})\;, &
E^{L^1}_{\gamma}(\boldsymbol{\mu})&:=
\int\nolimits_{\mathbb{R}^n}
\gamma(\boldsymbol{x})\,\lVert \boldsymbol{\mu}-\boldsymbol{x}\rVert
\,\mathrm{d}\boldsymbol{x}
\;.
\label{L1medd-E}
\end{align}
\end{definition}

\noindent
A generalisation to distribution-valued densities $\gamma$ is again 
possible. We do not detail this here but remark that in the case if
$\gamma$ is concentrated on a straight line (i.e.\ essentially a single-layer
distribution along this line), also the continuous $L^1$ median reproduces
its univariate counterpart.

In \cite{Haldane-Biomet48} the $L^1$ median with Euclidean norm
$\lVert\,\cdot\,\rVert$ as defined above is called \emph{geometric median,}
whereas the componentwise median (thus, the $L^1$ median with city-block 
metric) is called \emph{arithmetic median.} 

\runinhead{Equivariance.}
By its definition via distances, the $L^1$ median (with Euclidean distances
as discussed above) is obviously equivariant w.r.t.\ Euclidean transformations
of the data. This equivariance is extended to similarity transformations
by noticing that scalings just entail scalings of the objective function
$E^{L^1}_{\mathcal{X}}$ or $E^{L^1}_{\mathcal{\gamma}}$, respectively. 
However, the $L^1$
median is not equivariant under any larger transformation group,
which is a substantial difference to the univariate median with its very
general monotonic equivariance.

As a consequence, the use of the $L^1$ median always relies on a Euclidean
structure in the data space. Wherever this structure is not naturally
given, it is implicitly imposed on the data and not justified. For example,
if the dimensions of the data space $\mathbb{R}^n$ actually refer to
incommensurable quantities (such as physical quantities measured by different
units of measurement), $L^1$ median processing is not sound from a modelling
perspective as its results depend on arbitrary choices of the relative scales
between these dimensions. The componentwise median ($L^1$ median with
city-block distances) would be a more plausible option in such a case,
see also \cite{Haldane-Biomet48}.

\runinhead{Algorithmic aspects.}
The basic idea of algorithms for the efficient calculation of $L^1$ 
medians is an iterative weighted means computation. 
In each step, an estimate for the median is improved by calculating
weights for the data points as their reciprocal distances to the estimate;
the weighted mean of the data points with these weights becomes the
next estimate. Special treatment is required if the estimate coincides
with one of the data points. 

The original version of this algorithm was
found by Weiszfeld in 1937 \cite{Weiszfeld-TMJ37}. 
Reformulations and variants of this procedure, with corrections to
the treatment of the special cases, are found in
\cite{Austin-Met59,Kuhn-JRS62,Kuhn-MP73,Ostresh-TS78,Ostresh-OR78,
Seymour-Met70}. 
A complete correct algorithm was published by Vardi and Zhang in 2000
\cite{Vardi-PNAS00,Vardi-MP01}. 

The algorithmic complexity of this procedure
is $\mathcal{O}(N)$ per iteration; thus the overall computational cost
depends on the number of iterations. 
Further algorithmic improvements are therefore
directed at increasing the convergence speed, see
e.g.\ \cite{Beck-JOT15,Drezner-AOR92}.

\subsection{Oja Median}
\label{ssec-ojamed}

The restrictive equivariance properties of the $L^1$ median motivated
researchers since the 1970s to consider alternative ways to generalise
the median concept to multivariate data.
In $\mathbb{R}^n$-valued
data, including such of incommensurable (physical) dimensions, affine
structure is in most cases justifiable from a modelling point of view, such
that equivariance under affine transformations is an appropriate choice
to overcome the main weakness of the $L^1$ median.

Despite not being the first concept of this kind (in fact, the half-space
and convex-hull-stripping medians discussed later in this section preceded
it), the \emph{simplex median} introduced by Oja \cite{Oja-StPL83,Oja-JRSSB85},
nowadays mostly termed \emph{Oja median,}
is particularly close to the $L^1$ concept discussed before as it generalises
the same minimisation property \eqref{med-E}, \eqref{medd-E} of the univariate
median but in a different way.

\runinhead{Discrete data.}
The essential observation underlying Oja's definition is that the distances
$\lvert\mu-x_i\rvert$ in \eqref{med-E} can be interpreted as interval lengths,
thus, one-dimensional simplex volumes.
Transferring this interpretation to
higher dimensions yields the following definition of a multivariate median.

\begin{definition}[Discrete Oja median]
Let $\mathcal{X}=\{\boldsymbol{x}_1,\ldots,\boldsymbol{x}_N\}$ be a finite
multiset of data points $\boldsymbol{x}_i\in\mathbb{R}^n$ ($n\ge2$).
The \emph{simplex median} or \emph{Oja median} of $\mathcal{X}$ is defined as
\begin{align}
\operatorname{med}_{\mathrm{Oja}}(\mathcal{X}) &:=
\mathop{\operatorname{argmin}}\limits_{\boldsymbol{\mu}\in\mathbb{R}^n}
E^{\mathrm{Oja}}_{\mathcal{X}}(\boldsymbol{\mu})\;, 
\notag\\*
E^{\mathrm{Oja}}_{\mathcal{X}}(\boldsymbol{\mu}) &:=
\sum\nolimits_{1\le i_1<\ldots<i_n\le N}
\bigl\lvert 
[ \boldsymbol{\mu},\boldsymbol{x}_{i_1},\ldots,\boldsymbol{x}_{i_n}
] \bigr\rvert
\;,
\label{ojamed-E}
\end{align}
where 
$[\boldsymbol{\mu},\boldsymbol{x}_{i_1},\ldots,\boldsymbol{x}_{i_n}]$
denotes the simplex spanned by the $n+1$ points 
$\boldsymbol{\mu},\boldsymbol{x}_{i_1},\ldots,\boldsymbol{x}_{i_n}\in
\mathbb{R}^n$ and $\bigl\lvert[\ldots]\bigr\rvert$ its volume.{\sloppy\par}
\end{definition}

\noindent
To generalise \eqref{ojamed-E} to weighted countable sets,
simplices must be weighted with the
product of the weights of all participating data points.
The following considerations are written for the unweighted case but
can be generalised to the weighted case.

In the bivariate case ($n=2$), the median $\boldsymbol{\mu}$ thus
minimises the area sum of all triangles 
$[\boldsymbol{\mu},\boldsymbol{x}_i,\boldsymbol{x}_j]$; for $n=3$ a sum
of tetrahedron volumes is minimised etc.

Unlike the $L^1$ median, the Oja median is not defined anymore if the
data in $\mathcal{X}$ lie on a hyperplane, since in this case all simplices
degenerate to zero volume for whatever $\boldsymbol{\mu}$ on the same
hyperplane. Still, if in such a case the data set $\mathcal{X}$ is enlarged
by adding to each data point of $\mathcal{X}$ a duplicate which is just
shifted by $\varepsilon\boldsymbol{v}$, with the same $\varepsilon>0$ for
all data points and $\boldsymbol{v}$ being a vector transversal to
the hyperplane containing $\mathcal{X}$ (such as its normal vector), i.e.\
\begin{align}
\mathcal{X}_{\varepsilon\boldsymbol{v}} := 
\mathcal{X} \cup \bigl (\mathcal{X}+\varepsilon\boldsymbol{v}\bigr)\;,
\label{oja-extenddegenerated}
\end{align}
then the Oja median of $\mathcal{X}_{\varepsilon\boldsymbol{v}}$ approaches 
for $\varepsilon\to0$ the $(n-1)$-dimensional Oja median in the hyperplane 
of $\mathcal{X}$.
In particular, the univariate median of data lying on a straight line is 
approximated from the $2$-dimensional Oja median by this procedure.

For data that span $\mathbb{R}^n$, the Oja median will still
sometimes be non-unique: For example, for three non-collinear points in the
plane, the entire triangle with the data points as vertices consists of
minimisers of $E^{\mathrm{Oja}}_{\mathcal{X}}$. 

As observed in \cite{Niinimaa-JRSSC92},
each of the simplex volume functions $v_{i_1,\ldots,i_n}:=
\left\lvert[\boldsymbol{\mu},\boldsymbol{x}_{i_1},\ldots,\boldsymbol{x}_{i_n}]
\right\rvert$ is a
convex function and consists of two linear regions separated by
the hyperplane $H_{i_1,\ldots,i_n}$ spanned by 
$\boldsymbol{x}_{i_1},\ldots,\boldsymbol{x}_{i_n}$,
where it has a kink.
Therefore, $E^{\mathrm{Oja}}_{\mathcal{X}}$ is also convex, and linear within
each of the regions into which $\mathbb{R}^n$ is split by 
the hyperplanes $H_{i_1,\ldots,i_n}$ for all $(i_1,\ldots,i_n)$.
As an interesting consequence there exists always at least one minimiser
of $E^{\mathrm{Oja}}_{\mathcal{X}}$ that is the intersection of $n$ such
hyperplanes. All minimisers with this property are vertices of a convex
polygon, and the entire set of minimisers is their convex hull, i.e.\ the
entire polygon.
This is analogous to the fact
that the univariate median set from \eqref{med-E} always contains at
least one data point and is the convex hull of all data points with
this property (at most two points in the univariate case).
{\sloppy\par}

\begin{figure}[t!]
\unitlength0.01\textwidth
\begin{picture}(100,73)
\put(0,41){
\put(0,0){
\put(0,33){\rotatebox{270}{\includegraphics[height=32\unitlength]
  {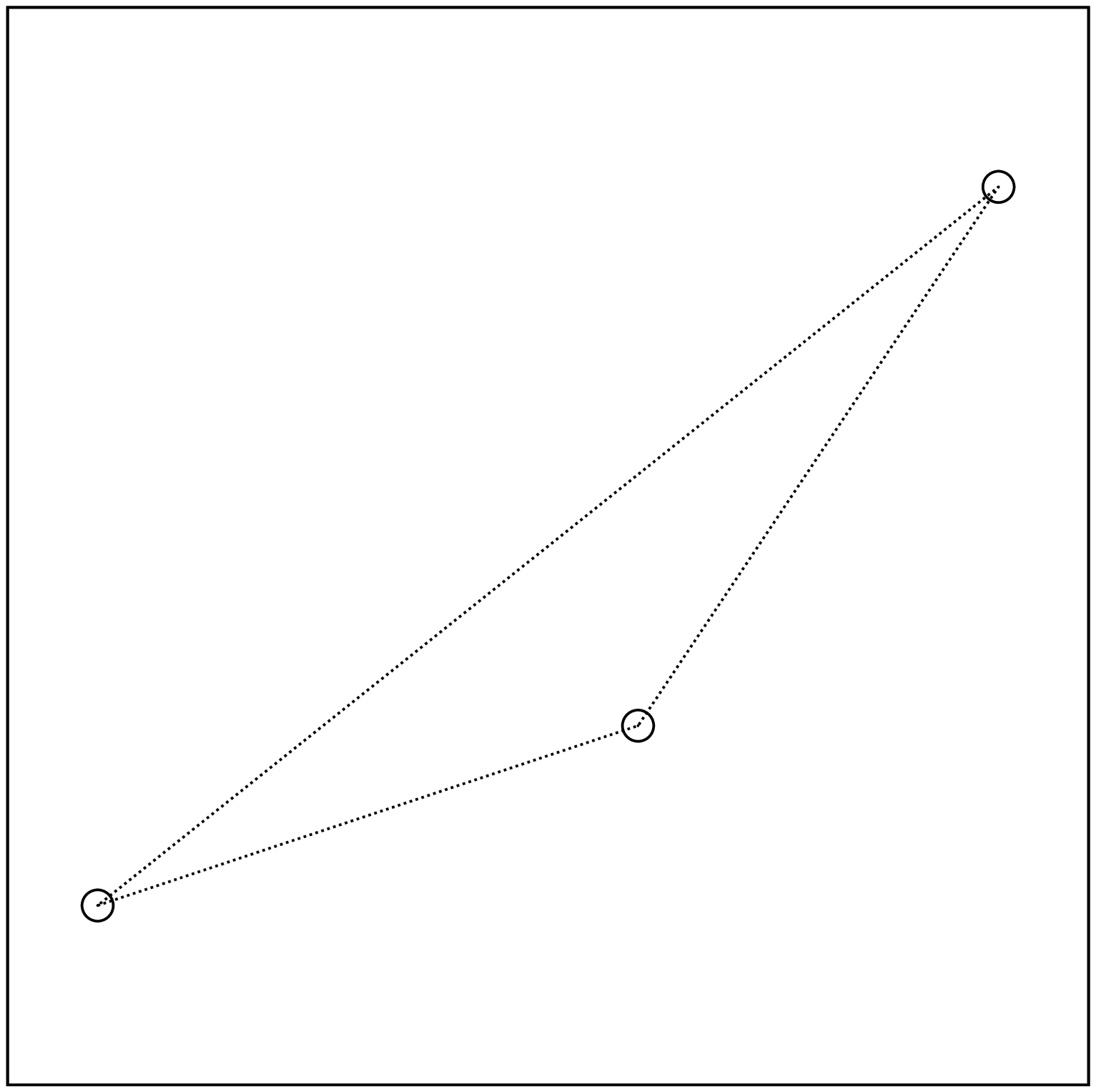}}}
\put(14,0){(a)}
}
\put(34,0){
\put(0,33){\rotatebox{270}{\includegraphics[height=32\unitlength]
  {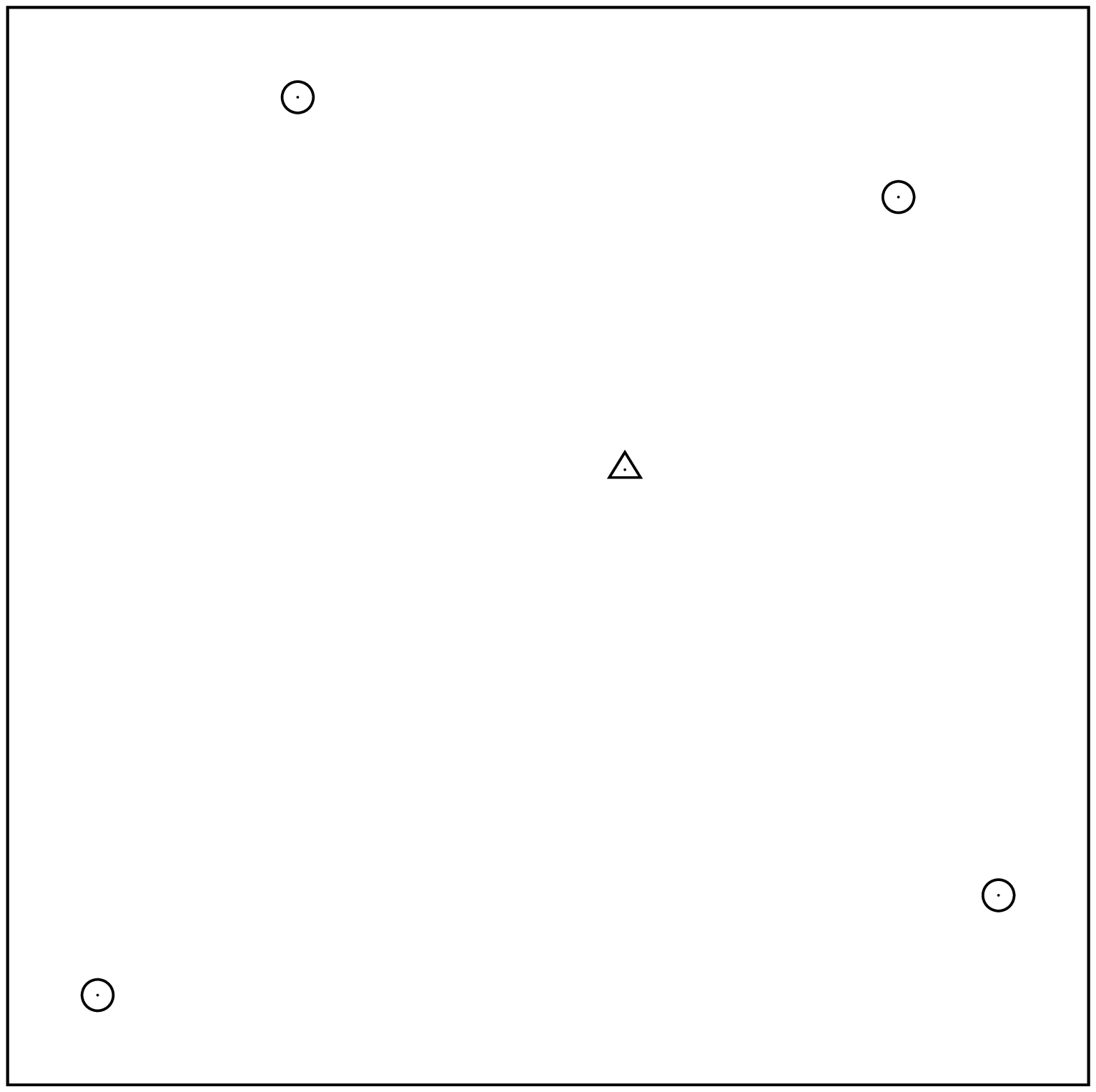}}}
\put(14,0){(b)}
}
\put(68,0){
\put(0,33){\rotatebox{270}{\includegraphics[height=32\unitlength]
  {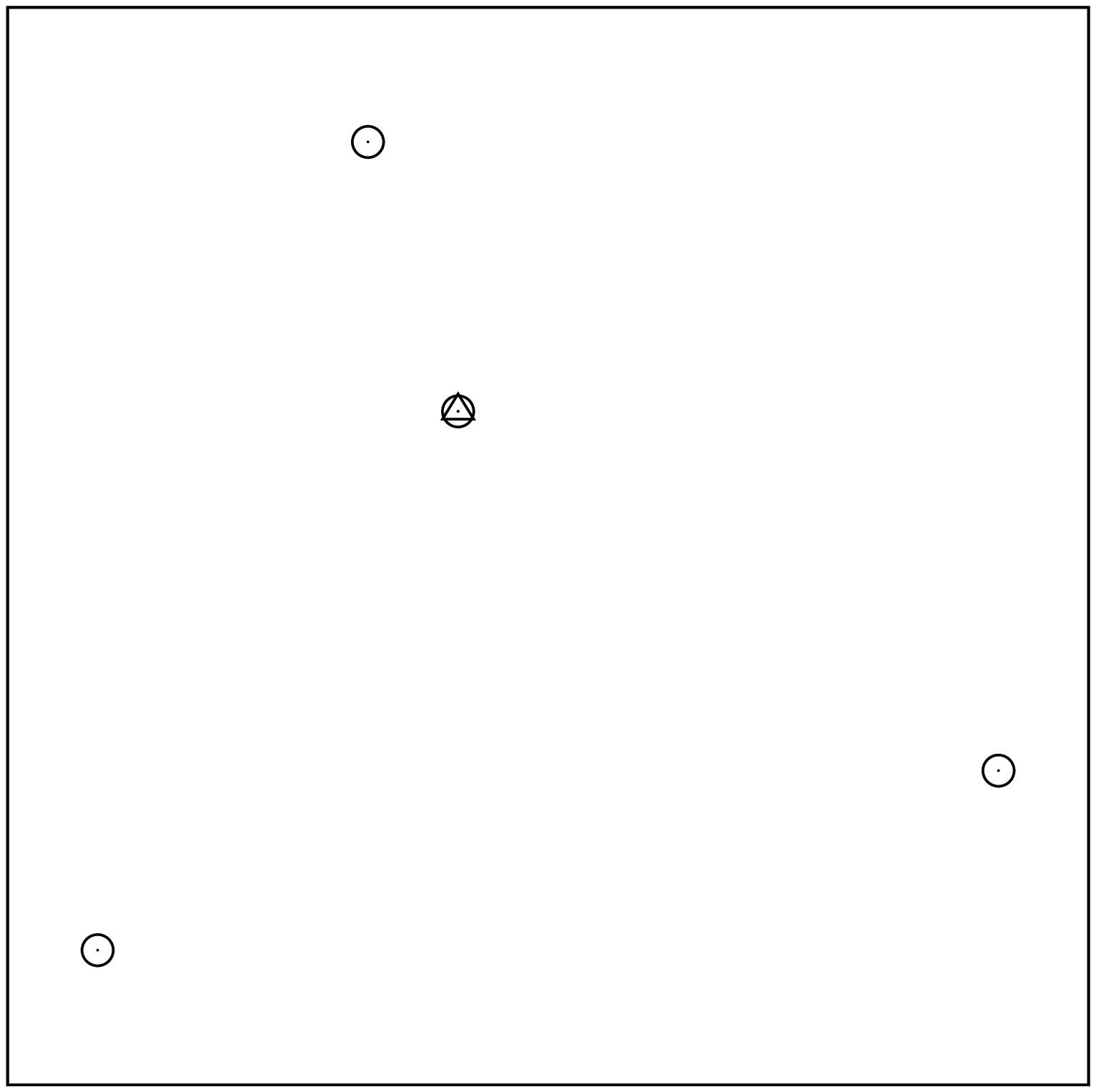}}}
\put(14,0){(c)}
}
}
\put(0,6){
\put(0,0){
\put(0,33){\rotatebox{270}{\includegraphics[height=32\unitlength]
  {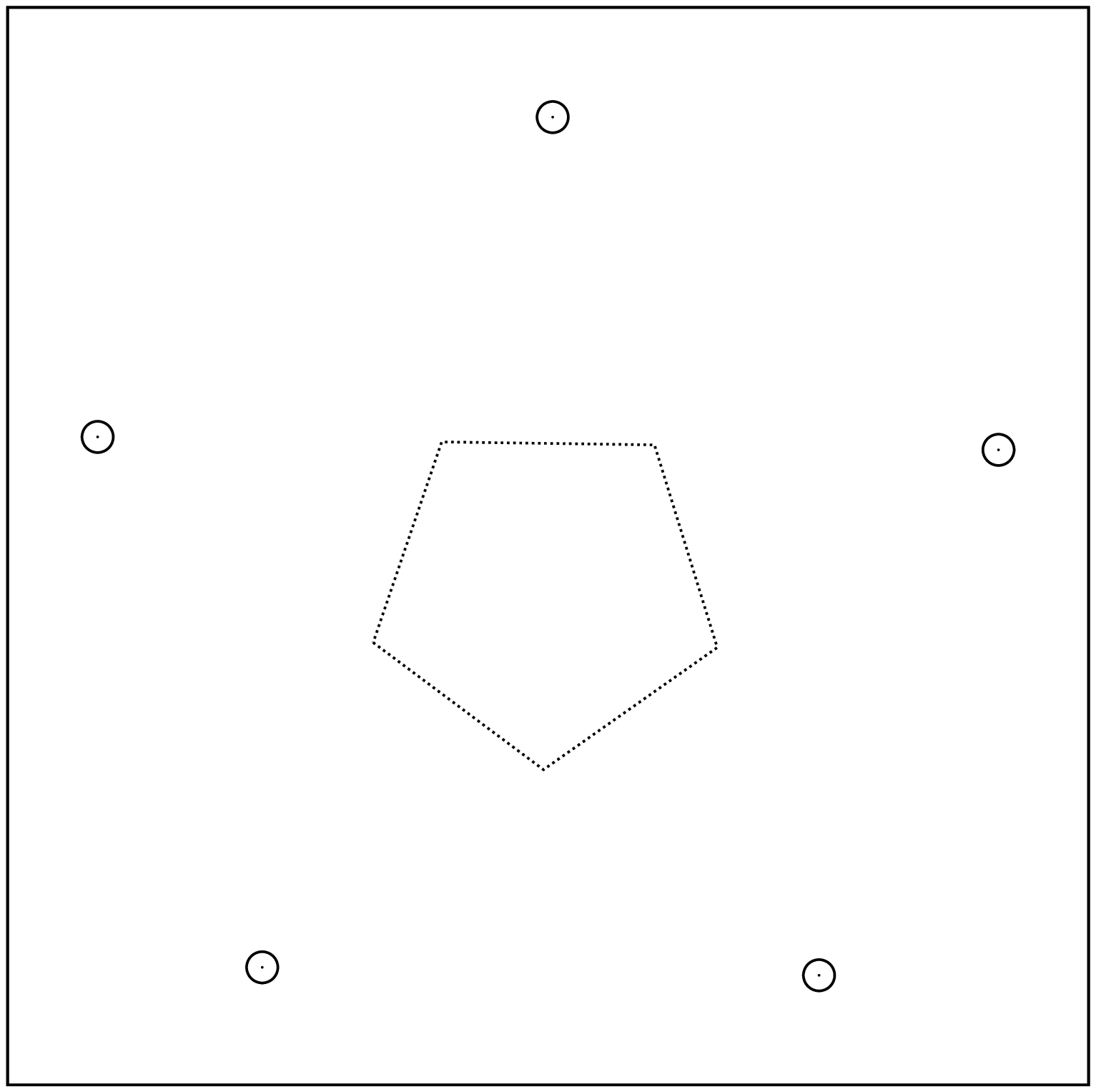}}}
\put(14,0){(d)}
}
\put(34,0){
\put(0,33){\rotatebox{270}{\includegraphics[height=32\unitlength]
  {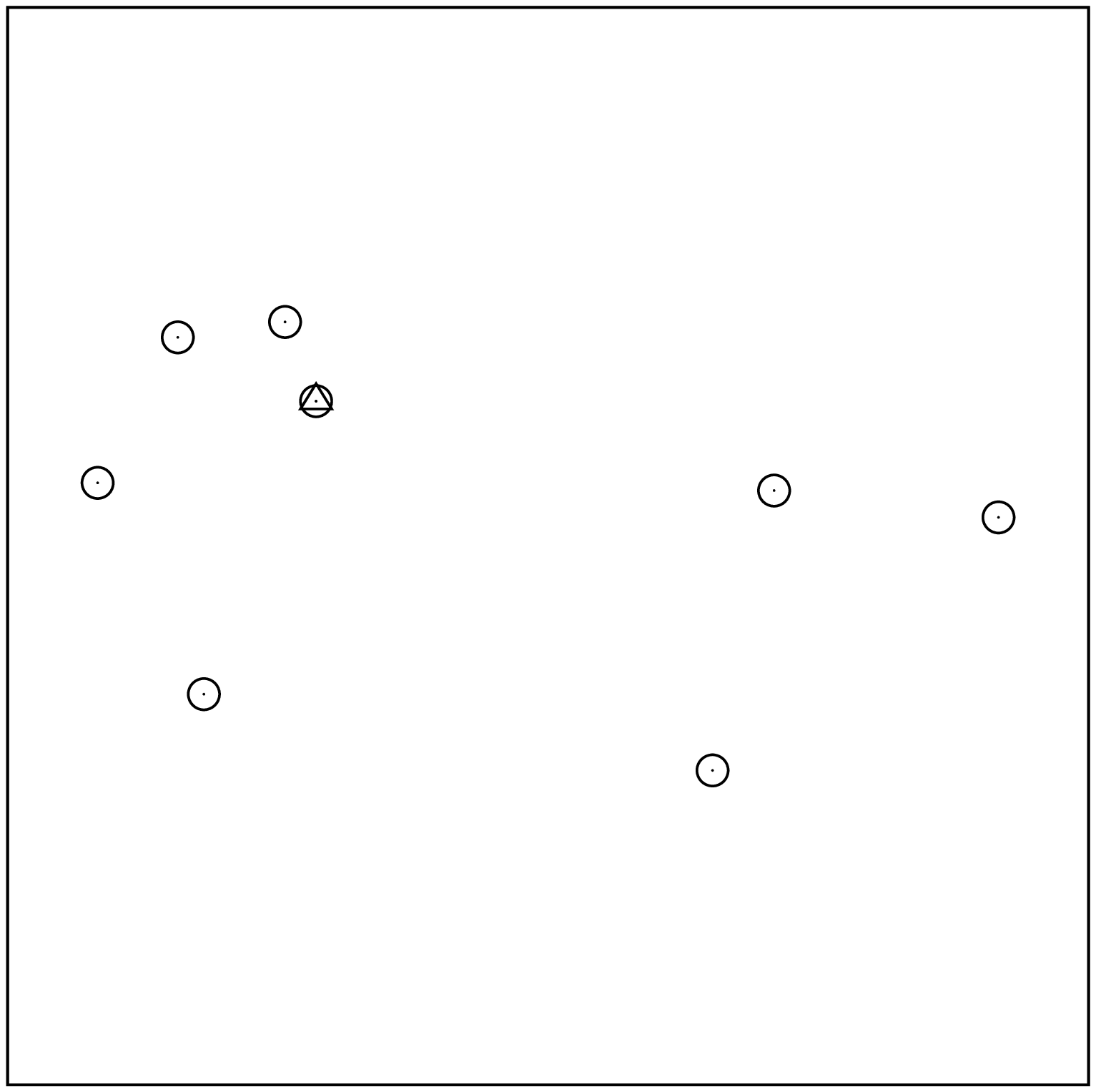}}}
\put(14,0){(e)}
}
\put(68,0){
\put(0,33){\rotatebox{270}{\includegraphics[height=32\unitlength]
  {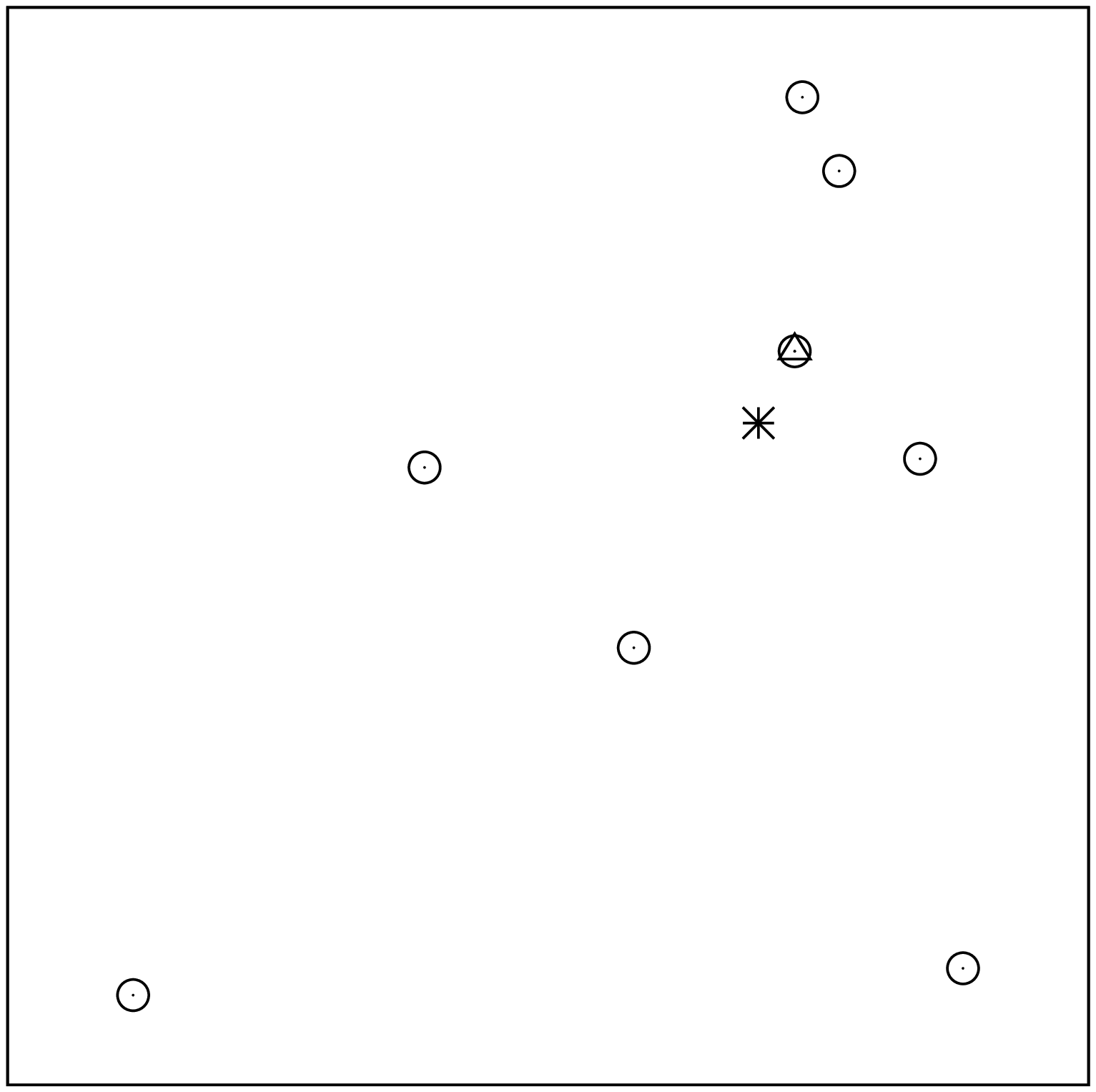}}}
\put(14,0){(f)}
}
}
\put(0,2){
\parbox[t]{100\unitlength}{\small\raggedright Legend:
\raisebox{2\unitlength}{\rotatebox{270}{\includegraphics[width=2\unitlength]{images/legend_datapt.eps}}}
Data points \quad
\raisebox{2\unitlength}{\rotatebox{270}{\includegraphics[width=2\unitlength]{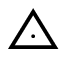}}}
\raisebox{2\unitlength}{\rotatebox{270}{\includegraphics[width=2\unitlength]{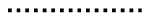}}}
Oja median (single point/outline)\quad
\raisebox{2\unitlength}{\rotatebox{270}{\includegraphics[width=2\unitlength]{images/legend_l1pt.eps}}}
$L^1$ median 
}
}
\end{picture}
\caption{Simple point sets with their Oja medians. 
\textbf{(a)} For three data 
points, the Oja median set is the entire triangle spanned by these points. --
\textbf{(b)} In a convex quadrilateral, the
intersection point of the diagonals is the Oja median. --
\textbf{(c)} From four points with a triangle as convex hull, the inner data 
point is the Oja median. -- 
\textbf{(d)} Regular pentagon; the intersection points of the
diagonals with the enclosed smaller pentagon form the Oja median set. --
\textbf{(e)} Same configuration of $8$ points as in 
Figure~\ref{fig-l1med-examples}(f).
The Oja median coincides with the $L^1$ median in one of the data points. --
\textbf{(f)} Another configuration of $8$ data points with Oja median.
In this case, the Oja median is one of the data points where the $L^1$ median
(also shown) is not.
}
\label{fig-ojamed-examples}
\end{figure}

In Figure~\ref{fig-ojamed-examples}, Oja medians of some simple point
configurations are shown.

\runinhead{Continuous data.}
The definition of the Oja median transfers in an obvious way to the continuous
setting.

\begin{definition}[Continuous Oja median]
Let $\gamma$ be a normalised regular density on $\mathbb{R}^n$. The
\emph{Oja median} of $\gamma$ is defined by
\begin{align}
\operatorname{med}_{\mathrm{Oja}}(\gamma) &:=
\mathop{\operatorname{argmin}}\limits_{\boldsymbol{\mu}\in\mathbb{R}^n}
E^{\mathrm{Oja}}_{\gamma}(\boldsymbol{\mu})\;, 
\notag\\*
E^{\mathrm{Oja}}_{\gamma}(\boldsymbol{\mu}) &:=
\int\nolimits_{\mathbb{R}^n}\ldots\int\nolimits_{\mathbb{R}^n}
\bigl\lvert 
[ \boldsymbol{\mu},\boldsymbol{x}_{1},\ldots,\boldsymbol{x}_{n}
] \bigr\rvert
\,\mathrm{d}\boldsymbol{x}_1\ldots\mathrm{d}\boldsymbol{x}_n
\;.
\label{ojamedd-E}
\end{align}
\end{definition}

\runinhead{Equivariance.}
Affine equivariance of the Oja median is guaranteed by its construction.

\runinhead{Algorithmic aspects.}
The computational complexity of the Oja median computation increases
with dimension: Already an evaluation of the objective function
$E^{\mathrm{Oja}}_{\mathcal{X}}(\boldsymbol{\mu})$ takes 
$\mathcal{O}(N^n)$ operations. Building on the geometric property 
mentioned above, \cite{Niinimaa-JRSSC92,Ronkainen-drs03} develop
algorithms that allow to compute the exact Oja median of $N$ data points in
$n$ dimensions with $\mathcal{O}(nN^n\log N)$, or an approximation
with stochastic accuracy guarantee in $\mathcal{O}(5^n/\varepsilon^2)$
where $\varepsilon$ is a confidence radius in $L^\infty$ sense.
An $\mathcal{O}(N\log^3N)$ algorithm for the bivariate Oja median is
stated in \cite{Aloupis-cccg01,Aloupis-CG03}.

\subsection{Transformation--Retransformation $L^1$ Median}

As an alternative to the computational expensive Oja median,
Chakraborty and Chaudhuri \cite{Chakraborty-PAMS96} proposed 
a standardisation procedure that allows to combine the computational
efficiency of the $L^1$ median with affine equivariance. Their idea
was to standardise a given data set by an affine transformation to a
configuration that is uniquely determined up to Euclidean transformations.

The resulting affine equivariant median was
further analysed in \cite{Hettmansperger-Biomet02}. The standardisation
procedure was discussed in a broader context of equivariance concepts
in \cite{Serfling-JNS10}.

\runinhead{Discrete data.}
Applying the affine normalisation procedure to finite multisets in
$\mathbb{R}^n$ ($n\ge2$) yields the following definition.

\begin{definition}[Discrete transformation--retransformation $L^1$ median]
Let $\mathcal{X}=\{\boldsymbol{x}_1,\ldots,\boldsymbol{x}_N\}$ be a finite
multiset of data points $\boldsymbol{x}_i\in\mathbb{R}^n$ ($n\ge2$).
The \emph{transformation--retransformation $L^1$ median} of $\mathcal{X}$ is
\begin{align}
\operatorname{med}_{\mathrm{TR}L^1}(\mathcal{X}) &:=
T^{-1}\operatorname{med}_{L^1}\bigl(T(\mathcal{X})\bigr)\;,
\end{align}
where $T$ denotes an affine transformation, and $T(\mathcal{X})$ its
element-wise application to $\mathcal{X}$, that is chosen such that
$T(\mathcal{X})$ has a unit covariance matrix $\boldsymbol{I}$. 
\end{definition}

\noindent
If the symmetric positive semidefinite covariance matrix
\begin{align}
C(\mathcal{X}) &:= 
\sum\nolimits_{i=1}^N\sum\nolimits_{j=1}^N 
(\boldsymbol{x}_i-\boldsymbol{x}_j)
(\boldsymbol{x}_i-\boldsymbol{x}_j)^{\mathrm{T}} \;,
\end{align}
is regular (thus, positive definite),
it is sufficient to choose $T:=C(\mathcal{X})^{-1/2}$
because one has then
\begin{align}
C\bigl(T(\mathcal{X})\bigr) &=
\sum\nolimits_i\sum\nolimits_j
(T\boldsymbol{x}_i-T\boldsymbol{x}_j)
(T\boldsymbol{x}_i-T\boldsymbol{x}_j)^\mathrm{T}
\notag\\*
&=T\,C(\mathcal{X})\,T^\mathrm{T} 
= C(\mathcal{X})^{-1/2}C(\mathcal{X})C(\mathcal{X})^{-1/2}
=\boldsymbol{I}\;.
\end{align}
It is obvious that the transformation--retransformation $L^1$ median
is undefined if all $\boldsymbol{x}_i$ lie on the same hyperplane,
as the covariance matrix is singular in this case. (The same limiting
procedure as for the Oja median may be used to recover from this problem.)
In all other cases, the covariance matrix is positive definite,
and the transformation--retransformation $L^1$ median is uniquely defined, 
based on the same property of the standard $L^1$ median.

\begin{figure}[t!]
\unitlength0.01\textwidth
\begin{picture}(100,38)
\put(0,6){
\put(0,0){
\put(0,33){\rotatebox{270}{\includegraphics[height=32\unitlength]
  {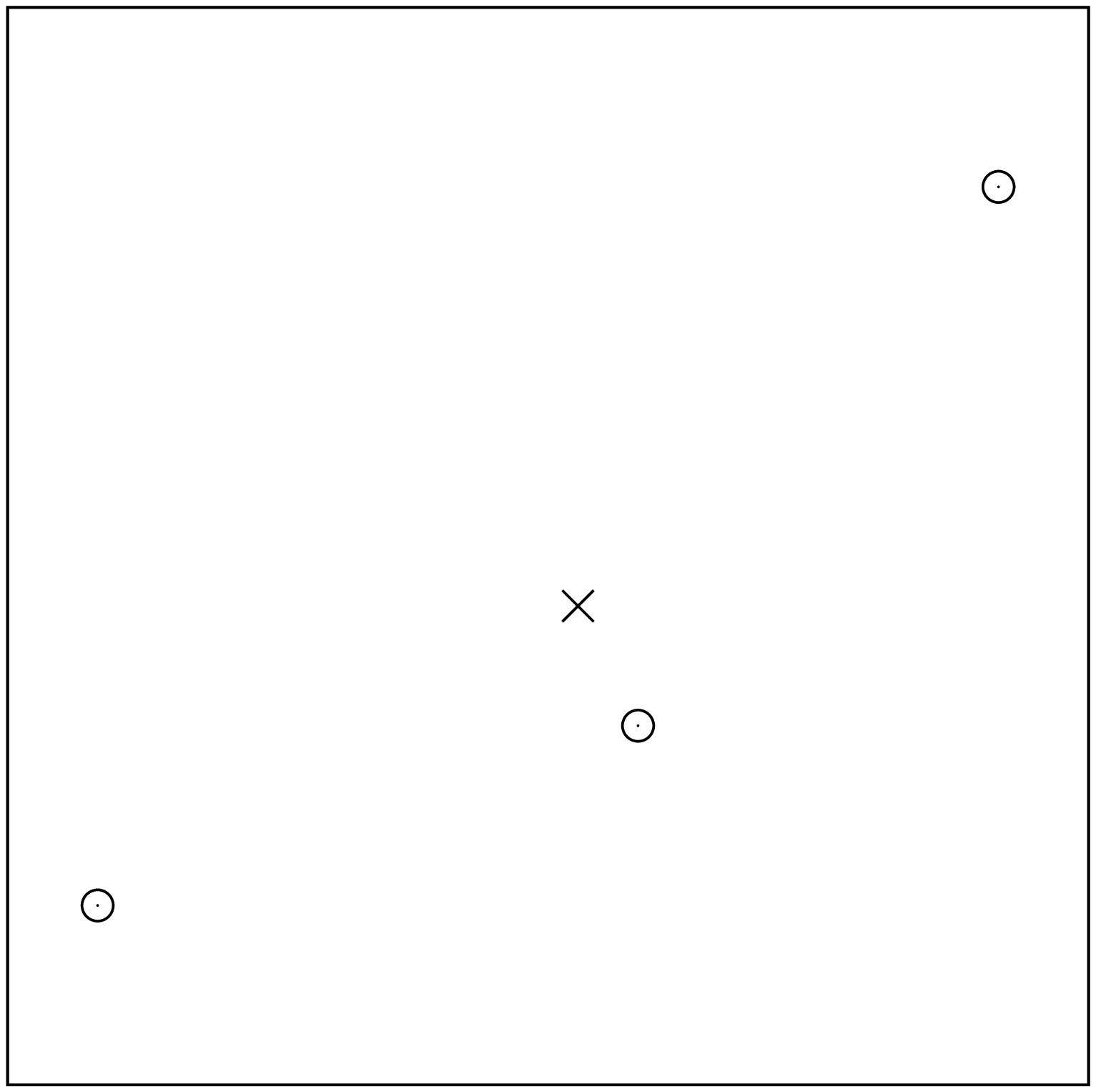}}}
\put(14,0){(a)}
}
\put(34,0){
\put(0,33){\rotatebox{270}{\includegraphics[height=32\unitlength]
  {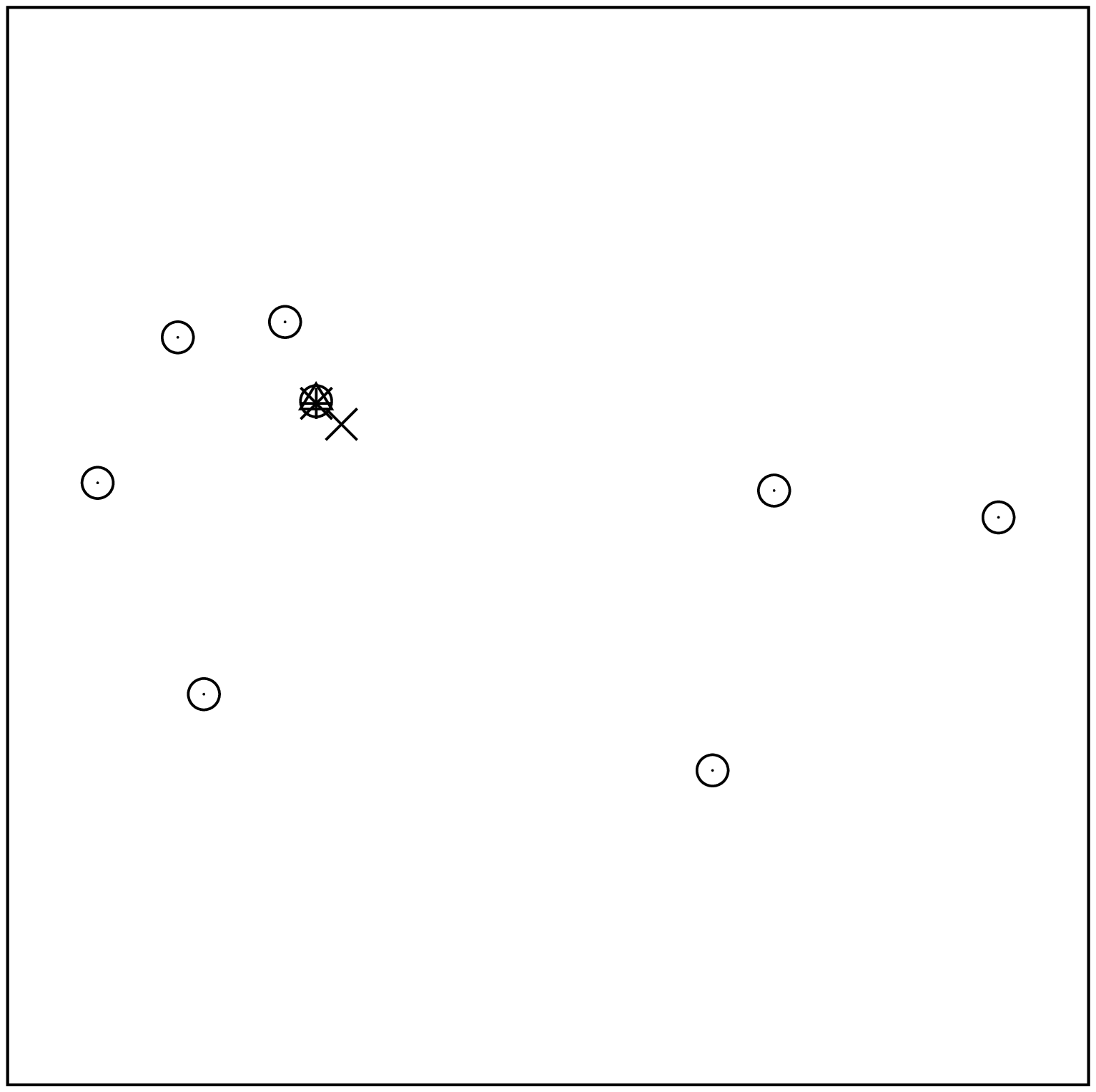}}}
\put(14,0){(b)}
}
\put(68,0){
\put(0,33){\rotatebox{270}{\includegraphics[height=32\unitlength]
  {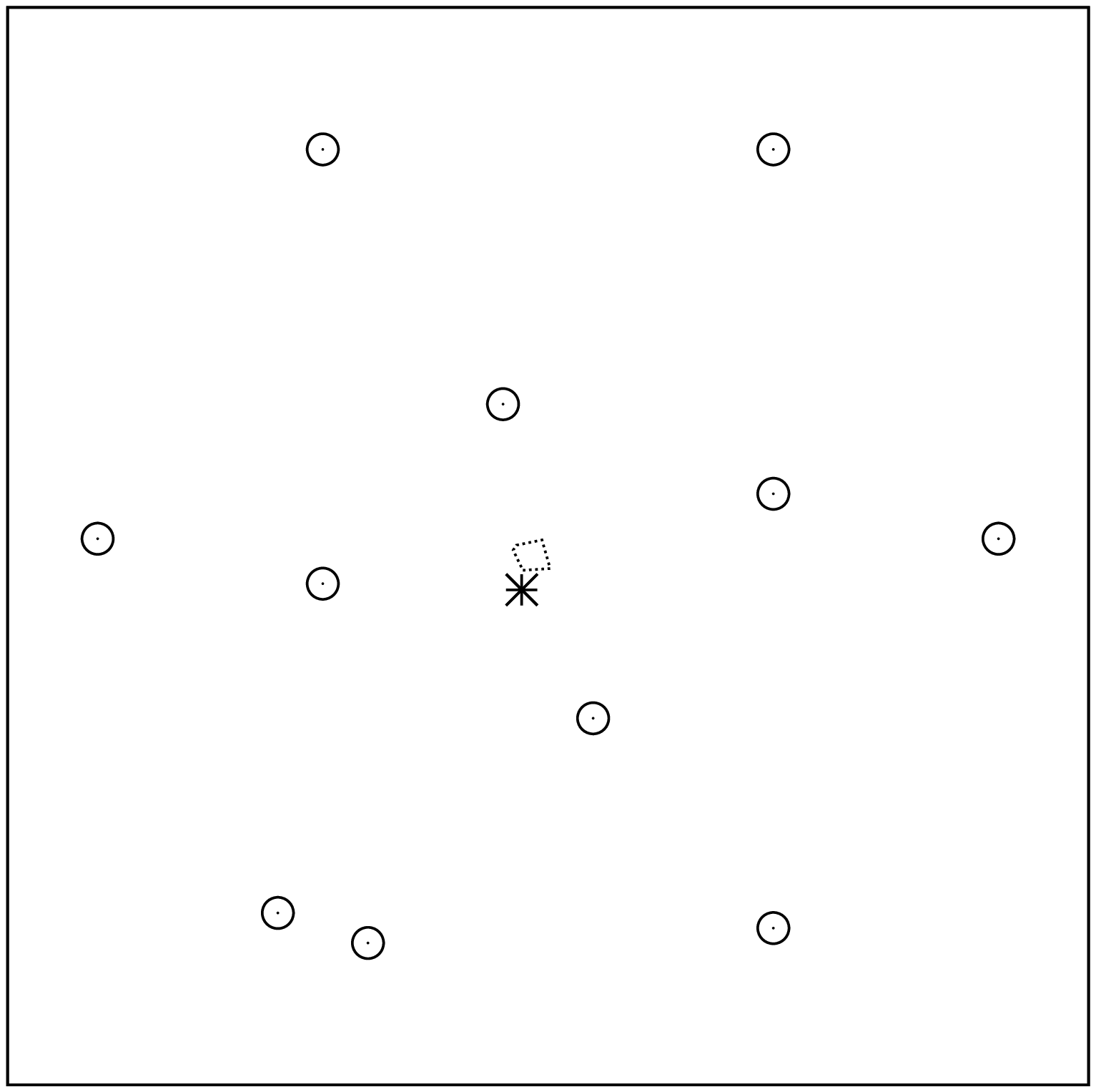}}}
\put(14,0){(c)}
}
}
\put(0,2){
\parbox[t]{100\unitlength}{\small\raggedright Legend:
\raisebox{2\unitlength}{\rotatebox{270}{\includegraphics[width=2\unitlength]{images/legend_datapt.eps}}}
Data points \quad
\raisebox{2\unitlength}{\rotatebox{270}{\includegraphics[width=2\unitlength]{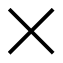}}}
TR-$L^1$ median \quad
\raisebox{2\unitlength}{\rotatebox{270}{\includegraphics[width=2\unitlength]{images/legend_l1pt.eps}}}
$L^1$ median \\
\phantom{Legend:}
\raisebox{2\unitlength}{\rotatebox{270}{\includegraphics[width=2\unitlength]{images/legend_ojapt.eps}}}
\raisebox{2\unitlength}{\rotatebox{270}{\includegraphics[width=2\unitlength]{images/legend_ojaln.eps}}}
Oja median (single point/outline)\quad
}}
\end{picture}
\caption{Simple point sets with their transformation--retransformation $L^1$
medians. 
\textbf{(a)} For three data 
points, the transformation--retransformation $L^1$ median is always in their
barycentre. --
\textbf{(b)} Same configuration of $8$ points as in 
Figures~\ref{fig-l1med-examples}(f) and~\ref{fig-ojamed-examples}(e).
Unlike the $L^1$ and Oja median, the transformation--retransformation $L^1$
median is none of the data points. --
\textbf{(c)} A sample configuration of $11$ data points with 
transformation--retransformation $L^1$, standard $L^1$ and Oja median.
The first two medians are visually indistinguishable here.
}
\label{fig-trl1med-examples}
\end{figure}

In Figure~\ref{fig-trl1med-examples}, some examples of 
transformation--retransformation $L^1$ medians are shown.

\runinhead{Continuous data.}
Transfer of the procedure to the space-continuous case depends on the
existence of the covariance matrix of the density, which is certainly
fulfilled if $\gamma$ is compactly supported.

\begin{definition}[Continuous transformation--retransformation $L^1$ median]
Let $\gamma$ be a normalised regular density for which
the covariance matrix
\begin{align}
C(\gamma) &:= 
\int\nolimits_{\mathbb{R}^n}
\int\nolimits_{\mathbb{R}^n}
\gamma(\boldsymbol{x})\,\gamma(\boldsymbol{y})\,
(\boldsymbol{x}-\boldsymbol{y})(\boldsymbol{x}-\boldsymbol{y})^{\mathrm{T}}
\,\mathrm{d}\boldsymbol{x}\,\mathrm{d}\boldsymbol{y}
\end{align}
exists and is positive definite,
the \emph{transformation--retransformation $L^1$ median} 
of $\gamma$ is defined as
\begin{align}
\operatorname{med}_{\mathrm{TR}L^1}(\gamma) &:=
T^{-1}\operatorname{med}_{L^1}\bigl(\gamma^T\bigr)\;,&
T&:=C(\gamma)^{-1/2}
\end{align}
with 
\begin{align}
\gamma^T(\boldsymbol{x}):=\operatorname{det}\,T\cdot\gamma(T\boldsymbol{x})
\;.
\end{align}
\end{definition}

\noindent
As for the $L^1$ median, we do not detail the case of distribution-valued 
densities but notice that for $\gamma$ concentrated on a straight line the
covariance matrix will again become singular, preventing the definition
of the transformation--retransformation $L^1$ median.

\runinhead{Equivariance.}
Affine equivariance of the transformation--retransformation $L^1$ median
is guaranteed by its construction.

\runinhead{Algorithmic aspects.}
The transformation--retransformation median of a finite multiset
can be computed using the
efficient algorithms for $L^1$ medians, with the additional effort
of computing the covariance matrix ($\mathcal{O}(N^2)$) and applying
it to the data ($\mathcal{O}(N)$).

\subsection{Half-Space Median}
\label{ssec-hsmed}

The \emph{half-space median} was first conceived by Tukey \cite{Tukey-icm74}.
It generalises the half-line characterisation of the univariate median
as given by \eqref{med-hs}, \eqref{medw-hs}, \eqref{medd-hs}.

\runinhead{Discrete data.}
Following \cite{Small-ISR90,Tukey-icm74}, we introduce the half-space depth
$d^{\mathrm{HS}}_\mathcal{X}(\boldsymbol{y})$
of a point $\boldsymbol{y}\in\mathbb{R}^n$ w.r.t.\ a finite multiset 
$\mathcal{X}$ of data points in $\mathbb{R}^n$ as the minimal number of 
data points contained in a closed half-space containing 
$\boldsymbol{y}$. In fact, it is sufficient to minimise over half-spaces
the boundary hyperplane of which goes through $\boldsymbol{y}$.
Further following \cite{Small-ISR90,Tukey-icm74}, the half-space median
then is the point of maximal half-space depth.

Characterising oriented hyperplanes by their unit normal vectors
$\boldsymbol{z}\in\mathrm{S}^{n-1}=\{\boldsymbol{x}\in\mathbb{R}^n~|~
\lVert\boldsymbol{x}\rVert=1\}$, we have the following definition.

\begin{definition}[Discrete half-space median]
Let $\mathcal{X}$ be a finite multiset in $\mathbb{R}^n$.
For any $\boldsymbol{y}\in\mathbb{R}^n$, the
\emph{half-space depth} of $\boldsymbol{y}$ w.r.t.\ $\mathcal{X}$ is
\begin{align}
d^{\mathrm{HS}}_\mathcal{X}(\boldsymbol{y}) &:=
\min\limits_{\boldsymbol{z}\in\mathrm{S}^{n-1}}
\#\bigl\{\boldsymbol{x}\in\mathcal{X}~|~
\boldsymbol{z}^\mathrm{T}(\boldsymbol{x}-\boldsymbol{y})\ge0\bigr\}
\;.
\label{hsdepth}
\end{align}
The \emph{half-space median} of $\mathcal{X}$ is given by
\begin{align}
\operatorname{med}_{\mathrm{HS}}(\mathcal{X}) &:=
\mathop{\operatorname{argmax}}\limits_{\boldsymbol{\mu}\in\mathbb{R}^n}
d^{\mathrm{HS}}_\mathcal{X}(\boldsymbol{\mu})\;.
\label{hsmed}
\end{align}
\end{definition}

\noindent
A generalisation of this definition to countable sets $\mathcal{X}$
with weights $w$ is straightforward by replacing the cardinality $\#$ with
the sum of weights in the respective sets.

Similar as for the Oja median, the half-space median set always contains
at least one intersection point of $n$ hyperplanes $H_{i_1,\ldots,i_n}$
going through $n$ data points, and the entire set of minimisers is the convex
hull of all intersection points with this property. 
However, unlike for the Oja median, some intersection points may also
lie in the interior of the set of minimisers.

If all data of $\mathcal{X}$ lie on a hyperplane in $\mathbb{R}^{n}$, their
half-space median automatically coincides with their half-space median
in $n-1$ dimensions.

\begin{figure}[t!]
\unitlength0.01\textwidth
\begin{picture}(100,73)
\put(0,41){
\put(0,0){
\put(0,33){\rotatebox{270}{\includegraphics[height=32\unitlength]
  {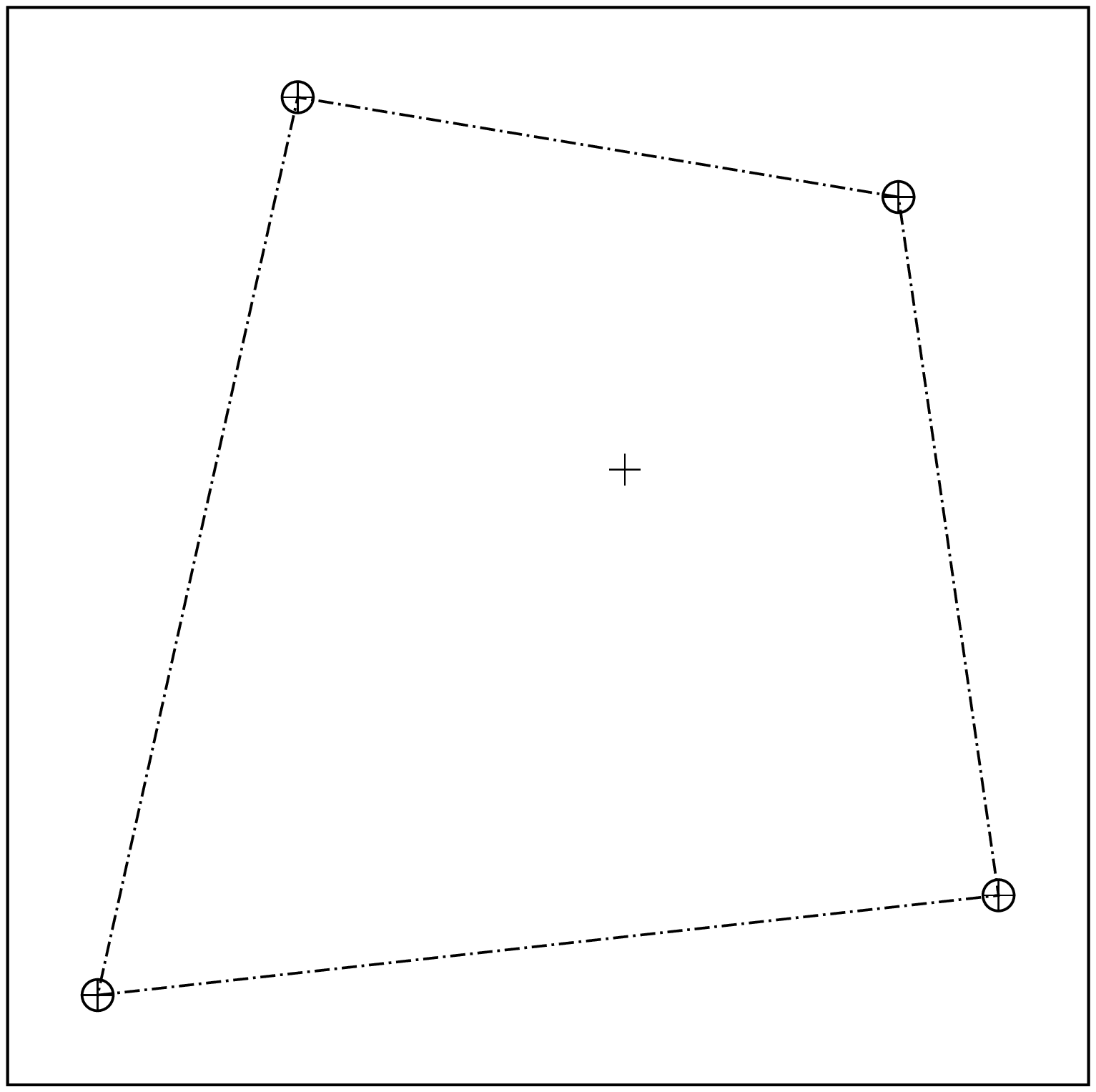}}}
\put(14,0){(a)}
}
\put(34,0){
\put(0,33){\rotatebox{270}{\includegraphics[height=32\unitlength]
  {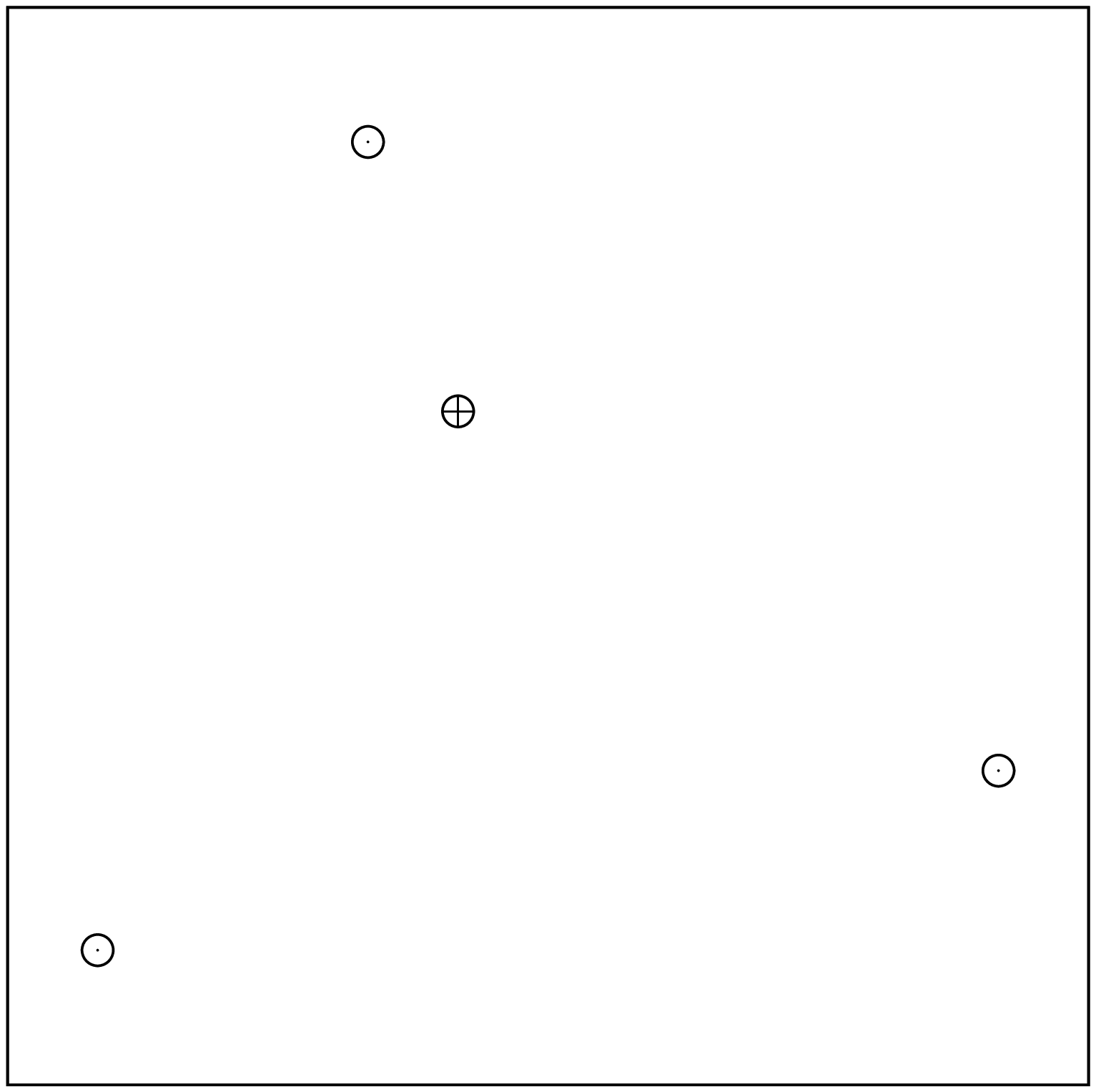}}}
\put(14,0){(b)}
}
\put(68,0){
\put(0,33){\rotatebox{270}{\includegraphics[height=32\unitlength]
  {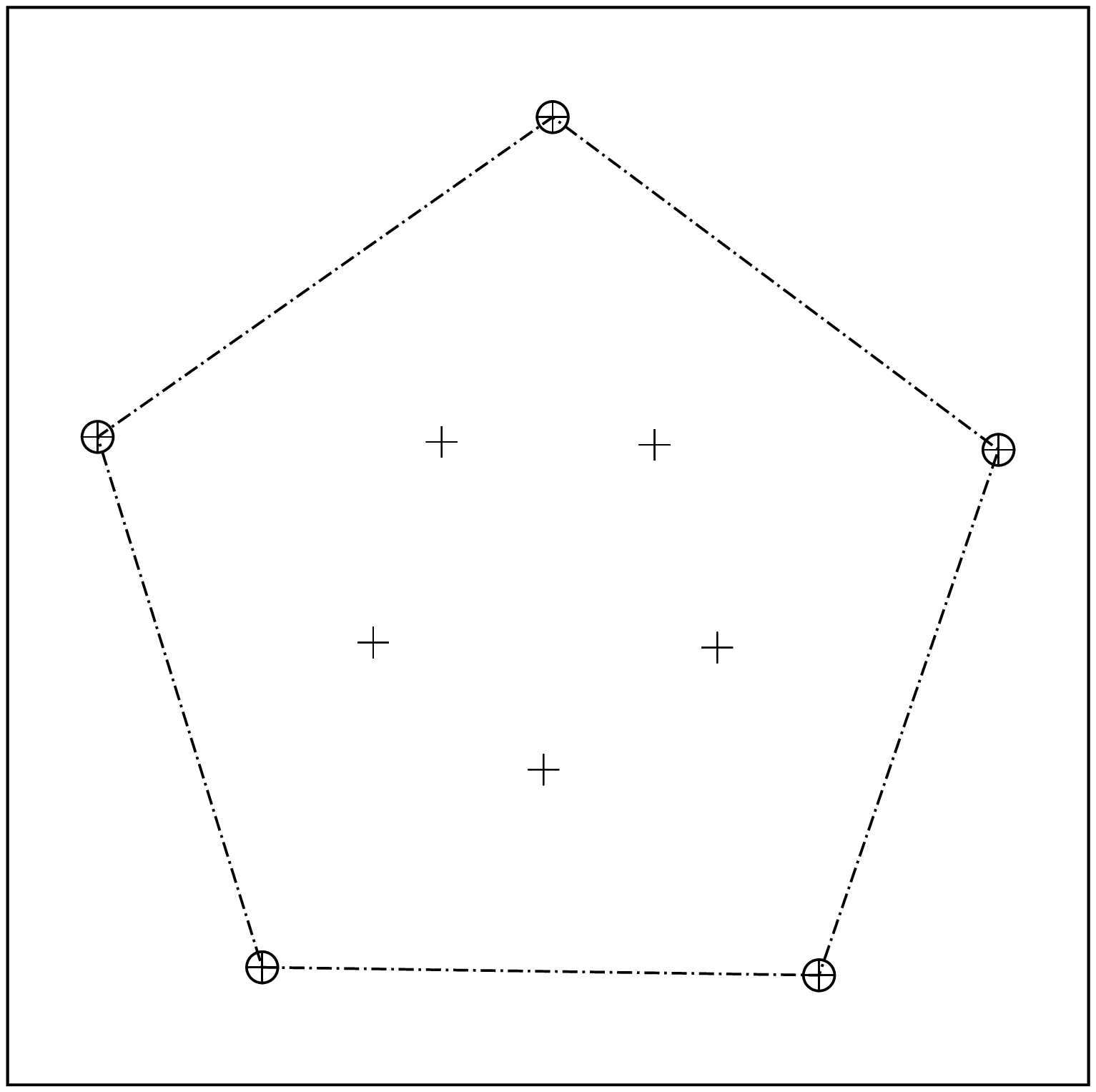}}}
\put(14,0){(c)}
}
}
\put(0,6){
\put(0,0){
\put(0,33){\rotatebox{270}{\includegraphics[height=32\unitlength]
  {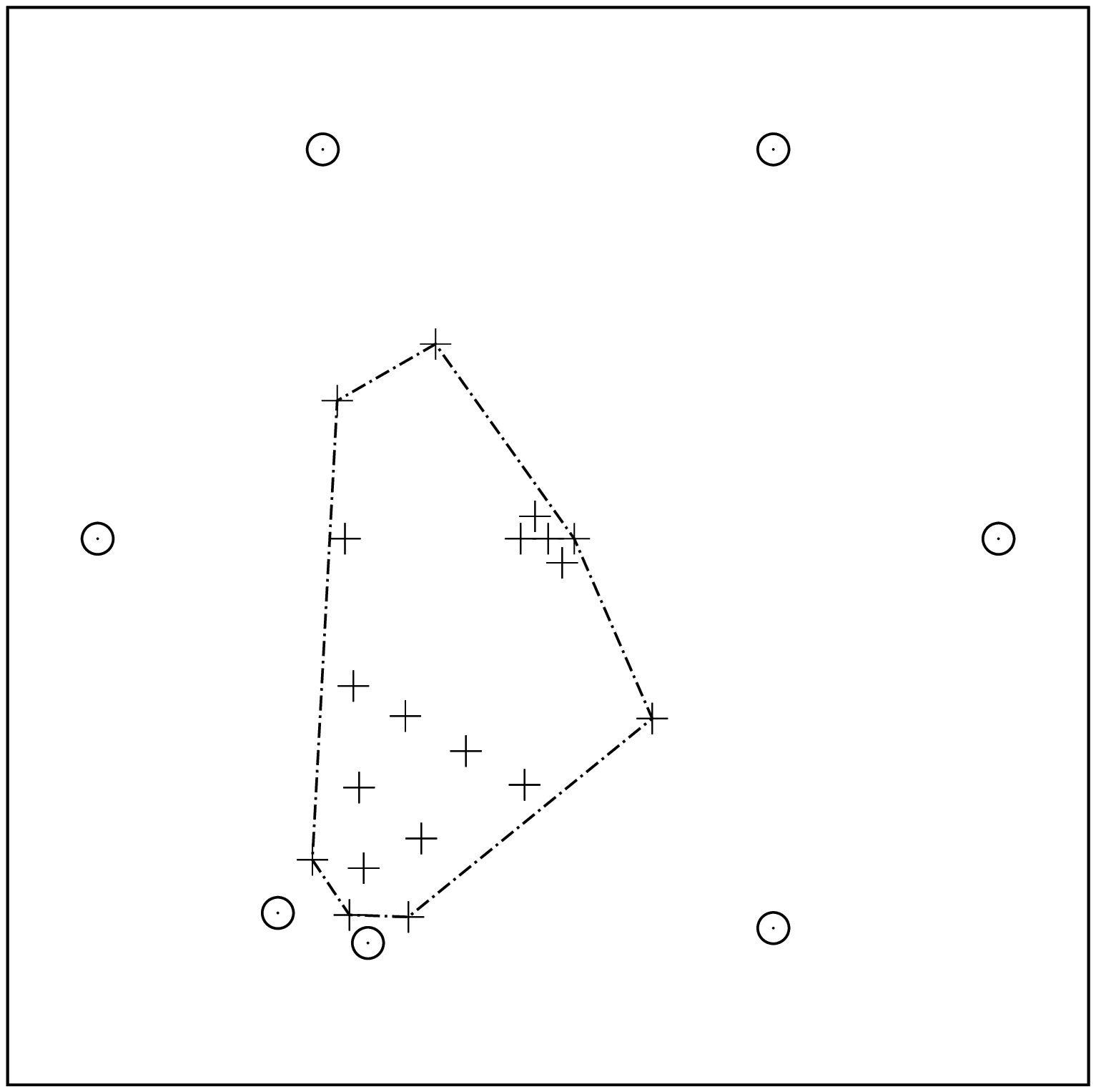}}}
\put(14,0){(d)}
}
\put(34,0){
\put(0,33){\rotatebox{270}{\includegraphics[height=32\unitlength]
  {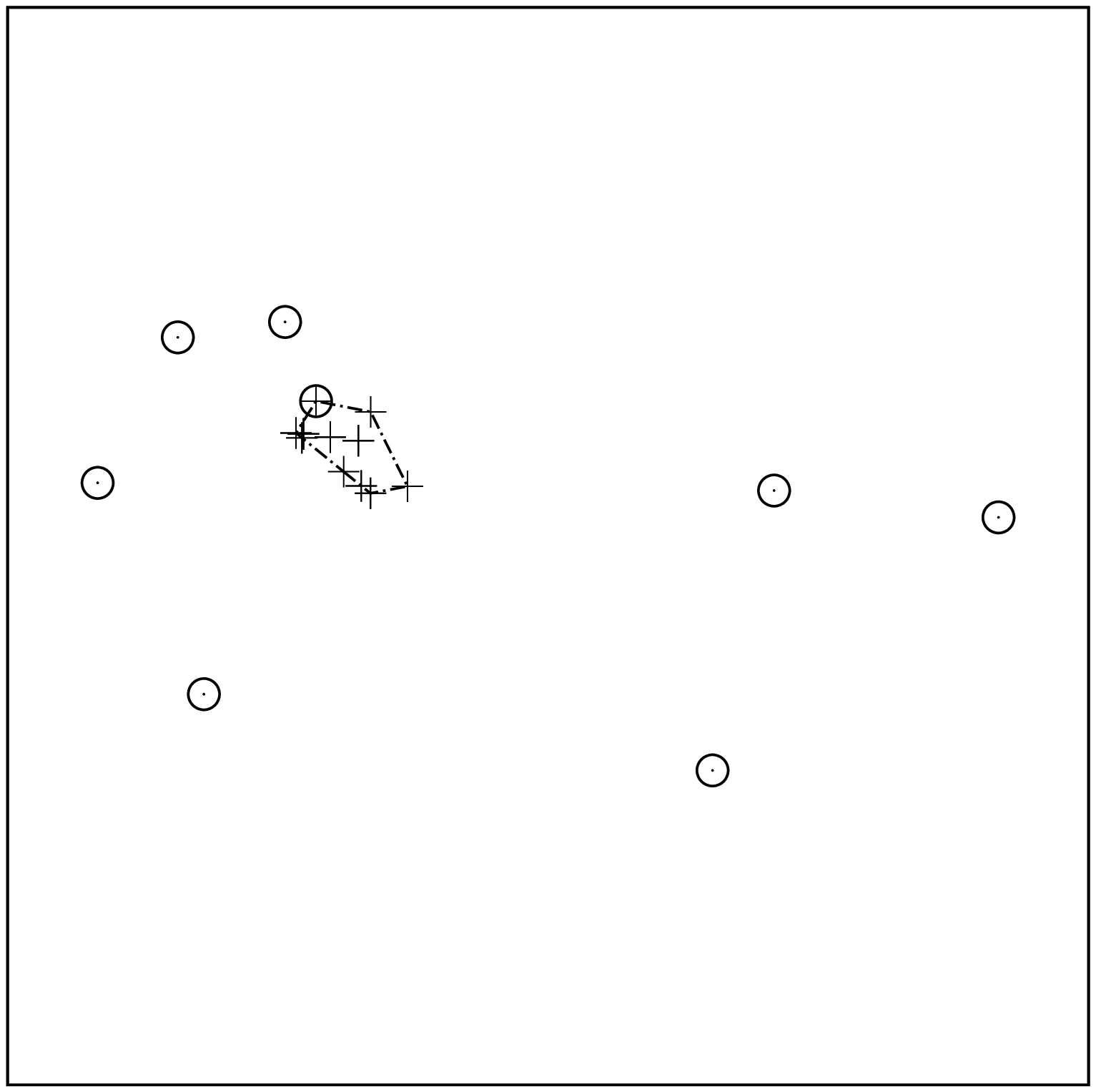}}}
\put(14,0){(e)}
}
\put(68,0){
\put(0,33){\rotatebox{270}{\includegraphics[height=32\unitlength]
  {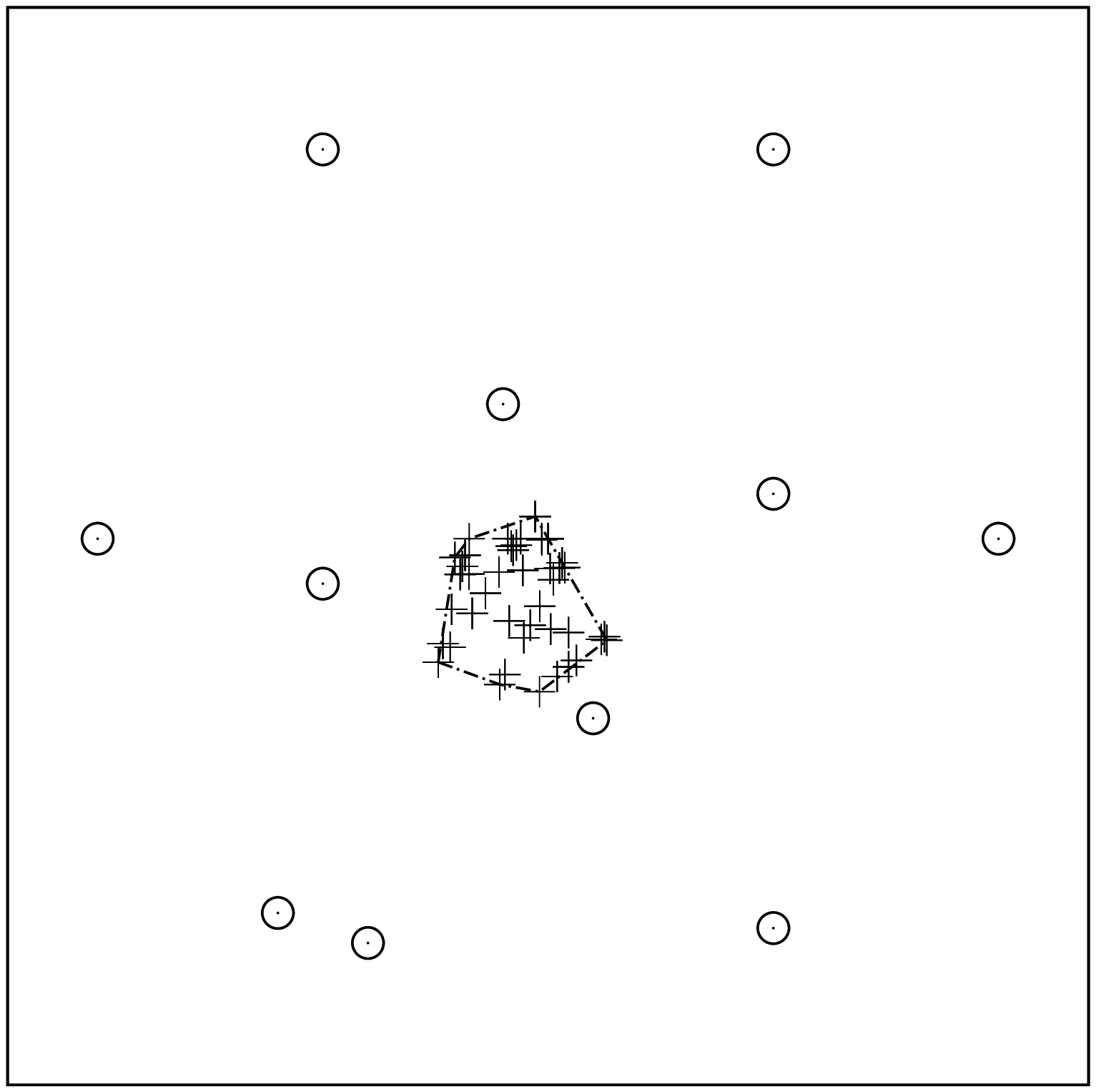}}}
\put(14,0){(f)}
}
}
\put(0,2){
\parbox[t]{100\unitlength}{\small\raggedright Legend:
\raisebox{2\unitlength}{\rotatebox{270}{\includegraphics[width=2\unitlength]{images/legend_datapt.eps}}}
Data points \quad
\raisebox{2\unitlength}{\rotatebox{270}{\includegraphics[width=2\unitlength]{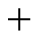}}}
\raisebox{2\unitlength}{\rotatebox{270}{\includegraphics[width=2\unitlength]{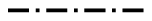}}}
Half-space median (single point/outline)
}
}
\end{picture}
\caption{Simple point sets with their half-space medians. 
If the half-space median set consists of more than one point, we show
additionally all those intersection points of lines between data points that
are contained in the half-space median set.
\textbf{(a)} For four points spanning a convex quadrilateral, the entire
quadrilateral is the half-space median set. --
\textbf{(b)} From four points with a triangle as convex hull, the inner data 
point is the half-space median. -- 
\textbf{(c)} Regular pentagon; the entire pentagon is the half-space median
set. --
\textbf{(d)} Sample configuration of $7$ points with their half-space median.
--
\textbf{(e)} Same configuration of $8$ points as in 
Figure~\ref{fig-l1med-examples}(f).
The half-space median set is a region including one of the data points
(which is also the $L^1$ and Oja median) on its boundary. --
\textbf{(f)} Configuration of $11$ data points (same as in
Figure~\ref{fig-trl1med-examples}(c)) with half-space median set.
}
\label{fig-hsmed-examples}
\end{figure}

In Figure~\ref{fig-hsmed-examples}, half-space medians of some simple point
configurations are shown.

\runinhead{Continuous data.}
The half-space median of continuous data has been considered e.g.\ in
\cite{Rousseeuw-MET99}.

\begin{definition}[Continuous half-space median]
Let $\gamma$ be a normalised regular density in $\mathbb{R}^n$.
For any $\boldsymbol{y}\in\mathbb{R}^n$, the
\emph{half-space depth} of $\boldsymbol{y}$ w.r.t.\ $\gamma$ is
\begin{align}
d^{\mathrm{HS}}_\gamma(\boldsymbol{y}) &:=
\min\limits_{\boldsymbol{z}\in\mathrm{S}^{n-1}}
\int\nolimits_{H^+_{\boldsymbol{y},\boldsymbol{z}}}
\gamma(\boldsymbol{x})\,\mathrm{d}\boldsymbol{x}
\label{hsdepthd}
\end{align}
where 
$H^+_{\boldsymbol{y},\boldsymbol{z}}:=
\{\boldsymbol{x}\in\mathbb{R}~|~
\boldsymbol{z}^\mathrm{T}(\boldsymbol{x}-\boldsymbol{y})\ge0\}$
denotes the half-space separated by a hyperplane with normal vector
$\boldsymbol{z}$ through $\boldsymbol{y}$.
The \emph{half-space median} of $\gamma$ is then defined as
\begin{align}
\operatorname{med}_{\mathrm{HS}}(\gamma) &:=
\mathop{\operatorname{argmax}}\limits_{\boldsymbol{\mu}\in\mathbb{R}^n}
d^{\mathrm{HS}}_\gamma(\boldsymbol{\mu})\;.
\label{hsmedd}
\end{align}
\end{definition}

\runinhead{Equivariance.}
In discussions of the equivariance properties of the half-space median in
the literature, emphasis is mostly laid on the affine equivariance
of the half-space median. However, in \cite{Small-CJS87}
equivariance of the half-space median under a larger transformation class
is proven, see also \cite{Small-ISR90}. The transformations discussed
there are characterised by the requirement that images of half-spaces
containing $\boldsymbol{\mu}$, and pre-images of half-spaces not
containing $\boldsymbol{\mu}$ must be convex.

Whereas this class of transformations is hard to describe in closed form,
we point out a smaller class of transformations under which the
half-space median is equivariant: Notice that the definition of
half-space depth relies on nothing but the splitting of $\mathbb{R}^n$
into half-spaces by hyperplanes, and the belonging of data points to these
half-spaces. The preservation of hyperplanes is what characterises
projective transformations. The caveat is that these transformations
act on the larger (and non-orientable) projective space $\mathbb{P}^n$,
which extends $\mathbb{R}^n$ by a hyperplane at infinity.
Nevertheless, projective transformations
do not change the relation between points and the hyperplanes relevant
for defining the half-space median as long as no point in the convex hull
of the data points is taken to infinity.

Thus, the half-space median of a data multiset is equivariant under all
projective transformations that do not take any point in its convex
hull to the infinite hyperplane. Note that this restricted class of
projective transformations is not a group; nevertheless, the closed-form
representation of projective transformations can be used.

On $\mathbb{R}^n$, a general projective transformation $T$ is given by 
a regular matrix $\hat{P}\in\mathrm{GL}(n+1,\mathbb{R})$ which we write as
\begin{align}
\hat{P} &= 
\begin{pmatrix}P&\boldsymbol{q}\\\boldsymbol{r}^\mathrm{T}&s\end{pmatrix}
\end{align}
with an $n\times n$-matrix $P$, two vectors 
$\boldsymbol{q},\boldsymbol{r}\in\mathbb{R}^n$ and a number $s\in\mathbb{R}$,
and transforms points $\boldsymbol{x}\in\mathbb{R}^n$ with
$\boldsymbol{r}^\mathrm{T}\boldsymbol{x}+s\ne0$ via
\begin{align}
T\boldsymbol{x} &= 
\frac{P\,\boldsymbol{x}+\boldsymbol{q}}
     {\boldsymbol{r}^\mathrm{T}\boldsymbol{x}+s} \;.
\end{align}
Any matrix $\lambda\hat{P}$ with $\lambda\in\mathbb{R}\setminus\{0\}$
describes the same transformation $T$. (Alternatively, a unique $\hat{P}$
can be associated to each $T$ by choosing
$\hat{P}\in\mathrm{SL}(n+1,\mathbb{R})$.)

The half-space median of a finite multiset $\mathcal{X}$
is then equivariant w.r.t.\ $T$ if
$\boldsymbol{r}^\mathrm{T}\boldsymbol{x}+s$ vanishes nowhere in the
convex hull $[\mathcal{X}]$.

\runinhead{Algorithmic aspects.}
Like the computation of the Oja median, that of the half-space median is
algorithmically not convenient.
In the bivariate case the half-space
depth $d^{\mathrm{HS}}_\mathcal{X}(\boldsymbol{y})$ of a point can be
computed by angular sorting of the data points w.r.t.\ $\boldsymbol{y}$
in $\mathcal{O}(N\log N)$ time, followed by a linear-time computation
to find the half-plane orientation with minimal number of points, see
\cite{Rousseeuw-JRSSC96,Ruts-CSDA96}. 
Using a strategy for restricting the core part of the computation to a 
smaller set, \cite{Bremner-SC08} obtains an algorithm with
complexity $\mathcal{O}(N+K\log K)$, where $K$ is the actual depth.

To compute the half-space median by a direct approach
requires up to $\mathcal{O}(N^4)$ evaluations of the depth (for all
intersection points of lines connecting two data points), making the
total complexity $\mathcal{O}(N^5\log N)$ 
\cite{Rousseeuw-JRSSC96}
with the $\mathcal{O}(N\log N)$ depth algorithm.
By an improved selection procedure for intersection points, 
\cite{Rousseeuw-SS98,Ruts-CSDA96} reduce the complexity to 
$\mathcal{O}(N^2\log N)$. 

Another approach to computing the half-space median uses algorithms that
compute the convex sets of points with half-space depth $\ge K$ for
given $K$. In \cite{Matousek-Cg91}, a divide-and-conquer algorithm
is described that allows to achieve this goal in $\mathcal{O}(N\log^4N)$.
Combined with a bisection strategy to find the largest $K$ for which the
respective set is non-empty, this yields an $\mathcal{O}(N\log^5 N)$ 
algorithm for the bivariate half-space median \cite{Matousek-Cg91}.
In \cite{Langerman-stacs03} an $\mathcal{O}(N\log^3 N)$ algorithm for
the bivariate half-space median is stated.

An extension of the angular sorting idea to $n>2$ via projections
yields $\mathcal{O}(N^{n-1}\log N)$ complexity for the depth computation
\cite{Rousseeuw-JRSSC96,Rousseeuw-SC98}; 
however, also the number of depth evaluations
increases since $\mathcal{O}(N^{n^2})$ intersections of $n$ hyperplanes
occur. In \cite{Rousseeuw-SC98} approximative algorithms for the depth
computation in higher dimensions with lower complexity are proposed.
A stochastic algorithm by Chan \cite{Chan-soda04} based on linear programming
reaches an average (expected) $\mathcal{O}(N\log N + N^{n-1})$ computation
time for the half-space median computation in dimension $n$.

\subsection{Convex-Hull-Stripping Median}
\label{ssec-chsmed}

Generalising the extremum-stripping formulations for the multivariate
median, Barnett's 1976 paper \cite{Barnett-JRSSA76} together with its
subsequent discussion by Seheult et al.\ \cite{Seheult-JRSSA76}
establish another geometric concept of a multivariate median. 
Extremal points of the convex hull of the data herein take the role of the
extrema in the univariate process described in Definition~\ref{def-med-extr}.

\runinhead{Discrete data.}
For simplicity, we state the following definition for a set $\mathcal{X}$,
i.e.\ without multiplicities.

\begin{definition}[Discrete convex-hull-stripping median]
Let $\mathcal{X}$ be a finite set of data points from $\mathbb{R}^n$.
Define a series $(\mathcal{X}^k)_{k=0,1,2,\ldots}$ of sets as follows:
Start with the given data $\mathcal{X}^0:=\mathcal{X}$.
Continue by deleting in each step $k\ge1$ those data points from 
$\mathcal{X}^{k-1}$ that are vertices of its convex hull boundary, i.e.\
with $V^{k-1}$ denoting the set of boundary vertices, set
$\mathcal{X}^k:=\mathcal{X}^{k-1}\setminus V^{k-1}$. This is repeated until
(after finitely many steps)
an empty set is obtained. If $\mathcal{X}^k=\varnothing\ne\mathcal{X}^{k-1}$,
the \emph{convex-hull-stripping median} of $\mathcal{X}$ is defined as the 
convex hull of $\mathcal{X}^{k-1}$,
\begin{align}
\operatorname{med}_{\mathrm{CHS}}(\mathcal{X}) &:=
[\mathcal{X}^{k-1}]\;.
\end{align}
\end{definition}

\noindent
When generalising this definition to multisets $\mathcal{X}$, one has to 
decide whether multiple copies of a data point are deleted in one step, 
or one copy per step, when the convex hull boundary reaches this point. 
The latter might appear closer to what is done in the univariate 
extremum-stripping procedure but whatever choice is made makes the process 
discontinuous with regard to the input data, i.e.\ arbitrary small variations
of the data can lead to the convex-hull-stripping median jumping to another
value (set). Discontinuities also occur at configurations where one data 
point crosses a face of the convex hull in some step during the procedure.
Thus, the convex-hull-stripping median does not depend continuously on the
given data points.

Again, when all data points of $\mathcal{X}$ are located on a hyperplane,
their convex-hull-stripping median in $n$ dimensions
coincides with its counterpart in $n-1$ dimensions:
Although in such a case in each step all data points are located on the
boundary of the convex hull, the stripping process is designed to delete
only vertices but not those data points lying somewhere on the facets of
the (then degenerated) polytope.

\begin{figure}[t!]
\unitlength0.01\textwidth
\begin{picture}(100,41)
\put(0,9){
\put(0,0){
\put(0,33){\rotatebox{270}{\includegraphics[height=32\unitlength]
  {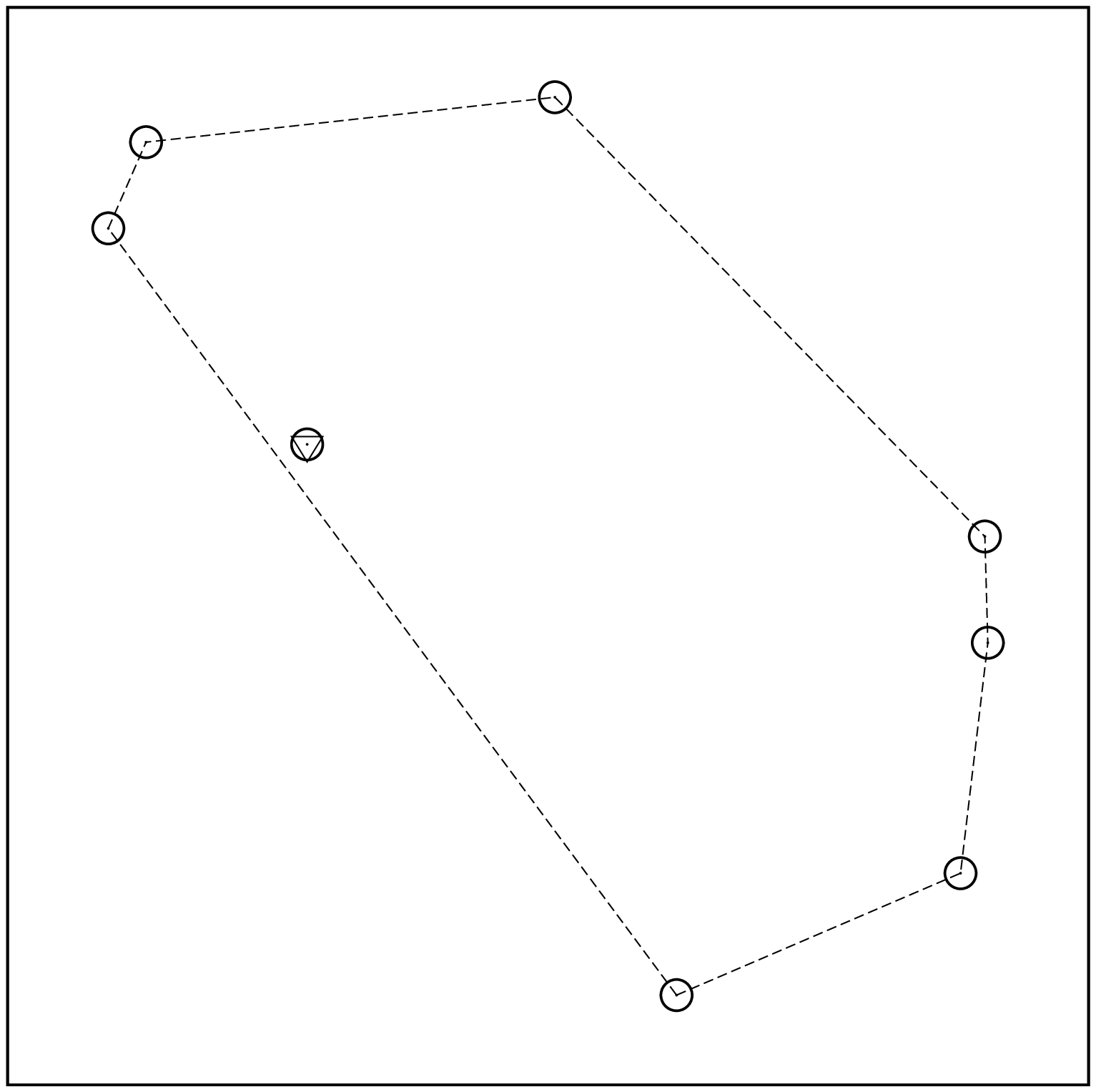}}}
\put(14,0){(a)}
}
\put(34,0){
\put(0,33){\rotatebox{270}{\includegraphics[height=32\unitlength]
  {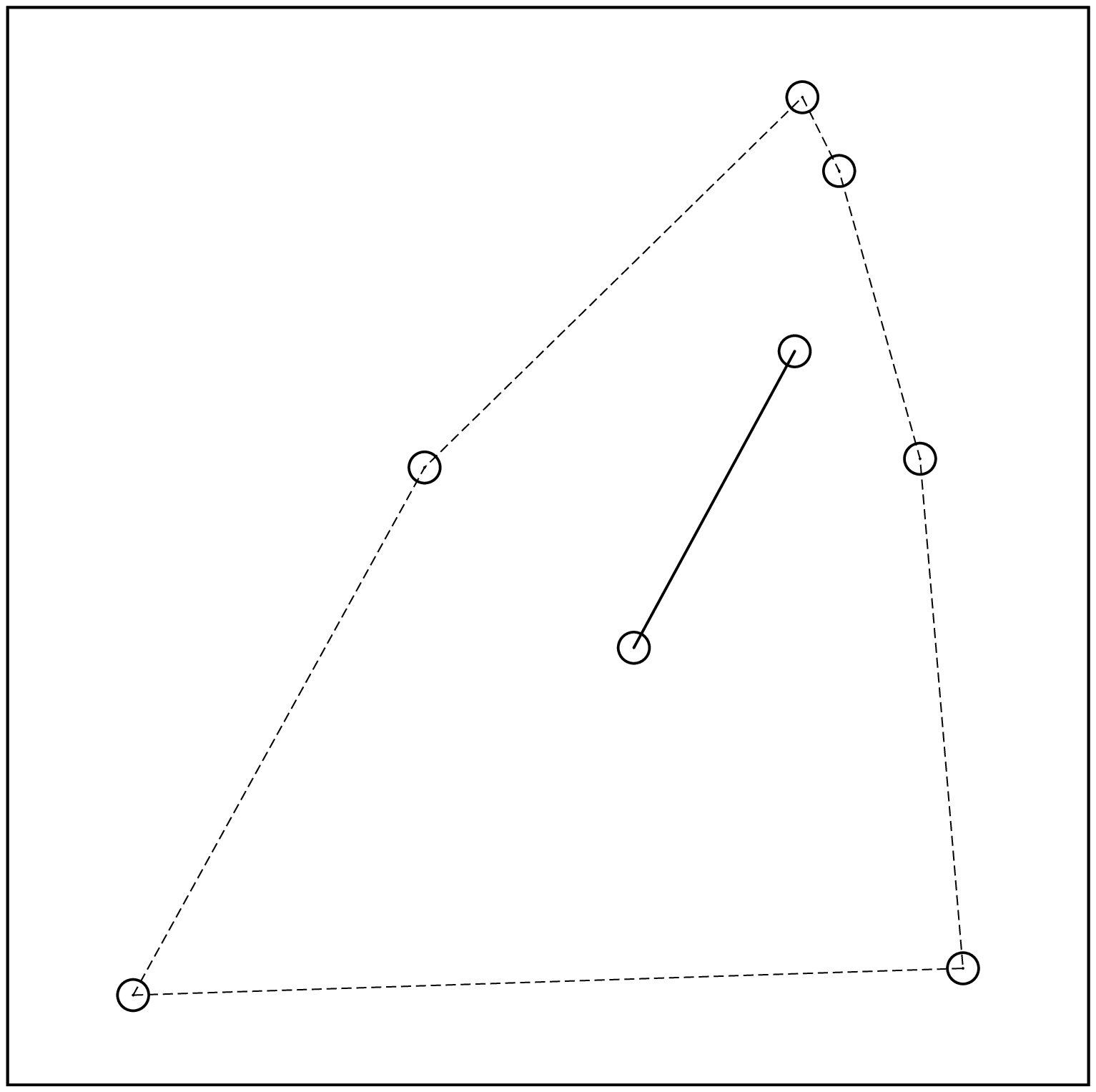}}}
\put(14,0){(b)}
}
\put(68,0){
\put(0,33){\rotatebox{270}{\includegraphics[height=32\unitlength]
  {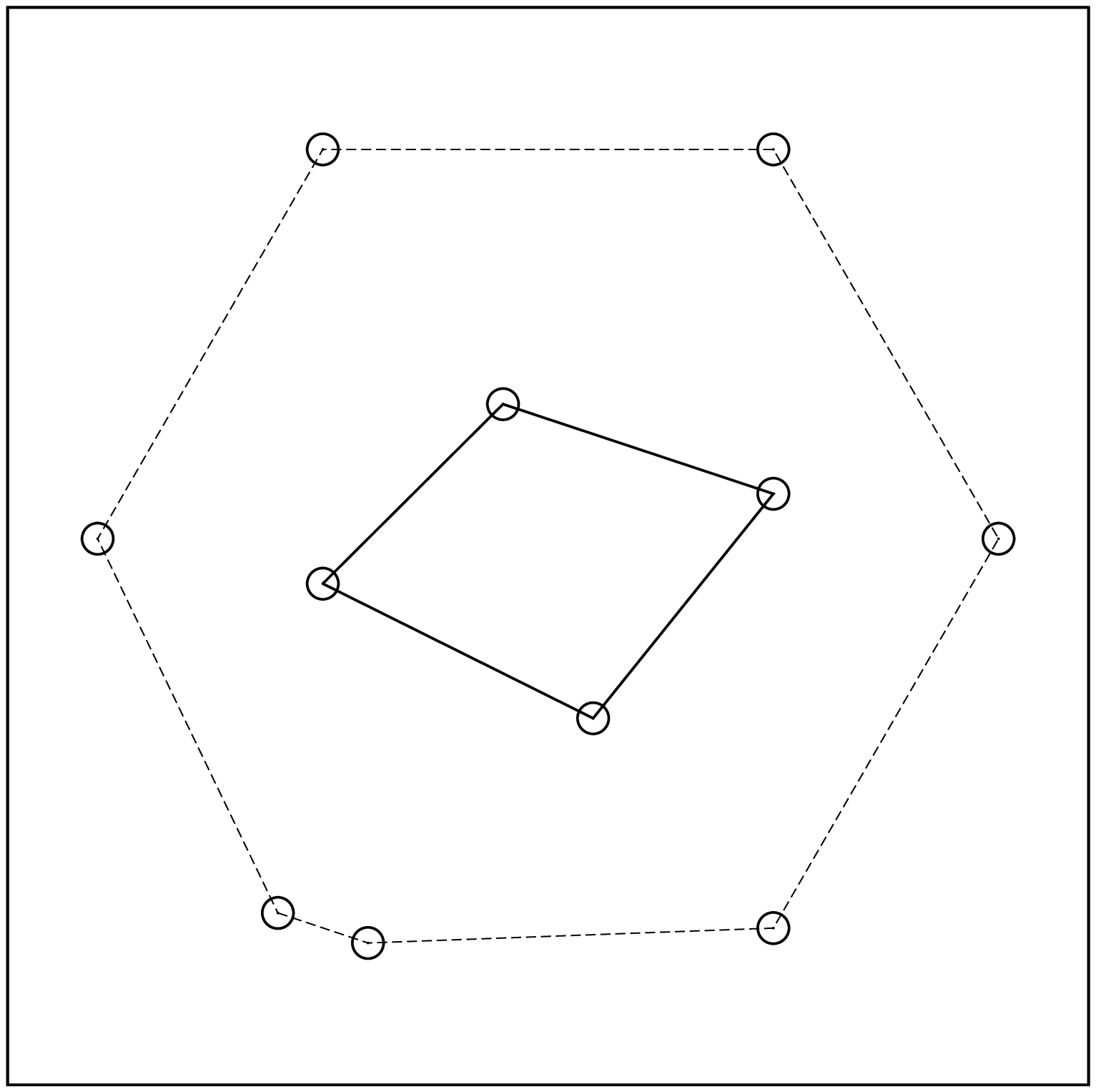}}}
\put(14,0){(c)}
}
}
\put(0,5){
\parbox[t]{12\unitlength}{\small\raggedright Legend:}
}
\put(12,5){
\parbox[t]{88\unitlength}{\small\raggedright
\raisebox{2\unitlength}{\rotatebox{270}{\includegraphics[width=2\unitlength]{images/legend_datapt.eps}}}
Data points \quad
\raisebox{2\unitlength}{\rotatebox{270}{\includegraphics[width=2\unitlength]{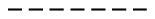}}}
initial convex hull\\
\raisebox{2\unitlength}{\rotatebox{270}{\includegraphics[width=2\unitlength]{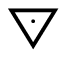}}}
\raisebox{2\unitlength}{\rotatebox{270}{\includegraphics[width=2\unitlength]{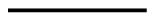}}}
Convex-hull-stripping median (single point/outline)
}
}
\end{picture}
\caption{Simple point sets with their convex-hull-stripping medians. 
\textbf{(a)} A configuration of $8$ data points with a unique
convex-hull-stripping median. --
\textbf{(b)} Same configuration of $8$ points as in 
Figure~\ref{fig-ojamed-examples}(f); the convex-hull-stripping median set
is a line segment.
--
\textbf{(c)} Same configuration of $11$ data points as in 
Figure~\ref{fig-hsmed-examples}(f). Here, the median set is spanned by
four points.
}
\label{fig-chsmed-examples}
\end{figure}

\begin{figure}[t!]
\unitlength0.01\textwidth
\begin{picture}(100,50)
\put(0,12){
\put(0,0){
\put(0,33){\rotatebox{270}{\includegraphics[height=32\unitlength]
  {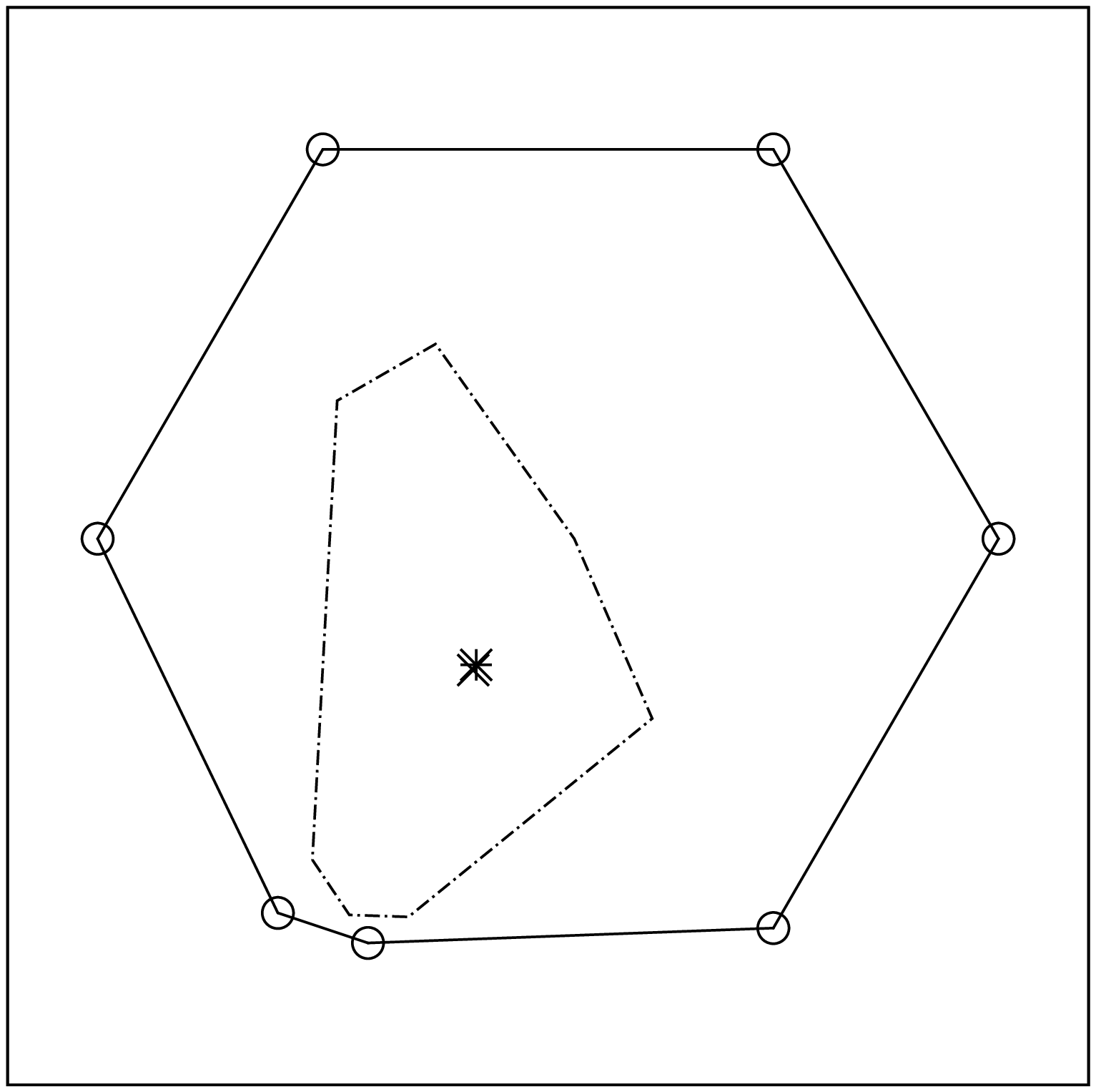}}}
\put(14,0){(a)}
}
\put(34,0){
\put(0,33){\rotatebox{270}{\includegraphics[height=32\unitlength]
  {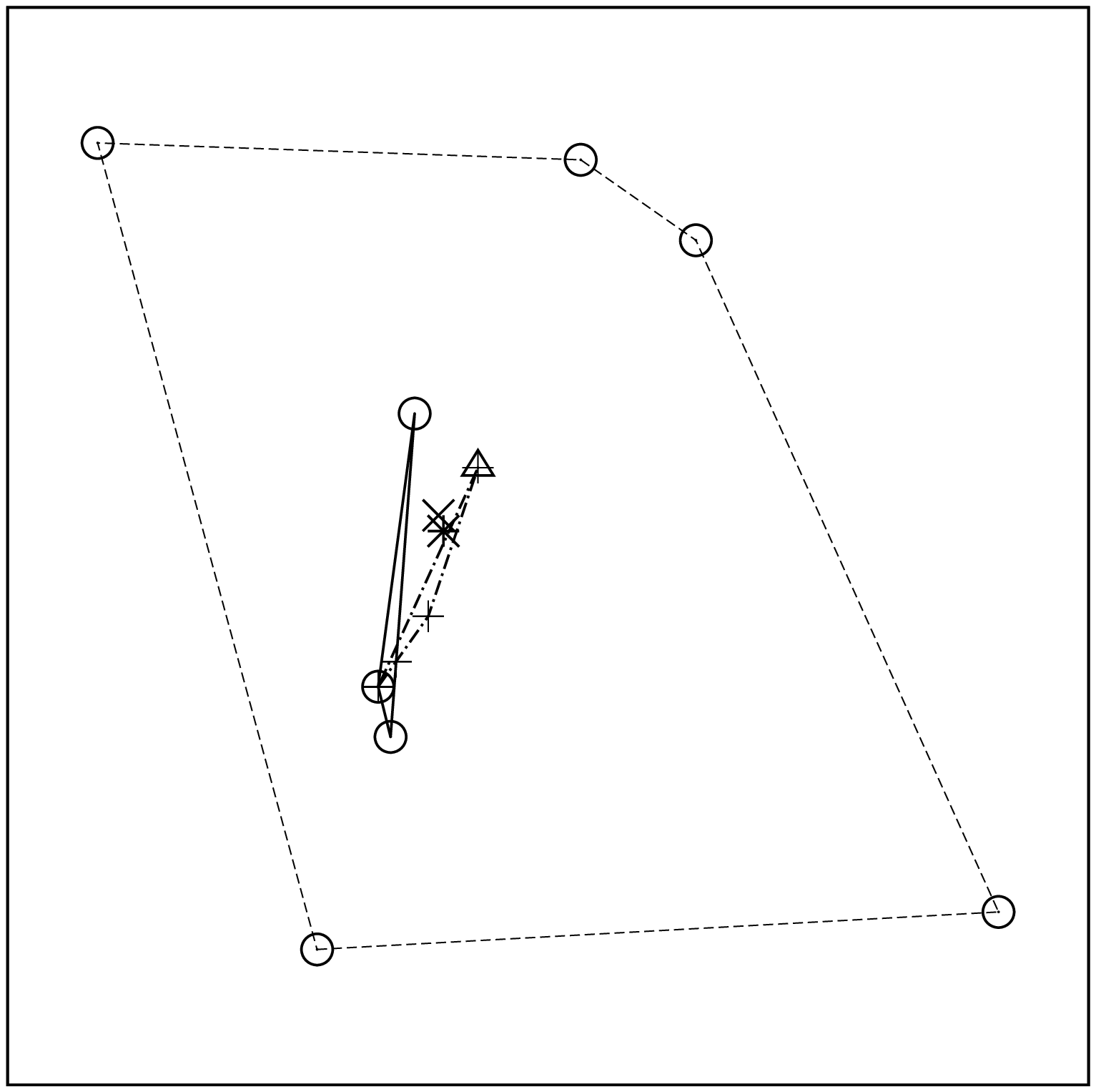}}}
\put(14,0){(b)}
}
\put(68,0){
\put(0,33){\rotatebox{270}{\includegraphics[height=32\unitlength]
  {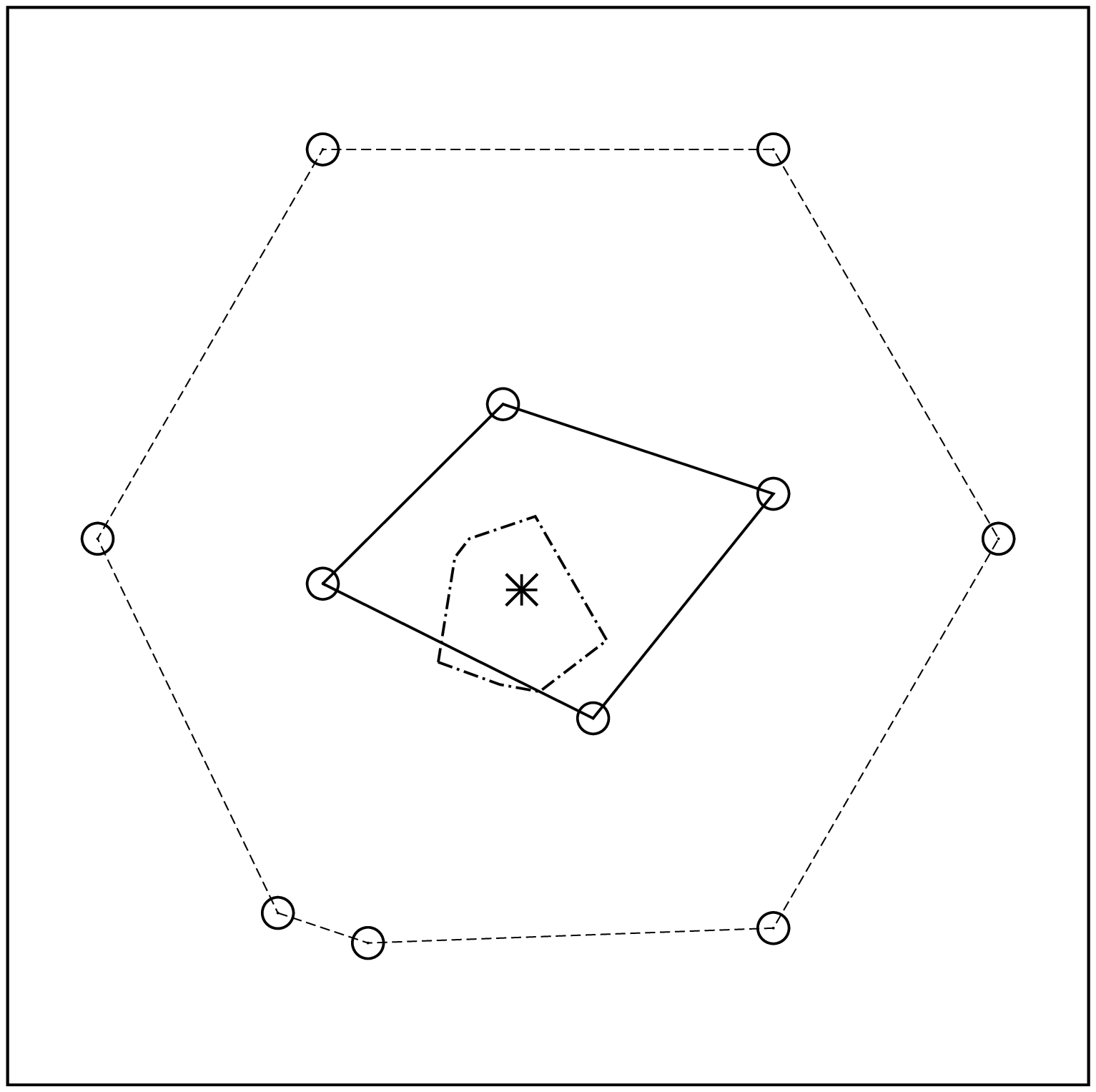}}}
\put(14,0){(c)}
}
}
\put(0,8){
\parbox[t]{12\unitlength}{\small\raggedright Legend:}
}
\put(12,8){
\parbox[t]{88\unitlength}{\small\raggedright
\raisebox{2\unitlength}{\rotatebox{270}{\includegraphics[width=2\unitlength]{images/legend_datapt.eps}}}
Data points \quad
\raisebox{2\unitlength}{\rotatebox{270}{\includegraphics[width=2\unitlength]{images/legend_l1pt.eps}}}
$L^1$ median \quad
\raisebox{2\unitlength}{\rotatebox{270}{\includegraphics[width=2\unitlength]{images/legend_trl1pt.eps}}}
TR-$L^1$ median \quad \\
\raisebox{2\unitlength}{\rotatebox{270}{\includegraphics[width=2\unitlength]{images/legend_ojapt.eps}}}
Oja median (single point) \quad
\raisebox{2\unitlength}{\rotatebox{270}{\includegraphics[width=2\unitlength]{images/legend_hsln.eps}}}
Oja and half-space median (outline) \quad \\
\raisebox{2\unitlength}{\rotatebox{270}{\includegraphics[width=2\unitlength]{images/legend_chsstepln.eps}}}
initial convex hull \quad
\raisebox{2\unitlength}{\rotatebox{270}{\includegraphics[width=2\unitlength]{images/legend_chsln.eps}}}
Convex-hull-stripping median (outline)
}
}
\end{picture}
\caption{Comparison of multivariate median concepts. 
\textbf{(a)} Same data set of $7$ points as in
Figure~\ref{fig-hsmed-examples}(d) with multivariate medians. 
The Oja and half-space median sets coincide.
The $L^1$ and
transformation--retransformation $L^1$ medians differ but their distance is
extremely small.
-- 
\textbf{(b)} A configuration of $8$ points as in 
Figure~\ref{fig-ojamed-examples}(f); the convex-hull-stripping median set
is a line segment.
--
\textbf{(c)} Same configuration of $11$ data points as in 
Figure~\ref{fig-hsmed-examples}(f). 
Again, the 
Oja and half-space median sets coincide;
$L^1$ and 
transformation--retransformation $L^1$ medians differ below
visibility distance.
}
\label{fig-mvmed-comp}
\end{figure}

In Figure~\ref{fig-chsmed-examples}, we show some point configurations
with their convex-hull-stripping medians. An additional comparison of
all five multivariate median concepts discussed is shown in 
Figure~\ref{fig-mvmed-comp}.

\runinhead{Continuous data.}
Unlike for the previously discussed multivariate medians, a space-continuous
formulation for the convex-hull-stripping median is not at all obvious.
As in the univariate case, where a system of ordinary differential
equations \eqref{medd-exstrip-ode} could be used to describe a continuous
extrema-stripping process, one needs a process that moves the convex hull
boundary in inward direction in a time-continuous manner, described by
a partial differential equation.

In \cite{Welk-ssvm19} a stochastic sampling procedure was proposed to derive
a continuous process that can be considered the adequate continuous
counterpart of the discrete convex-hull-stripping process. The results in
\cite{Welk-ssvm19} are restricted to the bivariate case, with the main
finding being the following proposition.
The curve evolution PDE in this result evolves a closed curve in
inward direction, reducing it in finite time to a point which is called its
vanishing point.

\begin{proposition}[{\cite[Proposition~1]{Welk-ssvm19}}]
\label{prop-fp-chsmed}
Let a piecewise smooth density $\gamma:\mathbb{R}^2\to\mathbb{R}$ with compact
support
$\varOmega_0\subset\mathbb{R}^2$ in the Euclidean plane $\mathbb{R}^2$ be 
given.
Assume that the boundary of $\varOmega_0$ is regular, and $\gamma$ is
differentiable on $\varOmega_0$.
Let the point set $\mathcal{X}$ be a stochastic sampling of this density
with sampling density $1/h^2$, i.e.\ there is on average one sampling point in
an area in which the density integrates to $h^2$.
For $h\to0$, the convex-hull-stripping median of the set $\mathcal{X}$
asymptotically coincides with the vanishing point of the curve evolution
\begin{equation}
\boldsymbol{c}_t(p,t) = \begin{cases}
\gamma(\boldsymbol{c}(p,t))^{-2/3}\kappa(p,t)^{1/3}
\boldsymbol{n}(p,t) \;,&
\boldsymbol{c}(p,t)\in\partial\,[\boldsymbol{c}(\cdot,t)]\;,\\
0&\text{else}
\end{cases}
\label{cpde-chsmed}
\end{equation}
where $\boldsymbol{c}:[0,L]\times[0,T]\to\mathbb{R}^2$ is a curve evolution
of closed curves
with curve parameter $p\in[0,L]$ and evolution time $t\in[0,T]$, which is
initialised at time $t=0$ with the boundary of the
support set, $\boldsymbol{c}_0:=\partial\varOmega_0$.
Furthermore, $\kappa(p,t)$ denotes the curvature and
$\boldsymbol{n}(p,t)$ the inward normal vector
of $\boldsymbol{c}$ at $(p,t)$.
At any time $t$, the evolution acts only on the part of
$\boldsymbol{c}$ that is on the boundary
$\partial\,[\boldsymbol{c}]$ of the convex hull of $\boldsymbol{c}$.
\end{proposition}

\noindent
By virtue of this proposition, one can define the convex-hull-stripping
median of a density of compact support as follows.

\begin{definition}[Continuous convex-hull-stripping median]
Let $\gamma$ be a normalised regular density with compact support.
Define the curve evolution $\boldsymbol{c}$ as in 
Proposition~\ref{prop-fp-chsmed}. The vanishing point of this 
curve evolution is the \emph{convex-hull-stripping median} of $\gamma$.
\end{definition}

\noindent
It is interesting to notice that \eqref{cpde-chsmed} is a modification of the
\emph{affine curvature flow} PDE \cite{Alvarez-ARMA93,Sapiro-IJCV93}
\begin{align}
\boldsymbol{c}_t(p,t) &= 
\kappa(p,t)^{1/3}
\boldsymbol{n}(p,t) \;.
\label{acm}
\end{align}
Note that affine curvature flow is an affine equivariant curve shrinking
evolution. Its appearance in this context appears natural given the
affine equivariant nature of the discrete convex-hull-stripping process.

The modification of \eqref{cpde-chsmed} compared to \eqref{acm} is twofold: 
On one hand, the evolution speed in the plane
is weighted to account for the varying density $\gamma$; on the other hand,
its effect is restricted to those parts of the boundary curve of the
evolving support set that belong to the boundary of its convex hull.

The latter restriction is based on a non-local condition which adds
complication to possible numeric approximations of the process. In
\cite{Welk-ssvm19} therefore a regularisation is proposed that allows to
work without such a restriction and only on convex boundary curves:

\begin{proposition}[{\cite[Corollary~1]{Welk-ssvm19}}]
\label{prop-chsconv}
Let the density $\gamma$ with support $\varOmega_0$, and the sampled point set
$\mathcal{X}$ with sampling density $1/h^2$ be given as in
Proposition~\ref{prop-fp-chsmed}.
Let $\tilde{\varOmega}_0\supseteq\varOmega_0$ be a convex compact set with
regular boundary. For $\varepsilon>0$, define
a piecewise smooth density $\gamma_\varepsilon$ on $\tilde{\varOmega}_0$ by
$\gamma_\varepsilon(\boldsymbol{x})=\gamma(\boldsymbol{x})$ for
$\boldsymbol{x}\in\varOmega_0$, 
$\gamma_\varepsilon(\boldsymbol{x})=\varepsilon$
for $\boldsymbol{x}\in\tilde{\varOmega}_0\setminus\varOmega_0$.

For $h\to0$, the convex-hull-stripping median of the set $\mathcal{X}$
asymptotically coincides with the limit for $\varepsilon\to0$
of the vanishing point of the curve evolution
\begin{equation}
\tilde{\boldsymbol{c}}_t(p,t) =
\gamma_\varepsilon(\tilde{\boldsymbol{c}}(p,t))^{-2/3}
\kappa(p,t)^{1/3}
\boldsymbol{n}_{\tilde{\boldsymbol{c}}}(p,t)
\label{cpde-chsmed-1}
\end{equation}
where $\tilde{\boldsymbol{c}}$ is initialised at $t=0$ with the convex closed
curve $\partial\tilde{\varOmega}_0$.
\end{proposition}

\runinhead{Algorithmic aspects.}
The core component of the convex-hull-stripping procedure for finite
multisets is the computation of the convex hull. This problem is
related to the half-space median computation as the convex hull consists
of the points of non-zero half-space depth. Some sources mentioned
above in the paragraph on half-space median algorithms therefore provide
also algorithms that can be used here.

In the bivariate case, \cite{Matousek-Cg91} shows that the convex hull
of a finite multiset can be computed in $\mathcal{O}(N\log^4N)$. 
Repeating this procedure at most $\mathcal{O}(N)$ times during the
stripping procedure, the bivariate convex-hull-stripping median can
be computed in $\mathcal{O}(N^2\log^4N)$.

For higher dimensions, the computation of halfspace depth in 
$\mathcal{O}(N^{n-1}\log N)$ can be used to compute the depth of
all data points (and thereby the convex hull) in $\mathcal{O}(N^n\log N)$.
The convex-hull-stripping median can then be computed with
$\mathcal{O}(N^{n+1}\log N)$ effort.

\runinhead{Equivariance.}
In \cite{Small-ISR90} only affine equivariance is mentioned for the
convex-hull-stripping median. 
However, our argument for the half-space median above stays valid for
the convex-hull-stripping median: The convex-hull-stripping procedure,
too, relies solely on the splitting of $\mathbb{R}^n$
into half-spaces by hyperplanes, and the incidence of data points with these
half-spaces.

As a consequence, the convex-hull-stripping median is equivariant
under the same sub-class of projective transformations as the half-space
median.

\begin{figure}[p!]
\unitlength0.0083\textwidth 
\begin{picture}(120,136)
\put(10,0){
\put(1,137){\rotatebox{270}{\includegraphics[height=46\unitlength]{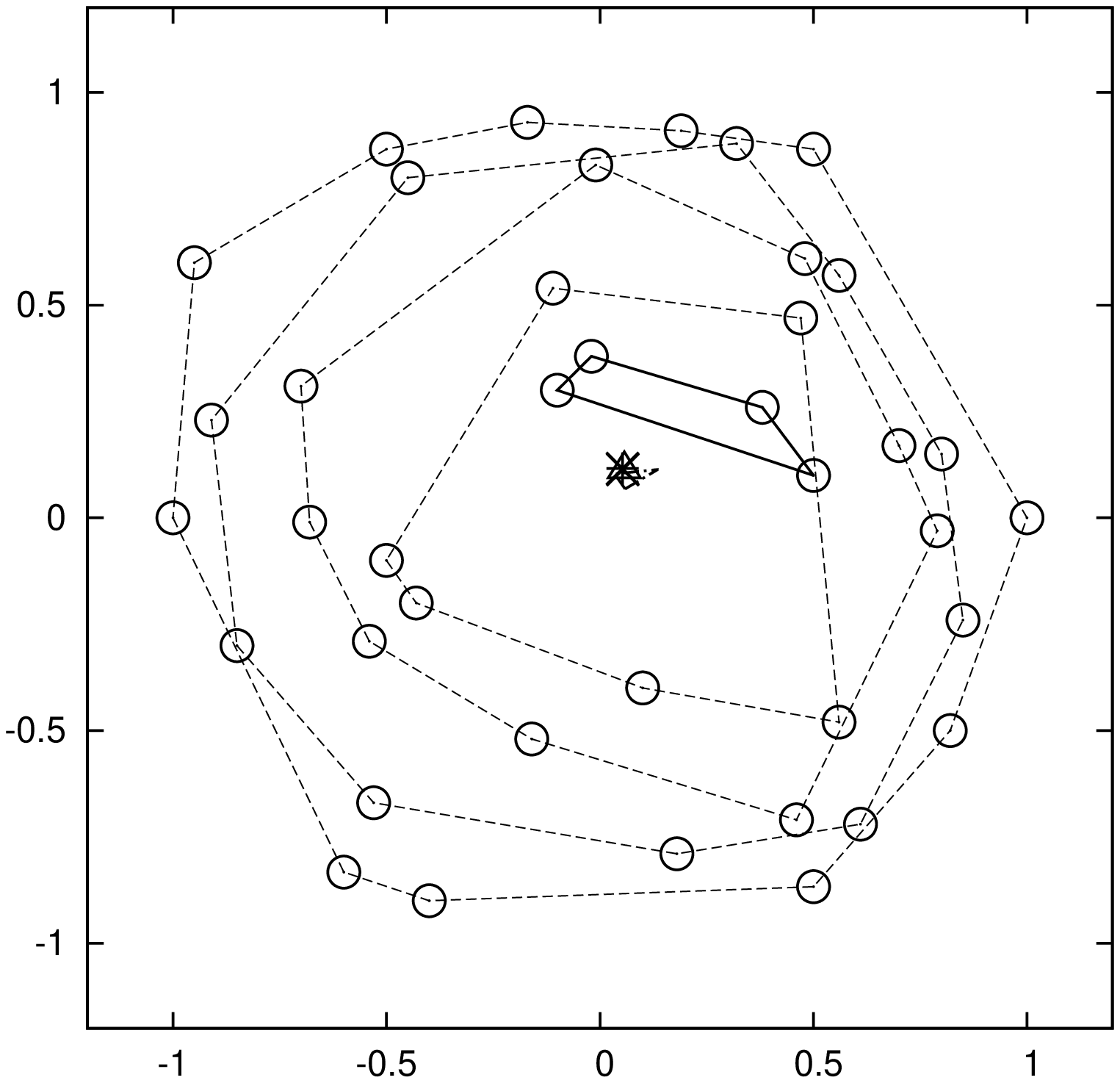}}}
\put(1,95){\rotatebox{270}{\includegraphics[height=46\unitlength]{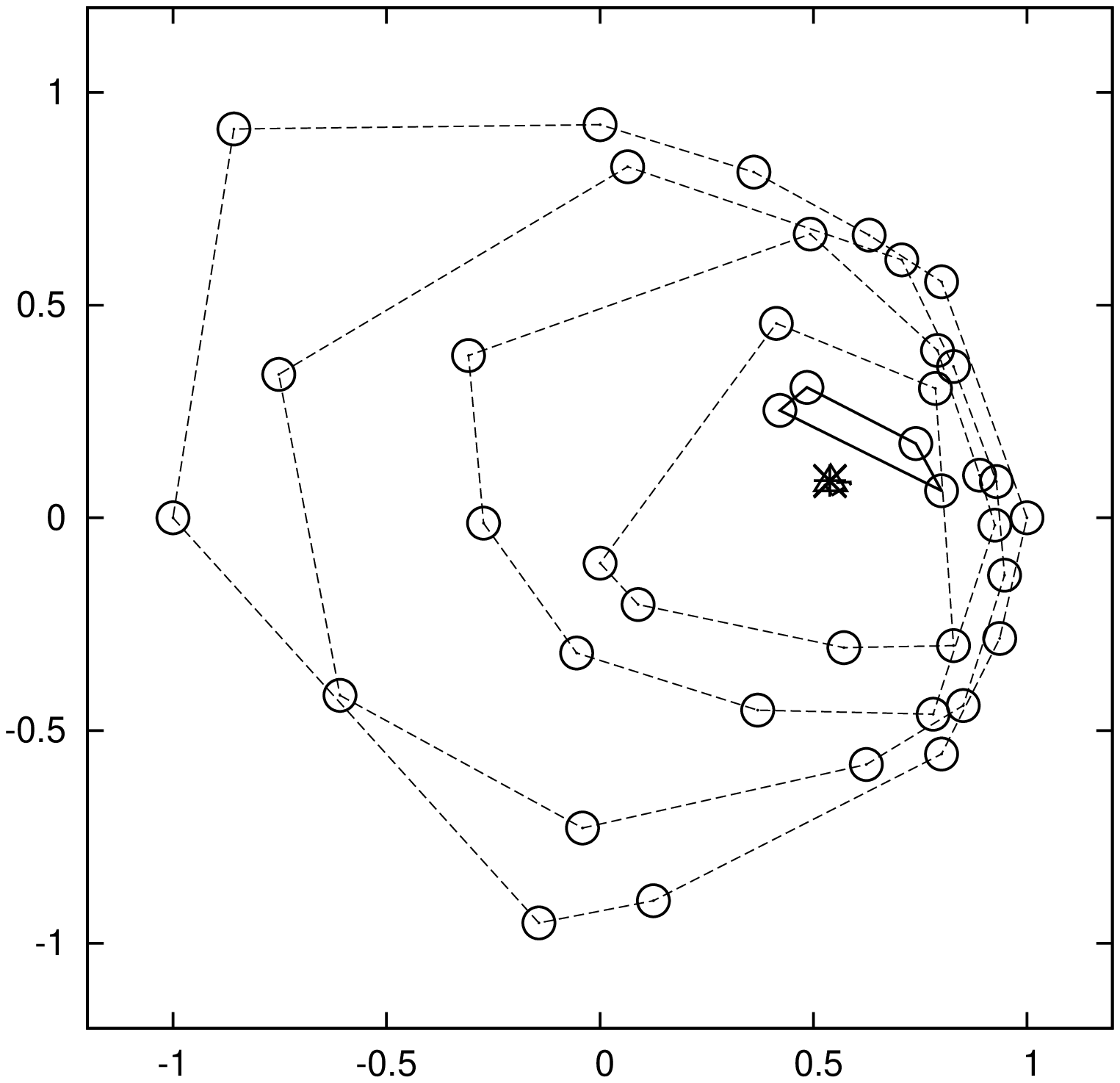}}}
\put(1,53){\rotatebox{270}{\includegraphics[height=46\unitlength]{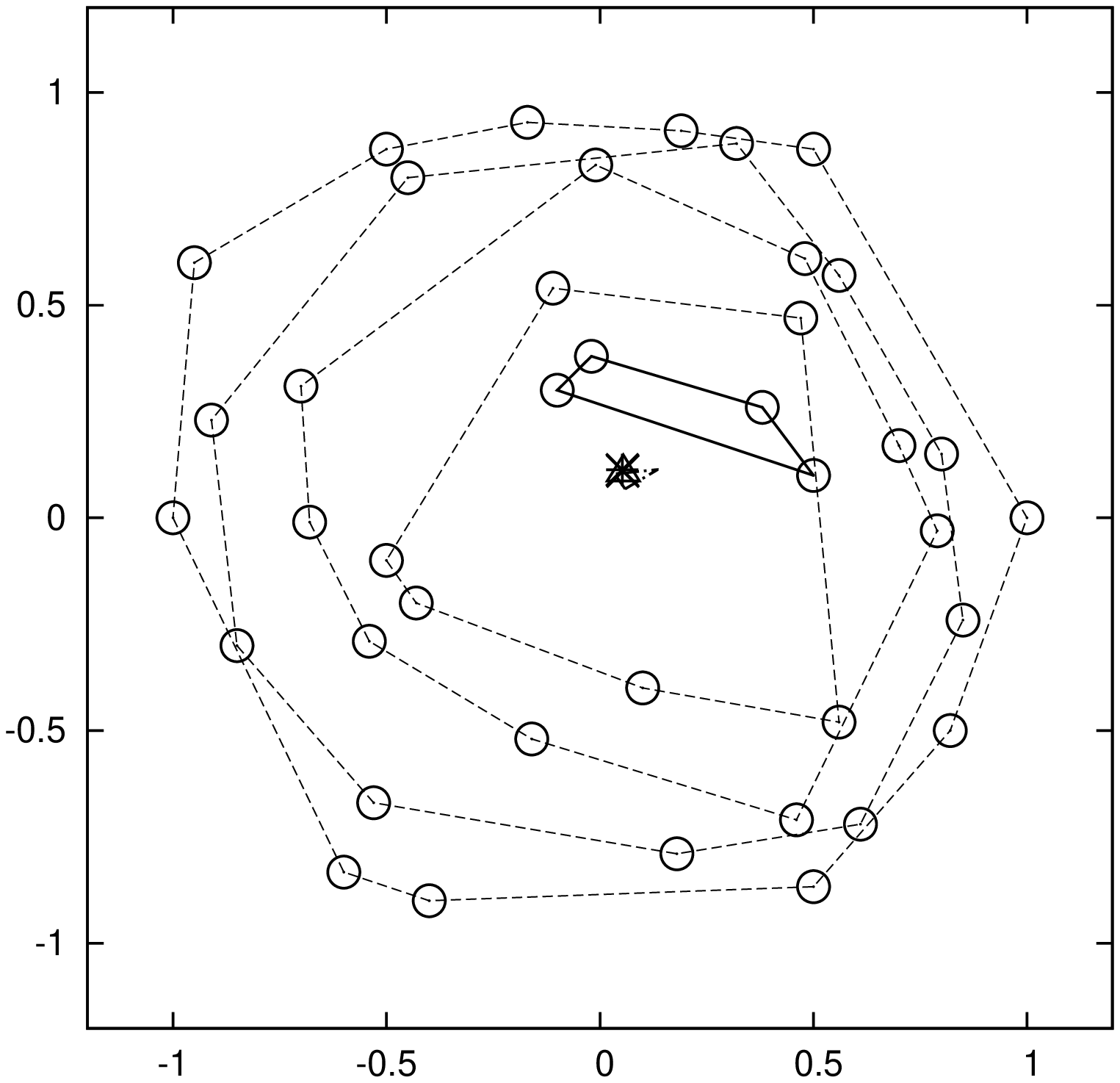}}}
\put(53,137){\rotatebox{270}{\includegraphics[height=46.85\unitlength]{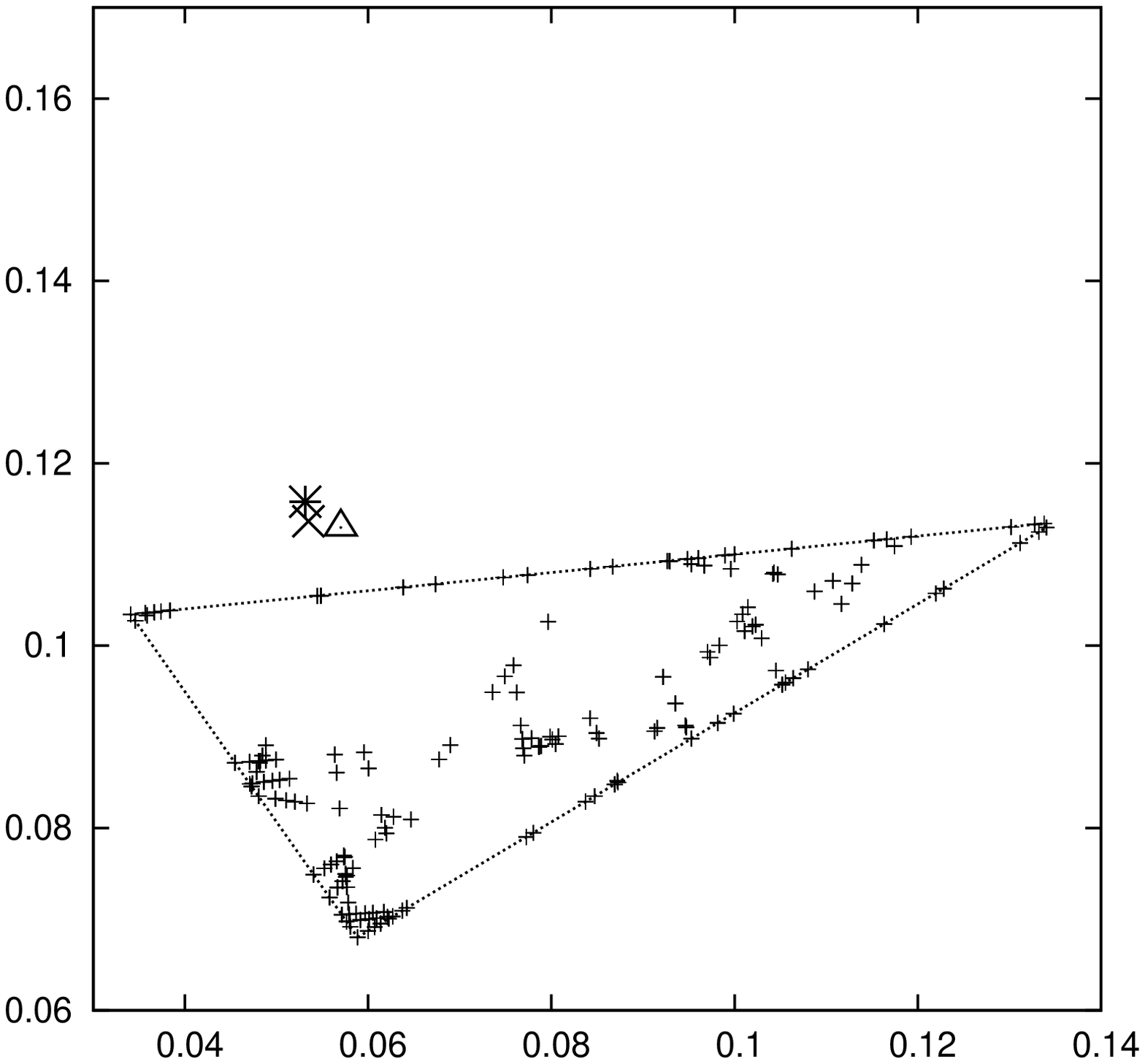}}}
\put(53,95){\rotatebox{270}{\includegraphics[height=46.85\unitlength]{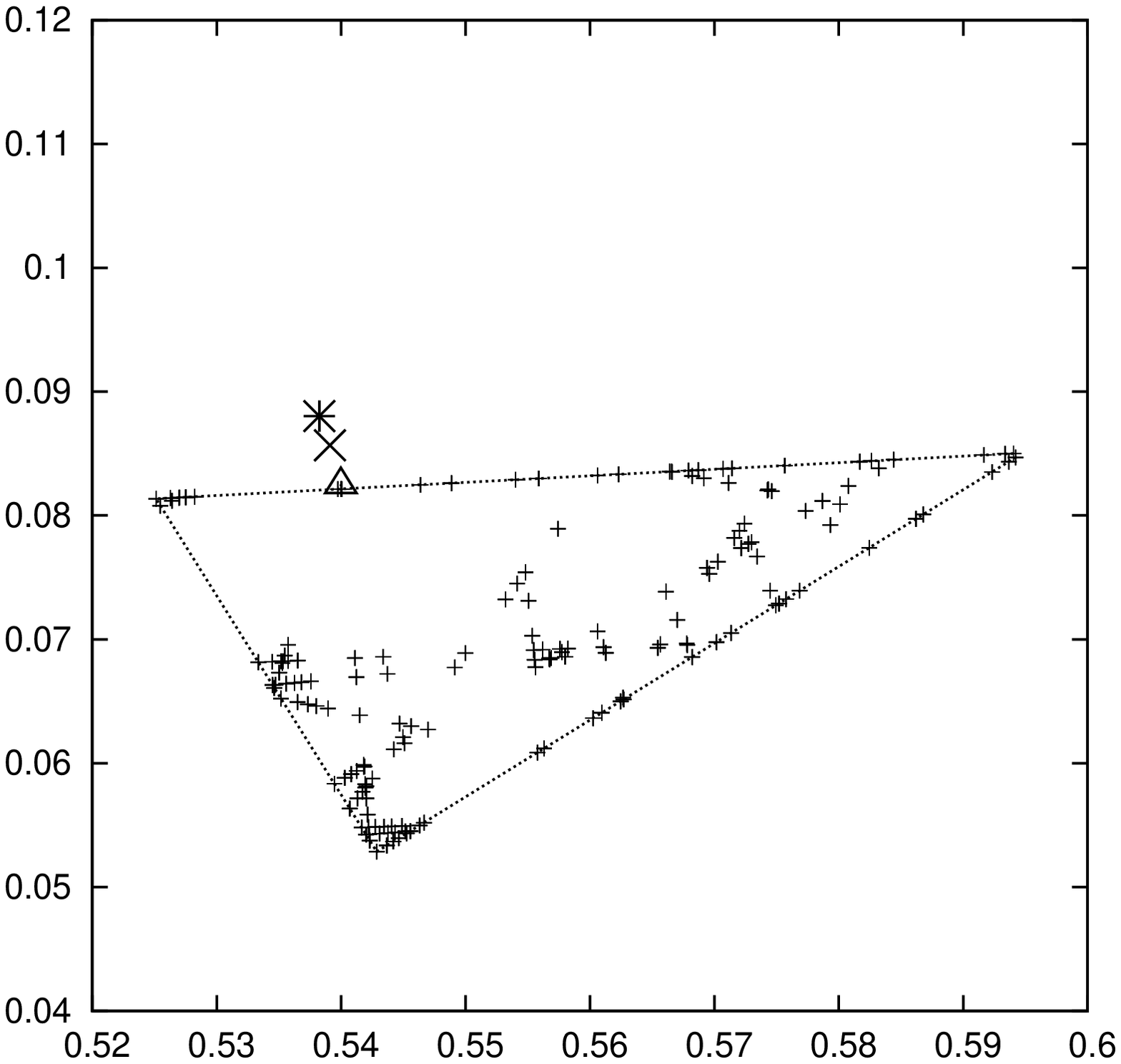}}}
\put(53,53){\rotatebox{270}{\includegraphics[height=46.85\unitlength]{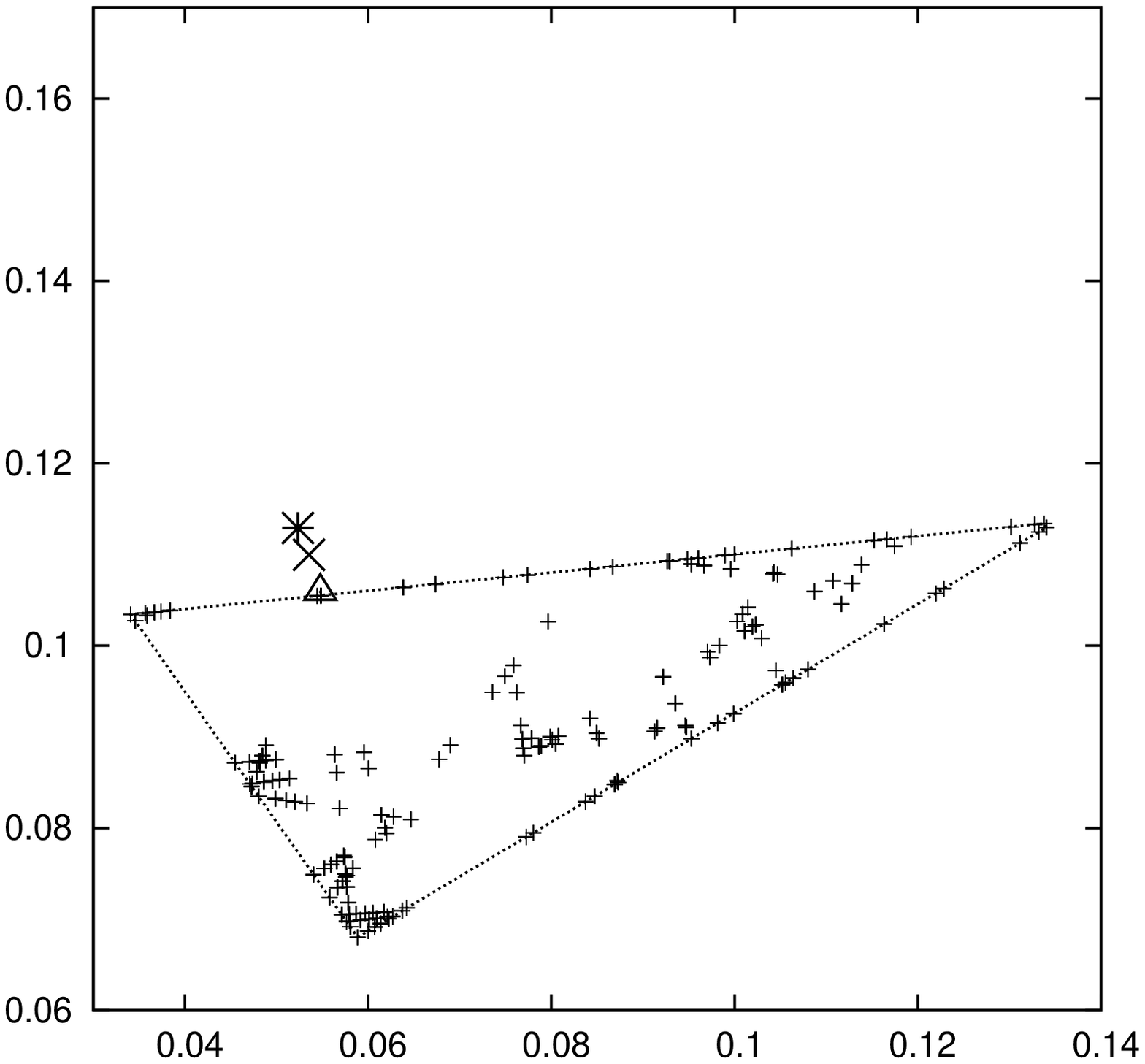}}}
\put(0,96){(a)}
\put(50,96){(b)}
\put(0,54){(c)}
\put(50,54){(d)}
\put(0,12){(e)}
\put(50,12){(f)}
\put(0,7){\parbox[t]{100\unitlength}{\small\raggedright Legend:
\raisebox{2\unitlength}{\rotatebox{270}{\includegraphics[width=2\unitlength]{images/legend_datapt.eps}}}
Data points \quad
\raisebox{2\unitlength}{\rotatebox{270}{\includegraphics[width=2\unitlength]{images/legend_l1pt.eps}}}
$L^1$ median \quad
\raisebox{2\unitlength}{\rotatebox{270}{\includegraphics[width=2\unitlength]{images/legend_trl1pt.eps}}}
TR-$L^1$ median \quad
\raisebox{2\unitlength}{\rotatebox{270}{\includegraphics[width=2\unitlength]{images/legend_ojapt.eps}}}
Oja median \\
\raisebox{2\unitlength}{\rotatebox{270}{\includegraphics[width=2\unitlength]{images/legend_hsln.eps}}}
\raisebox{2\unitlength}{\rotatebox{270}{\includegraphics[width=2\unitlength]{images/legend_hspt.eps}}}
half-space median (outline and intersection points within median region) \\
\raisebox{2\unitlength}{\rotatebox{270}{\includegraphics[width=2\unitlength]{images/legend_chsln.eps}}}
convex-hull-stripping median (outline) \quad
\raisebox{2\unitlength}{\rotatebox{270}{\includegraphics[width=2\unitlength]{images/legend_chsstepln.eps}}}
convex hull stripping steps}
}}
\end{picture}
\caption{Comparison of multivariate median concepts.
\textbf{(a)} Data set of $40$ points with multivariate medians. -- 
\textbf{(b)} Zoomed region
from (a) showing the $L^1$, transformation--retransformation $L^1$, Oja
and half-space median. Crosses in the half-space median area indicate further
intersection points contained. -- 
\textbf{(c)} Projective transformed data set from (a)
and medians of transformed data. -- 
\textbf{(d)} Zoomed region of (c). --
\textbf{(e)} Data and medians from (c) transformed back. -- 
\textbf{(f)} Zoomed region of (e).
}
\label{fig-mvmed40}
\end{figure}

In Figure~\ref{fig-mvmed40} we demonstrate the restricted projective
equivariance of half-space and convex-hull-stripping median and show
that the other medians do not share this property.
In this example, multivariate medians are computed from an original
data set, Figure~\ref{fig-mvmed40}(a, b) and a projectively transformed
version, Figure~\ref{fig-mvmed40}(c, d). After transforming the medians
back, Figure~\ref{fig-mvmed40}(e, f), the half-space and 
convex-hull-stripping medians coincide with those from (a, b) whereas
this is not true for the other three multivariate medians.

\subsection{Remarks on Further Concepts}

The list of multivariate median concepts discussed so far is not exhaustive.
In \cite{Small-ISR90} some more concepts are mentioned.
Several of these approaches can be formulated similarly to the half-space
median as maximisation of some depth of a median candidate point 
$\boldsymbol{\mu}$ w.r.t.\ the given data $\mathcal{X}$. 
Instead of the half-space depth, one can use e.g.\ the 
\emph{simplicial depth,} i.e.\ the number of simplices with vertices from
$\mathcal{X}$ that contain $\boldsymbol{\mu}$, which is conceptually
linked to the Oja median construction.
Also variants of the Oja median objective function can be considered in
which the simplex volumes are replaced by e.g.\ the height of 
$\boldsymbol{\mu}$ over the base of each simplex.
As a modification of the convex-hull-stripping procedure, 
one can consider a minimal ellipsoid enclosing $\mathcal{X}$ instead of
the convex hull, and establish a stripping procedure that deletes just
the data points on the boundary of this ellipsoid.
Another variant is the \emph{zonoid median} discussed in
\cite{Dyckerhoff-compstat96}.
For more details and further variants see also the references in 
\cite{Dyckerhoff-compstat96,Small-ISR90}.

\section{Image Filtering of Multivariate Images}
\label{sec-mvmedf}

The first approaches to construct median filters for multivariate images
go back to around 1990. In \cite{Astola-PIEEE90}, 
by the name \emph{vector median filter} a minimisation of the  
objective function of the $L^1$ median but restricted to the sample
values is introduced, see also \cite{Viero-TCSVT94} for an application
to video sequences. In modern nomenclature, compare \cite{Struyf-JSS97},
such a concept could 
rather be called a medoid. The actual $L^1$ median filter was
brought to image processing in \cite{Spence-icip07} for RGB colour images,
and in \cite{Welk-dagm03} for matrix-valued images. The latter work was
extended by some variants with other matrix norms instead of the
Euclidean (thus, Frobenius) norm for matrices in \cite{Welk-SP07} and
transferred to colour images in \cite{Kleefeld-cciw15} by mapping
the RGB colour space to symmetric matrices. Affine equivariant medians
as the basis for image filters were considered in \cite{Welk-JMIV16}.
 
\subsection{$L^1$ Median}

The application of the $L^1$ median to multivariate image data is fairly
straightforward. In particular, the $L^1$ median of data in $\mathbb{R}^n$
naturally decays into a lower-dimensional $L^1$ or univariate median if
the data lie on a hyperplane or submanifold, it works for any
dimensions of the image domain and data range.

\subsection{Affine Equivariant Medians}
\label{ssec-mvmedf-affeq}

The affine equivariant medians considered in Section~\ref{sec-mvmed} 
are in the one or other way sensitive to the dimension of the space
actually spanned by the data values. When applying these medians to
the filtering of multivariate images, the dimension $m$ of the image
domain and the dimension $n$ of the data space therefore need closer
consideration.

\runinhead{Dimensionality considerations.}
For bivariate planar images ($m=n=2$) as well as for trivariate volume images
($m=n=3$), and generally for every setting with $m\ge n$, the
data values within the neighbourhood of any grid location will in generally
span the data space, such that the affine equivariant medians can be 
applied straightforwardly.

This changes for $m<n$, which includes especially the practically relevant
case of trivariate planar images, such as RGB colour images, which we
will therefore discuss in the following for the Oja median.
For general $m$, $n$ with $m<n$ these considerations apply analogously.

A trivariate planar image in fact results from the discretisation of a 
function $\boldsymbol{u}: \mathbb{R}^2\supset\varOmega\to\mathbb{R}^3$, 
i.e.\ a parametrised surface in $\mathbb{R}^3$.
The values of $\boldsymbol{u}$ in the neighbourhood of a given location
are therefore sampled from a small surface patch in $\mathbb{R}^3$.
For a noise-free image, the function $\boldsymbol{u}$ can be assumed to be
smooth, resulting in almost planar surface patches.

This implies that e.g.\ the 3D Oja median applied to the
data points selected by a sliding window will be the minimiser of a sum of
simplex volumes where virtually all of the simplices are almost degenerated,
rendering the direct computation of the 3D Oja median unstable.
On the other hand, the 2D Oja median, which minimises a sum of triangle
areas, can easily be applied to these data. From the remarks in
Section~\ref{ssec-ojamed} it is evident that the 2D Oja median 
is also the proper limit case of the 3D Oja median when data collapse to
a plane. 
Oja median filtering of trivariate images over planar domains
should therefore be carried out by using the two-dimensional
Oja objective function. This is proposed in
\cite[Section~3.3]{Welk-JMIV16}.

Spelling out the two-dimensional Oja median for three-dimensional (but 
almost planar) data yields
\begin{equation}
\operatorname{med}_{\mathrm{Oja}(2,3)}(\mathcal{X}):=
\mathop{\operatorname{argmin}}\limits_{\boldsymbol{\mu}\in\mathbb{R}^3}
\sum\nolimits_{1\le i<j\le N}
\bigl\lvert[\boldsymbol{\mu},\boldsymbol{x}_i,\boldsymbol{x}_j]\bigr\rvert \;.
\label{mOja23}
\end{equation}
Of course, the 2D Oja median $\operatorname{med}_{\mathrm{Oja}(2,3)}$ 
for general 3D data is not equivariant
under affine transformations of $\mathbb{R}^3$. However, the 2D Oja median
of co-planar data from $\mathbb{R}^3$ is affine equivariant even
with respect to affine transformations of $\mathbb{R}^3$. Since the
RGB triples being filtered are almost co-planar, it can be expected
that a 2D Oja median filter for planar RGB images will display a good
approximation to affine equivariance.

Alternatively, one could enlarge the data (multi-) set
by adding shifted copies of all data points analogous to 
\eqref{oja-extenddegenerated} to ensure full dimensionality of the data
samples, compare \cite[Section~2.1]{Welk-JMIV16}, and apply the 
three-dimensional Oja median.

For the transformation--retransformation $L^1$ median, the affine normalisation
via the covariance matrix should be restricted to the plane of main variation
of the data values.

\section{PDE Limits of Multivariate Median Filters}
\label{sec-mvmedpde}

In this section, we collect known results on PDE limits of 
space-continuous multivariate median filters that extend the result by 
Guichard and Morel \cite{Guichard-sana97} mentioned in
Section~\ref{ssec-meddf-pde}.

\subsection{General Definitions}

To represent the results from \cite{Welk-JMIV16,Welk-ismm17} in a more 
compact form, we start by generalising our definitions from
Section~\ref{ssec-meddf} to multivariate images over domains of
different dimensions.

\begin{definition}[Smooth multivariate image]
\label{def-mvimg}
Let $\varOmega\subset\mathbb{R}^m$ be compact and equal to the closure of
its interior.
Let $\boldsymbol{u}:\varOmega\to\mathbb{R}^n$ be a bounded smooth function on
$\varOmega$. We call $\boldsymbol{u}$ 
\emph{$n$-variate $m$-dimensional smooth image with domain $\varOmega$}, 
or shortly \emph{$(m,n)$-image.}

We denote by $\mathrm{D}\boldsymbol{u}$ the Jacobian of $\boldsymbol{u}$,
i.e.\ the $n\times m$-tensor of its first-order derivatives,
and by $\mathrm{D}^2\boldsymbol{u}$ its Hessian, i.e.\ the
$n\times n\times m$-tensor of its second-order derivatives.
We call $\boldsymbol{x}\in\varOmega$ \emph{regular point of $\boldsymbol{u}$} 
if $\mathrm{D}\boldsymbol{u}(\boldsymbol{x})$ has maximal rank.
\end{definition}

\begin{definition}[Multivariate image evolution]
Let $\varOmega\subset\mathbb{R}^m$ be as in Definition~\ref{def-mvimg}, 
and $[0,T]\subset\mathbb{R}$ an interval.
The coordinates of $\Omega$ will be called \emph{spatial coordinates,}
the additional coordinate from $[0,T]$ \emph{time coordinate.}
Let $\boldsymbol{u}:\varOmega\times[0,T]\to\mathbb{R}^n$ be a bounded smooth 
function. We call $\boldsymbol{u}$ an 
\emph{$(m,n)$-image evolution with spatial domain $\varOmega$.}

We denote by $\mathrm{D}\boldsymbol{u}$ the spatial Jacobian of 
$\boldsymbol{u}$, i.e.\ the $n\times m$-tensor of its first-order derivatives
w.r.t.\ the spatial coordinates,
and by $\mathrm{D}^2\boldsymbol{u}$ its spatial Hessian, i.e.\ the
$n\times n\times m$-tensor of its second-order derivatives w.r.t\ the
spatial coordinates.

By $\boldsymbol{u}(t^*)$ for $t^*\in[0,T]$ we denote the $(m,n)$-image
with $\boldsymbol{u}(t^*)(\boldsymbol{x}):=\boldsymbol{u}(\boldsymbol{x},t^*)$
for all $\boldsymbol{x}\in\Omega$.
\end{definition}

\noindent
The definitions for selectors and scaled selectors from 
Section~\ref{ssec-meddf} carry over verbatim if just the scalar-valued
$u$ in these definitions is replaced with the multivariate $\boldsymbol{u}$
and the image domain $\varOmega\subset\mathbb{R}^2$ with 
$\varOmega\subset\mathbb{R}^m$.

The planar selector family $\mathcal{D}$ from Section~\ref{ssec-meddf}
can easily be generalised to arbitrary dimensions as the family
$\mathcal{B}^m=\{B^m_\varrho~|~\varrho>0\}$ of
balls with radius $\varrho$ given by 
$B^m_\varrho(\boldsymbol{x})=\{\boldsymbol{y}\in\mathbb{R}^m~|~
\lVert\boldsymbol{y-x}\rVert\le\varrho\}$.
Also the amoeba families $\mathcal{A}_\beta$ could be generalised in a
similar way but this generalisation is not needed here.

Similarly, the definition of an aggregator can be generalised to 
integrable densities over $\mathbb{R}^n$.
In particular, each of the space-continuous median operators introduced in 
Section~\ref{sec-mvmed} is an aggregator.

We give here just the adapted version of
Definition~\ref{def-localfilter-2-1}.

\begin{definition}[Local filters]
\label{def-localfilter-m-n}
Let $\boldsymbol{u}$ be an $(m,n)$-image with domain $\varOmega$,
and $S$ a selector for $\boldsymbol{u}$.
Let $\boldsymbol{u}{\upharpoonright} S$
be the restriction of $\boldsymbol{u}$ to $\varOmega\cap S$,
i.e.\ $\boldsymbol{u}{\upharpoonright} S:\varOmega\cap S\to\mathbb{R}^n$,
$(\boldsymbol{u}{\upharpoonright} S)(\boldsymbol{x})
=\boldsymbol{u}(\boldsymbol{x})$ for $\boldsymbol{x}\in\varOmega\cap S$,

Let $A$ be a $\varGamma$-aggregator on $\mathbb{R}^n$.
If for each $\boldsymbol{x}\in\varOmega$, the density 
$\gamma(\boldsymbol{u}{\upharpoonright} (S(\boldsymbol{x})):
\mathbb{R}^n\to\mathbb{R}^+_0$ of the values of 
$\boldsymbol{u}{\upharpoonright}(S(\boldsymbol{x})$ belongs to $\varGamma$,
we define $A(\boldsymbol{u},S):\varOmega\to\mathbb{R}^n$ by
$A(\boldsymbol{u},S)(\boldsymbol{x}):=
A\bigl(\gamma(\boldsymbol{u}{\upharpoonright} (\boldsymbol{x}+S))\bigr)$ 
for all $\boldsymbol{x}\in\varOmega$
and call it \emph{$(A,S)$-filtering of $\boldsymbol{u}$.}

If $\varSigma$ is a scaled selector for $\boldsymbol{u}$, and the 
$(A,\varSigma_{\boldsymbol{u},\varrho}$-filtering
of $\boldsymbol{u}$ exists for all $\varrho>0$, we call the family
of $(A,\varSigma_{\boldsymbol{u},\varrho})$-filterings the 
\emph{$(A,\varSigma)$-filtering of $\boldsymbol{u}$.}
\end{definition}

\noindent
Adapted to the case of $(m,n)$-images, the definition of a regular
PDE limit takes the following form.

\begin{definition}[Regular PDE limit]
\label{def-regpdelimit-m-n}
Let $\varSigma$ be a scaled selector in $\mathbb{R}^m$, 
and $A$ a $\varGamma$-aggregator on $\mathbb{R}^n$.

The second-order PDE 
$\boldsymbol{u}_t=F(\mathrm{D}\boldsymbol{u},\mathrm{D}^2\boldsymbol{u})$
is the \emph{regular PDE limit of $(A,\varSigma)$-filtering} with 
\emph{time scale $\tau(\varrho)$}
if for each $(m,n)$-image $\boldsymbol{u}$ the $(A,\varSigma)$-filtering
exists, and 
if there exists an $\varepsilon>0$ such that
for each regular point $\boldsymbol{x}$ of $\boldsymbol{u}$
one has for $\varrho\to0$
\begin{align}
\frac{A(\boldsymbol{u},\varSigma_{\boldsymbol{u},\varrho})
(\boldsymbol{x})
-\boldsymbol{u}(\boldsymbol{x})}{\tau(\varrho)}
- F\bigl(\mathrm{D}\boldsymbol{u}(\boldsymbol{x}),
\mathrm{D}^2\boldsymbol{u}(\boldsymbol{x})\bigr) =
\mathcal{O}(\varrho^{\varepsilon}) \;.
\label{regpdelimit-m-n}
\end{align}
\end{definition}
 
\noindent
Since we will need a local version of this limit process in the following,
we add the following definition that was omitted in the univariate case
in Section~\ref{ssec-meddf-pde}.

\begin{definition}[Local PDE limit]
Let $\varSigma$ and $A$ be as in Definition~\ref{def-regpdelimit-m-n}.
Let $\boldsymbol{u}$ be an $(m,n)$-image with a regular point 
$\boldsymbol{x}_0$.

The second-order PDE 
$\boldsymbol{u}_t=F(\mathrm{D}\boldsymbol{u},\mathrm{D}^2\boldsymbol{u})$
is the \emph{local PDE limit of $(A,\varSigma)$-filtering} with
\emph{time scale $\tau(\varrho)$}
for $\boldsymbol{u}$ at $\boldsymbol{x}_0$ if
the $(A,\varSigma)$-filtering of $\boldsymbol{u}$ 
at $\boldsymbol{x}_0$ exists,
and 
if there exists an $\varepsilon>0$ such that
one has for $\varrho\to0$
\begin{align}
\frac{A(\boldsymbol{u},\varSigma_{\boldsymbol{u},\varrho})
(\boldsymbol{x}_0)
-\boldsymbol{u}(\boldsymbol{x}_0)}{\tau(\varrho)}
- F\bigl(\mathrm{D}\boldsymbol{u}(\boldsymbol{x}_0),
\mathrm{D}^2\boldsymbol{u}(\boldsymbol{x}_0)\bigr) =
\mathcal{O}(\varrho^{\varepsilon}) \;.
\end{align}
\end{definition}

\subsection{Bivariate Planar Images}

We start now by considering bivariate planar images 
($\mathbb{R}^2\to\mathbb{R}^2$). 
As colour images usually have at least three channels,
the practical relevance of bivariate images is limited, with
planar optical flow fields as the most prominent exception.
However, they are the most accessible case for theoretical analysis, and 
therefore the broadest range of asymptotic results are available for
this setting.

\runinhead{Normalisations.}
All results on PDE limits of multivariate median filters reported
in the following are obtained in two steps: 
First, an asymptotic analysis is carried out for
a specific geometric situation where the Jacobian made up by the first
derivatives of the multivariate image is aligned with coordinate directions.
Second, this result is generalised to general geometric situations by using
a transformation group depending on the equivariance properties of the
respective median filter, i.e.\ a similarity or affine transformation.

For filtering processes that are just Euclidean/similarity equivariant, 
similarity transforms will be used. 
As a result, the PDE in the general case
will naturally be stated in terms of
\emph{geometric coordinates} aligned with the principal directions of
the structure tensor of the image. 
This applies obviously to standard
$L^1$ median filtering. However, it happens also to be the case for amoeba 
median filtering even with affine equivariant multivariate medians because
the amoeba structuring elements depend on the Euclidean structure of the
image domain.
In contrast, filtering processes with affine equivariance are normalised in
the analysis by affine transformations. The general PDE in these cases
is stated in standard coordinates because the geometric coordinates
have no specific meaning in this context.

We give here the basic definitions for geometric coordinates and Euclidean
normalisation.

\begin{definition}[Geometric coordinates]
\label{def-geocoo22}
Let $\boldsymbol{u}$ be a $(2,2)$-image and $\boldsymbol{x}$ a regular
location of $\boldsymbol{u}$. 
Let
\begin{align}
\boldsymbol{J}&:=\boldsymbol{J}(\mathrm{D}\boldsymbol{u}(\boldsymbol{x})):=
\boldsymbol{\nabla}u\boldsymbol{\nabla}u^\mathrm{T}
+\boldsymbol{\nabla}v\boldsymbol{\nabla}v^\mathrm{T}=
\mathrm{D}\boldsymbol{u}^\mathrm{T}\mathrm{D}\boldsymbol{u}
\end{align}
where all derivatives are evaluated at $\boldsymbol{x}$.
Then $\boldsymbol{J}$ is called \emph{structure tensor} of $\boldsymbol{u}$
at $\boldsymbol{x}$.
At each regular location $\boldsymbol{x}$,
let $\boldsymbol{\eta}$, $\boldsymbol{\xi}$ be unit eigenvectors 
for the major and minor eigenvalues, respectively,
of $\boldsymbol{J}(\mathrm{D}\boldsymbol{u}(\boldsymbol{x}))$.
We call $\boldsymbol{\eta},\boldsymbol{\xi}$ \emph{geometric coordinates}.
\end{definition}

\noindent
The geometric coordinates are the spatial directions in the image domain
$\varOmega$ in which the function values of $\boldsymbol{u}$ show the
fastest and slowest change (measured by the Euclidean norm), and thus
the closest analoga to the gradient and level line directions of
univariate images, see \cite{Chung-SPL00}. In the univariate case, these
coordinates occur naturally when describing curvature-based image evolutions
such as (mean) curvature motion. The bivariate generalisations will be 
useful in representing some PDE evolutions arising from bivariate median
filters.

\begin{definition}[Euclidean normalisation matrix]
\label{def-eucnorm22}
Let $\boldsymbol{u}$ and $\boldsymbol{x}$ be as in 
Definition~\ref{def-geocoo22}.
Define the matrix $\boldsymbol{R}$ at location $\boldsymbol{x}$ using the
eigenvector matrix 
$\boldsymbol{P}:=\bigl(\boldsymbol{\eta}~|~\boldsymbol{\xi}\bigr)$
of the structure tensor, the Jacobian $\mathrm{D}\boldsymbol{u}$, and
the directional derivatives of $\boldsymbol{u}$ in the directions of the
geometric coordinates as
\begin{align}
\boldsymbol{R} &:= (\mathrm{D}\boldsymbol{u}^{-1})^\mathrm{T}\,
\boldsymbol{P}\,\mathrm{diag}\bigl(
\lVert\partial_{\boldsymbol{\eta}}\boldsymbol{u}\rVert,
\lVert\partial_{\boldsymbol{\xi}}\boldsymbol{u}\rVert
\bigr)\;.
\end{align}
We call $\boldsymbol{R}$ \emph{Euclidean normalisation matrix} of
$\boldsymbol{u}$ at $\boldsymbol{x}$.
\end{definition}

\noindent
The Euclidean normalisation matrix is a rotation matrix that
allows to transform a bivariate planar
image locally in such a way that its geometric coordinates are aligned to
coordinate axes in the image domain $\varOmega$, \emph{and} the 
directional derivatives along the geometric coordinates are aligned to
coordinate axes in the range of image values.

\runinhead{$L^1$ median.}
As equivariance of $L^1$ median filtering is restricted to similarity
transformations, its general PDE limit will be derived from a special
geometric situation that can be reached by Euclidean transformations
and is covered by the following lemma.

\begin{lemma}\cite[Lemma~1]{Welk-JMIV16}
\label{lem-l1pde22-aligned}
For a $(2,2)$-image $\boldsymbol{u}$ with a regular point $\boldsymbol{x}_0$
where $\mathrm{D}\boldsymbol{u}(\boldsymbol{x}_0)=\mathrm{diag}(u_x,v_y)$
with $u_x\ge v_y>0$ (i.e.\ $u_y=v_x=0$), the local PDE limit of
$(\operatorname{med}_{L^1},\mathcal{B}^2)$-filtering of $\boldsymbol{u}$ at
$\boldsymbol{x}_0$ with time scale $\varrho^2/6$ is
\begin{align}
u_t &= 
Q_1{\left(\frac{u_x}{v_y}\right)} u_{xx} 
+ Q_2{\left(\frac{v_y}{u_x}\right)} u_{yy}
- \frac{2u_x}{v_y}\,Q_1{\left(\frac{u_x}{v_y}\right)} v_{xy}
\label{l1pde22-aligned-u}
\;,\\
v_t &= 
Q_2{\left(\frac{u_x}{v_y}\right)} v_{xx} 
+ Q_1{\left(\frac{v_y}{u_x}\right)} v_{yy} 
- \frac{2v_y}{u_x}\,Q_1{\left(\frac{v_y}{u_x}\right)} u_{xy}
\label{l1pde22-aligned-v}
\;.
\end{align}
The functions $Q_1,Q_2:[0,\infty]\to\mathbb{R}$
are given by the quotients of elliptic integrals
\begin{align}
Q_1(\lambda) &= \frac{3 \iint_{D_1(\boldsymbol{0})} 
s^2t^2/(s^2+\lambda^2t^2)^{3/2}\,\mathrm{d}s\,\mathrm{d}t}
{\iint_{D_1(\boldsymbol{0})}
s^2/(s^2+\lambda^2t^2)^{3/2}\,\mathrm{d}s\,\mathrm{d}t}\;,\\
Q_2(\lambda) &= \frac{3 \iint_{D_1(\boldsymbol{0})} 
t^4/(s^2+\lambda^2t^2)^{3/2}\,\mathrm{d}s\,\mathrm{d}t}
{\iint_{D_1(\boldsymbol{0})}
t^2/(s^2+\lambda^2t^2)^{3/2}\,\mathrm{d}s\,\mathrm{d}t}
\end{align}
for $\lambda\in(0,\infty)$, together with the limits $Q_1(0)=Q_2(0)=1$.
\end{lemma}

\noindent
The elliptic integrals in the
coefficient expressions $Q_1(\lambda)$ and $Q_2(\lambda)$
can in general not be evaluated in closed form. However, they
are connected by
\begin{equation}
Q_2(\lambda)=1-Q_1(\lambda^{-1})\;.
\label{l1-22-q1q2}
\end{equation}
Note that for $\lambda\to\infty$, $\lambda\,Q_1(\lambda)$ goes to zero such
that the coefficients for $v_{xy}$ in \eqref{l1pde22-aligned-u}
and for $u_{xy}$ in \eqref{l1pde22-aligned-v} are globally bounded
for arbitrary $u_x$, $v_y$, and in the limit case $v_y=0$ one has the
decoupled PDEs $u_t=u_{yy}$, $v_t=v_{xx}$.

A general geometric situation can be normalised to the one described in the 
lemma by a Euclidean transformation in the image domain and data space which
is given by the Euclidean normalisation matrix. 
The general PDE in the following proposition is obtained by reverting this
transformation.

\begin{proposition}[{\cite[Prop.~1]{Welk-JMIV16}}]
\label{prop-l1pde22}
For $(2,2)$-images $\boldsymbol{u}$,
$(\operatorname{med}_{L^1},\mathcal{D})$-filtering with the $L^1$
median as aggregator and the disc family $\mathcal{D}$ as scaled selector 
has the regular PDE limit with time scale $\tau(\varrho)=\varrho^2/6$
given by
\begin{align}
\boldsymbol{u}_t &=
\boldsymbol{S}(\mathrm{D}\boldsymbol{u}) 
\boldsymbol{u}_{\boldsymbol{\eta}\boldsymbol{\eta}}
+\boldsymbol{T}(\mathrm{D}\boldsymbol{u})
\boldsymbol{u}_{\boldsymbol{\xi}\boldsymbol{\xi}}
-2\,\boldsymbol{W}(\mathrm{D}\boldsymbol{u})
\boldsymbol{u}_{\boldsymbol{\xi}\boldsymbol{\eta}}
\label{l1pde22}
\end{align}
where $\boldsymbol{\eta},\boldsymbol{\xi}$ are the geometric coordinates,
and 
the coefficient matrices
$\boldsymbol{S}(\mathrm{D}\boldsymbol{u})$, 
$\boldsymbol{T}(\mathrm{D}\boldsymbol{u})$
and $\boldsymbol{W}(\mathrm{D}\boldsymbol{u})$
are given by
\begin{align}
\boldsymbol{S}(\mathrm{D}\boldsymbol{u})
&:= \boldsymbol{R}\,
\mathrm{diag}\left(
Q_1{\left(\frac{\lVert\partial_{\boldsymbol{\eta}}\boldsymbol{u}\rVert}
{\lVert\partial_{\boldsymbol{\xi}}\boldsymbol{u}\rVert}\right)},
Q_2{\left(\frac{\lVert\partial_{\boldsymbol{\eta}}\boldsymbol{u}\rVert}
{\lVert\partial_{\boldsymbol{\xi}}\boldsymbol{u}\rVert}\right)}
\right)
\boldsymbol{R}^\mathrm{T}\;,
\label{l1pde22-S}
\\
\boldsymbol{T}(\mathrm{D}\boldsymbol{u})
&:= \boldsymbol{R}\,
\mathrm{diag}\left(
Q_2{\left(\frac{\lVert\partial_{\boldsymbol{\xi}}\boldsymbol{u}\rVert}
{\lVert\partial_{\boldsymbol{\eta}}\boldsymbol{u}\rVert}\right)},
Q_1{\left(\frac{\lVert\partial_{\boldsymbol{\xi}}\boldsymbol{u}\rVert}
{\lVert\partial_{\boldsymbol{\eta}}\boldsymbol{u}\rVert}\right)}\right)
\boldsymbol{R}^\mathrm{T}\;,
\label{l1pde22-T}
\\
\boldsymbol{W}(\mathrm{D}\boldsymbol{u})
&:= \boldsymbol{R}\,
\begin{pmatrix}
0&
\frac{\lVert\partial_{\boldsymbol{\eta}}\boldsymbol{u}\rVert}
{\lVert\partial_{\boldsymbol{\xi}}\boldsymbol{u}\rVert}\,
Q_1{\left(\frac{\lVert\partial_{\boldsymbol{\eta}}\boldsymbol{u}\rVert}
{\lVert\partial_{\boldsymbol{\xi}}\boldsymbol{u}\rVert}\right)}
\\
\frac{\lVert\partial_{\boldsymbol{\xi}}\boldsymbol{u}\rVert}
{\lVert\partial_{\boldsymbol{\eta}}\boldsymbol{u}\rVert}\,
Q_1{\left(\frac{\lVert\partial_{\boldsymbol{\xi}}\boldsymbol{u}\rVert}
{\lVert\partial_{\boldsymbol{\eta}}\boldsymbol{u}\rVert}\right)}
&
0
\end{pmatrix}
\boldsymbol{R}^\mathrm{T}\;,
\label{l1pde22-W}
\end{align}
where
$\boldsymbol{R}$ is the Euclidean normalisation matrix,
and the coefficient functions $Q_1$, $Q_2$ are those stated in
Lemma~\ref{lem-l1pde22-aligned}.
\end{proposition}

\noindent
Equivariance of the PDE \eqref{l1pde22}
with regard to Euclidean transformations of the $u$-$v$ plane
follows immediately from its derivation for a special
case and transfer to the general case by a Euclidean
transformation.

Univariate median filtering is contained in the statement of
Lemma~\ref{lem-l1pde22-aligned} when $v_y$ is sent to $0$. In this case,
the first PDE \eqref{l1pde22-aligned-u} becomes $u_t=u_{yy}$ by virtue
of $Q_1(\infty)=0$, $Q_2(0)=1$, and the previous remark.
This translates to $u_t=u_{\boldsymbol{\xi\xi}}$ in the general setting of
Proposition~\ref{prop-l1pde22}, i.e.\ the (mean) curvature motion
equation, thus reproducing exactly the result of \cite{Guichard-sana97}.

\runinhead{Oja and transformation--retransformation $L^1$ median.}
We turn now to the two mini\-mi\-sa\-tion-based affine equivariant medians.
As was shown in \cite{Welk-JMIV16},
Oja median filtering and transformation--retransformation $L^1$ median
filtering approximate the same PDE, i.e.\ they are asymptotically
equivalent. The following lemma states the PDE limit for Oja median filtering
in a normalised geometric setting. Since now affine transformations of the
data space can be used for normalisation, a simpler case -- with a unit 
Jacobian -- is sufficient to consider in this step.

\begin{lemma}\cite[Lemma~2]{Welk-JMIV16}
\label{lem-ojapde22-I}
For a $(2,2)$-image $\boldsymbol{u}$ with a regular point $\boldsymbol{x}_0$
where $\mathrm{D}\boldsymbol{u}(\boldsymbol{x}_0)=\mathrm{diag}(1,1)$,
the local PDE limit of
$(\operatorname{med}_{\mathrm{Oja}},\mathcal{D})$-filtering of 
$\boldsymbol{u}$ at $\boldsymbol{x}_0$ with time scale $\varrho^2/24$ is
\begin{align}
u_t &= u_{xx}+3u_{yy}-2v_{xy}\;, \label{ojapde22-I1} \\
v_t &= 3v_{xx}+v_{yy}-2u_{xy}\;. \label{ojapde22-I2}
\end{align}
\end{lemma}

\noindent
The setting described can be reached from a generic geometric situation by
an affine transform in the data space.
It is worth noting that the PDE system \eqref{ojapde22-I1}, \eqref{ojapde22-I2}
coincides exactly with the special case $u_x=v_y=1$ of 
Lemma~\ref{lem-l1pde22-aligned} (rewritten to the time scale $\varrho^2/24$
of Lemma~\ref{lem-ojapde22-I}).
Transformation--retransformation $L^1$ filtering coincides in this case
with standard $L^1$ filtering. Thereby, reverting the affine transform leading
to the setting of Lemma~\ref{lem-ojapde22-I} yields the general PDE limit
not only for Oja but also for transformation--retransformation $L^1$ filtering.
In this context it is important to notice that the covariance matrix of
the data within the patch $D_\varrho$ is (up to scaling) asymptotically equal
to $\mathrm{D}\boldsymbol{u}\,\mathrm{D}\boldsymbol{u}^{\mathrm{T}}$ such that
the affine transform based in the definition of 
the continuous transformation--retransformation $L^1$ median essentially
becomes $\mathrm{D}\boldsymbol{u}^{-1}$, compare \cite[Def.~1]{Welk-JMIV16},
where a continuous version of the transformation--retransformation $L^1$ median
is stated using this transform directly.

The general PDE limit result is stated in the following proposition that 
combines the two results that were stated in \cite{Welk-JMIV16} separately 
for the Oja median and the transformation--retransformation $L^1$ median.

\begin{proposition}[{\cite[Theorem~1 and Corollary~1]{Welk-JMIV16}}]
\label{prop-ojatrl1pde22}
\sloppy
For $(2,2)$-images $\boldsymbol{u}$, both
$(\operatorname{med}_{\mathrm{Oja}},\mathcal{D})$-filtering 
with the Oja median as aggregator 
as well as
$(\operatorname{med}_{\mathrm{TR}L^1},\mathcal{D})$-filtering 
with the transformation--retransformation $L^1$ median as aggregator
and the disc family $\mathcal{D}$ as scaled selector
have the regular PDE limit with time scale $\tau(\varrho)=\varrho^2/24$
given by
\begin{align}
\boldsymbol{u}_t &=
2\,\Delta\boldsymbol{u}
+
\boldsymbol{A}(\mathrm{D}\boldsymbol{u})
(\boldsymbol{u}_{yy}-\boldsymbol{u}_{xx})
+
\boldsymbol{B}(\mathrm{D}\boldsymbol{u})
\boldsymbol{u}_{xy}
\label{ojapde22}
\end{align}
with the coefficient matrices
\begin{align}
\boldsymbol{A}(\mathrm{D}\boldsymbol{u})
&:=
\frac{1}{u_xv_y-u_yv_x} 
\begin{pmatrix}u_xv_y+u_yv_x&-2u_xu_y\\2v_xv_y&-u_xv_y-u_yv_x\end{pmatrix}
\;,
\label{ojapde22-A}
\\
\boldsymbol{B}(\mathrm{D}\boldsymbol{u})
&:=
\frac{2}{u_xv_y-u_yv_x} 
\begin{pmatrix}u_xv_x-u_yv_y&-u_x^2+u_y^2\\v_x^2-v_y^2&-u_xv_x+u_yv_y
\end{pmatrix}
\;.
\label{ojapde22-B}
\end{align}
\end{proposition}

\noindent
The derivation of the PDE of Proposition~\ref{prop-ojatrl1pde22}
by affine transformation immediately implies its equivariance under
affine transformations of the data space. Interestingly,
the PDE itself is even equivariant under affine transformations of
the $x$-$y$ plane. Regarding the approximation of multivariate median
filtering, however, the Euclidean disc-shaped structuring element
allows only for Euclidean transformations of the $x$-$y$ plane.

To understand the effect of the PDE limit from 
Proposition~\ref{prop-ojatrl1pde22} as an image evolution, we remark that
it consists of three contributions. The first, 
$\Delta\boldsymbol{u}+\boldsymbol{A}(\mathrm{D}\boldsymbol{u})
(\boldsymbol{u}_{yy}-\boldsymbol{u}_{xx})$, 
corresponds to $2u_yy$ and $2v_{xx}$ in the 
equations~\eqref{ojapde22-I1}--\eqref{ojapde22-I2} of the lemma, and
can be understood 
as independent (mean) curvature motion in the two components $u$ and $v$,
which is in analogy with the evolution \eqref{mcm} approximated by
univariate median filtering. 
The second part, $\Delta\boldsymbol{u}$, corresponding to $u_{xx}+u_{yy}$
and $v_{xx}+v_{yy}$ in the lemma,
is a homogeneous diffusion process that blurs the image, and is not
present in the univariate process \eqref{mcm}.
The third component is the interaction term 
$\boldsymbol{B}(\mathrm{D}\boldsymbol{u})\boldsymbol{u}_{xy}$ that 
creates a cross-influence between the $u$ and $v$ components and has obviously
no analogue in the univariate case.

\runinhead{Half-space median.}
The previous results can be extended even to bivariate half-space median
filtering.

\begin{proposition}[{\cite{Welk-aapr20}}]
For a $(2,2)$-image $\boldsymbol{u}$ with a regular point 
$\boldsymbol{x}_0$
where $\mathrm{D}\boldsymbol{u}(\boldsymbol{x}_0)=\mathrm{diag}(1,1)$,
the local PDE limit of
$(\operatorname{med}_{\mathrm{HS}},\mathcal{D})$-filtering of
$\boldsymbol{u}$ at $\boldsymbol{x}_0$ with time scale $\varrho^2/24$ is
given by \eqref{ojapde22-I1}, \eqref{ojapde22-I2}.
\end{proposition}

\noindent
As a consequence, also the general PDE limit from 
Proposition~\ref{prop-ojatrl1pde22} carries over 
to half-space median filtering.

In the light of the equivariance discussion in Section~\ref{ssec-hsmed}
this has an interesting implication: The PDE limit \eqref{ojapde22}
is equivariant even under projective transformations of the data space,
as long as these do not take the bounded set of image values in the 
neighbourhood of the regular location under consideration to infinity.
This can also be verified directly by applying a projective transformation
to \eqref{ojapde22}.

In other words, \emph{although
Oja and transformation--retransformation $L^1$ median filtering
are merely affine equivariant, their infinitesimal limit is even 
projective equivariant.}

\runinhead{Convex-hull-stripping median.}
No infinitesimal analysis of the convex-hull-stripping median as filter
for multivariate images is available so far. Since with the results reported
in Section~\ref{ssec-chsmed} the space-continuous process as such can now
be described, such an analysis may be achieved in future work.
We conjecture that the PDE limit of this process coincides with the
one for Oja, transformation--retransformation $L^1$ and half-space
median filtering.

\runinhead{Oja and transformation--retransformation $L^1$ median filtering
with amoeba structuring elements.}
In the bivariate planar case, also a PDE limit for multivariate amoeba 
median filtering could be obtained \cite{Welk-ismm17}. This result is
restricted to Oja and transformation--retransformation $L^1$ medians;
the standard $L^1$ median case in combination with amoebas appears too 
complicated for analysis so far.

Unlike for the previous cases, image information now enters not only the
aggregation step by the multivariate median but also the computation of
the amoeba structuring elements. These influences are decomposed in the
analysis in \cite{Welk-ismm17}.
Whereas the aggregation step by median filtering can again be normalised
by affine transformations in the data space, and Lemma~\ref{lem-ojapde22-I}
for the Oja median together with its counterpart for the 
transformation--retransformation $L^1$ median can be reused, 
the amoeba computation only admits a Euclidean normalisation. 
The influence of the amoeba structuring elements is described for an
appropriate normalised case by the following lemma.

\begin{lemma}\cite[Lemma~2 and Proof of Theorem~2]{Welk-ismm17}
\label{lem-amoeoja2trl12-aff-asym}
Consider a $(2,2)$-image with domain $\varOmega$. Assume that
$\boldsymbol{x}_0=\boldsymbol{0}\in\varOmega$ is a regular location for
$\boldsymbol{u}$, and 
$\mathrm{D}\boldsymbol{u}(\boldsymbol{x})=\mathrm{diag}(1,1)$ for all
$\boldsymbol{x}\in\mathit{\Omega}$.
At $\mathbf{0}$, let an amoeba structuring element 
$\mathcal{A}(\mathbf{0})$
be given in polar coordinates $(r,\varphi)$ with $x=r\cos\varphi$,
$y=r\sin\varphi$ by its contour
$r(\varphi)=\varrho-a(\varphi)$,
$a(\varphi):=\frac12\varrho^2\beta^2(\alpha_1\cos^3\varphi+
\alpha_2\cos^2\varphi\sin\varphi+\alpha_3\cos\varphi\sin^2\varphi+
\alpha_4\sin^3\varphi)$.
Then one step of Oja or transformation--retransformation $L^1$ median 
filtering of $\boldsymbol{u}$ at $\mathbf{0}$ within $\mathcal{A}(\mathbf{0})$
approximates an explicit time step of size $\tau=\varrho^2/24$ of the PDE
system
\begin{align}
u_t &= -9\beta^2\alpha_1 -3\beta^2\alpha_3 \;, &
v_t &= -3\beta^2\alpha_2 -9\beta^2\alpha_4 \;.
\label{pde-amoeoja2-aff-asym}
\end{align}
\end{lemma}

\noindent
Reverting the normalisations and combining the influences of amoeba
computation and multivariate median filtering, the general PDE limit as
stated in the following proposition is obtained. Due to the Euclidean
normalisation for the amoeba computations, the geometric coordinates
and Euclidean normalisation matrix appear again in this result.

\begin{proposition}[{\cite[Theorems~1 and~2]{Welk-ismm17}}]
\label{prop-amoeoja2pde}
\sloppy
For $(2,2)$-images $\boldsymbol{u}$, both
$(\operatorname{med}_{\mathrm{Oja}},\mathcal{A}_\beta)$-filtering 
with the Oja median as aggregator 
as well as
$(\operatorname{med}_{\mathrm{TR}L^1},\mathcal{A}_\beta)$-filtering 
with the transformation--retransformation $L^1$ median as aggregator
and the amoeba family $\mathcal{A}_\beta$ for given $\beta>0$ as scaled 
selector
have the regular PDE limit with time scale $\tau(\varrho)=\varrho^2/24$
given by
\begin{align}
\partial_t\boldsymbol{u} &=
\boldsymbol{T}_1(\mathrm{D}\boldsymbol{u})
\partial_{\boldsymbol{\eta\eta}}\boldsymbol{u} +
\boldsymbol{T}_2(\mathrm{D}\boldsymbol{u})
\partial_{\boldsymbol{\xi\xi}}\boldsymbol{u} +
\boldsymbol{T}_3(\mathrm{D}\boldsymbol{u})
\partial_{\boldsymbol{\eta\xi}}\boldsymbol{u}
\label{pde-amoeoja2-generic}
\end{align}
where $\boldsymbol{\eta},\boldsymbol{\xi}$ are the geometric coordinates.
The coefficient matrices $\boldsymbol{T}_i(\mathrm{D}\boldsymbol{u})$
are given by
\begin{align}
\boldsymbol{T}_i(\mathrm{D}\boldsymbol{u}) &:=
\boldsymbol{R}\,\boldsymbol{\varTheta}_i
\bigl(\lVert\partial_{\boldsymbol{\eta}}\boldsymbol{u}\rVert,
\lVert\partial_{\boldsymbol{\xi}}\boldsymbol{u}\rVert\bigr) \,
\boldsymbol{R}^{\mathrm{T}}\;,
\qquad i=1,2,3\;,
\\
\boldsymbol{\varTheta}_1(r,s) &:=
\mathrm{diag}\bigl(\vartheta_1(r),\vartheta_2(r,s)\bigr)  \;, \\
\boldsymbol{\varTheta}_2(r,s) &:=
\mathrm{diag}\bigl(\vartheta_2(r,s),\vartheta_1(s)\bigr)  \;, \\
\boldsymbol{\varTheta}_3(r,s) &:=
-2 \begin{pmatrix} 0 & \vartheta_3(r,s) \\ \vartheta_3(s,r) & 0 \end{pmatrix}
, \\
\vartheta_1(z) &:= \frac{1-8\beta^2z^2}{(1+\beta^2z^2)^2} \;, \\
\vartheta_2(w,z)&:=\frac{3}{(1+\beta^2w^2)(1+\beta^2z^2)} \;, \\
\vartheta_3 (w,z) &:= \frac{w}{z}\,
\frac{1+4\beta^2z^2}{(1\!+\!\beta^2w^2)(1\!+\!\beta^2z^2)} \,,
\label{pde-amoeoja2-generic-coeff4}
\end{align}
with the Euclidean normalisation matrix $\boldsymbol{R}$.
\end{proposition}

\subsection{Trivariate Volume Images}

We continue with the case of trivariate volume images
($\mathbb{R}^3\to\mathbb{R}^3$). Again, this case is of limited practical
relevance as volume imaging methods often yield univariate images only
or, like diffusion-tensor MRI, already higher-dimensional data. 
Results are available for the Oja and transformation--retransformation
$L^1$ median.
In the following, $\boldsymbol{I}=\operatorname{diag}(1,1,1)$ denotes the
$3\times3$ unit matrix.

\begin{lemma}\cite[Lemmas~3 and~4]{Welk-JMIV16}
\label{lem-ojal1apde33-I}
For a $(3,3)$-image $\boldsymbol{u}$ with a regular point $\boldsymbol{x}_0$
where $\mathrm{D}\boldsymbol{u}(\boldsymbol{x}_0)=\boldsymbol{I}$,
the local PDE limit of
$(\operatorname{med}_{\mathrm{Oja}},\mathcal{B}^3)$-filtering 
as well as
$(\operatorname{med}_{L^1},\mathcal{B}^3)$-filtering
of $\boldsymbol{u}$ at
$\boldsymbol{x}_0$ with time scale $\varrho^2/20$ is
\begin{align}
u_t&=u_{xx}+2(u_{yy}+u_{zz})-(v_{xy}+w_{xz}) \label{ojapde33-I-1}\\
v_t&=v_{yy}+2(v_{xx}+v_{zz})-(u_{xy}+w_{yz}) \label{ojapde33-I-2}\\
w_t&=w_{zz}+2(w_{xx}+w_{yy})-(u_{xz}+v_{yz}) \label{ojapde33-I-3}\;.
\end{align}
\end{lemma}

\noindent
Again, the general result is obtained by reverting an affine transform
that takes a generic regular point to the special situation of the lemma.
Similarly as in the bivariate planar case, it is important to notice that
the transform by $\mathrm{D}\boldsymbol{u}^{-1}$ also coincides with
the asymptotic limit of the covariance-based transform used in the
transformation--retransformation $L^1$ median.

\begin{proposition}[{\cite[Theorem~2 and Prop.~2]{Welk-JMIV16}}]
\label{prop-ojatrl1pde33}
\sloppy
For $(3,3)$-images $\boldsymbol{u}$, 
$(\operatorname{med}_{\mathrm{Oja}},\mathcal{B}^3)$-filtering 
with the Oja median as aggregator 
as well as
$(\operatorname{med}_{\mathrm{TR}L^1},\mathcal{B}^3)$-filtering 
with the transformation--retransformation $L^1$ median as aggregator
and the ball family $\mathcal{B}^3$ as scaled selector in both cases
have the regular PDE limit with time scale $\tau(\varrho)=\varrho^2/60$
given by
\begin{align}
\boldsymbol{u}_t &=
5\,\Delta\boldsymbol{u}
+ \boldsymbol{A}_1(\mathrm{D}\boldsymbol{u}) 
(\boldsymbol{u}_{yy}-\boldsymbol{u}_{xx})
+ \boldsymbol{A}_2(\mathrm{D}\boldsymbol{u})
(\boldsymbol{u}_{zz}-\boldsymbol{u}_{xx})
\notag\\*&\quad{}
+\boldsymbol{B}_1(\mathrm{D}\boldsymbol{u})
\boldsymbol{u}_{xy}
+\boldsymbol{B}_2(\mathrm{D}\boldsymbol{u})
\boldsymbol{u}_{xz}
+\boldsymbol{B}_3(\mathrm{D}\boldsymbol{u})
\boldsymbol{u}_{yz}
\label{ojapde33}
\end{align}
where for $\boldsymbol{D}:=\mathrm{D}\boldsymbol{u}$ the coefficient matrices 
are given by
\begin{align}
\boldsymbol{A}_1(\boldsymbol{D}) &:= 
\boldsymbol{I} 
- 3\,\boldsymbol{D}\,
\mathrm{diag}(0,1,0)
\,\boldsymbol{D}^{-1}\;,
\label{ojapde33-A1}
\\
\boldsymbol{A}_2(\boldsymbol{D}) &:= 
\boldsymbol{I} 
- 3\,\boldsymbol{D}\,
\mathrm{diag}(0,0,1)
\,\boldsymbol{D}^{-1}\;,
\label{ojapde33-A2}
\\
\boldsymbol{B}_1(\boldsymbol{D}) &:= 
-3\,\boldsymbol{D}\,
{\small\begin{pmatrix}0&1&0\\1&0&0\\0&0&0\end{pmatrix}}
\,
\boldsymbol{D}^{-1}\;,
\label{ojapde33-B1}
\\
\boldsymbol{B}_2(\boldsymbol{D}) &:= 
-3\,\boldsymbol{D}\,
{\small\begin{pmatrix}0&0&1\\0&0&0\\1&0&0\end{pmatrix}}
\,\boldsymbol{D}^{-1}\;,
\label{ojapde33-B2}
\\
\boldsymbol{B}_3(\boldsymbol{D}) &:= 
-3\,\boldsymbol{D}\,
{\small\begin{pmatrix}0&0&0\\0&0&1\\0&1&0\end{pmatrix}}
\,\boldsymbol{D}^{-1}
\label{ojapde33-B3}
\;.
\end{align}
\end{proposition}

\subsection{Trivariate Planar Images}

The case of trivariate planar images 
($\mathbb{R}^2\to\mathbb{R}^3$)
is of high practical interest since
the vast majority of colour images falls into this category.
Asymptotic analysis results are available for Oja and 
transformation--retransformation $L^1$ median filtering.
A PDE limit stated in \cite{Welk-Aiep14} for standard $L^1$ median filtering
of $n$-variate planar images that could be applied to trivariate planar
images was incomplete; a corrected result still needs to be published.

Also in the case of trivariate planar images 
both Oja and transformation--retransformation $L^1$ median filtering
approximate the same PDE as stated in the following results.
By $\boldsymbol{I}^{3,2}={\small\begin{pmatrix}1&0\\0&1\\0&0\end{pmatrix}}$ 
we will abbreviate a ``truncated unit matrix''.

\begin{lemma}\cite[Lemma~5 and Remark~12]{Welk-JMIV16}
\label{lem-ojal1pde23-I}
For a $(2,3)$-image $\boldsymbol{u}$ with a regular point $\boldsymbol{x}_0$
where $\mathrm{D}\boldsymbol{u}(\boldsymbol{x}_0)=\boldsymbol{I}^{3,2}$,
the local PDE limit of
$(\operatorname{med}_{\mathrm{Oja}},\mathcal{D})$-filtering 
as well as of
$(\operatorname{med}_{L^1},\mathcal{D})$-filtering
of $\boldsymbol{u}$ at $\boldsymbol{x}_0$ with time scale $\varrho^2/24$ is
\begin{align}
u_t&=u_{xx}+3u_{yy}-2v_{xy} \label{ojapde23-I-1} \\
v_t&=3v_{xx}+v_{yy}-2u_{xy} \label{ojapde23-I-2} \\
w_t&=2w_{xx}+2w_{yy}        \label{ojapde23-I-3} \;.
\end{align}
\end{lemma}

\noindent
Generic regular points can be transformed to the situation of the lemma
using the affine transform $\boldsymbol{D}_3^{-1}$ where 
\begin{equation}
\boldsymbol{D}_3:= 
\left(\partial_x \boldsymbol{u}~\Big|~\partial_y \boldsymbol{u}~\Big|~
\partial_x \boldsymbol{u}\times\partial_y\boldsymbol{u}\right)
\label{D3}
\end{equation}
is a rank-three extension of the Jacobian $\mathrm{D}\boldsymbol{u}$.
Again, this transform can be obtained also from the covariance-matrix
construction in the transformation--retransformation $L^1$ median with
appropriate handling of the dimensionality issues discussed in
Section~\ref{ssec-mvmedf-affeq}, compare also 
\cite[Section~3.3.2]{Welk-JMIV16}.

\begin{proposition}[{\cite[Theorem~3 and Prop.~3]{Welk-JMIV16}}]
\label{prop-ojatrl1pde23}
\sloppy
For $(2,3)$-images $\boldsymbol{u}$, 
$(\operatorname{med}_{\mathrm{Oja}},\mathcal{D})$-filtering 
with the Oja median as aggregator 
as well as
$(\operatorname{med}_{\mathrm{TR}L^1},\mathcal{D})$-filtering 
with the transformation--retransformation $L^1$ median as aggregator
and the disc family $B^2$ as scaled selector in both cases
have the regular PDE limit with time scale $\tau(\varrho)=\varrho^2/24$
given by
\begin{align}
\boldsymbol{u}_t &=
2\,\Delta\boldsymbol{u}
+ \boldsymbol{A}(\mathrm{D}\boldsymbol{u})
(\boldsymbol{u}_{yy}-\boldsymbol{u}_{xx})
+ \boldsymbol{B}(\mathrm{D}\boldsymbol{u})
\boldsymbol{u}_{xy}
\label{ojapde23}
\end{align}
where for 
$\boldsymbol{D}:=\mathrm{D}\boldsymbol{u}
=(\partial_x \boldsymbol{u}~|~\partial_y \boldsymbol{u})$
and $\boldsymbol{D}_3$ from \eqref{D3}
the coefficient matrices are given by
\begin{align}
\boldsymbol{A}(\boldsymbol{D}) &:= 
\boldsymbol{D}_3\,
\mathrm{diag}(1,-1,0)
\,\boldsymbol{D}_3^{-1}\;,
\label{ojapde23-A}
\\
\boldsymbol{B}(\boldsymbol{D}) &:= 
-2\, \boldsymbol{D}_3\,
{\small\begin{pmatrix}0&1&0\\1&0&0\\0&0&0\end{pmatrix}}
\,\boldsymbol{D}_3^{-1}
\label{ojapde23-B}
\;.
\end{align}
\end{proposition}

\noindent
Note that $\boldsymbol{D}_3$, the $3\times3$ matrix obtained by enlarging the
$2\times3$ Jacobian $\mathrm{D}\boldsymbol{u}$ with a third column orthogonal 
to the first two ones, is regular if and only if $\mathrm{D}\boldsymbol{u}$ 
has rank 2 as required in the hypothesis of the proposition. The transformed
variables $\hat{\boldsymbol{u}}:=\boldsymbol{D}_3^{-1}\boldsymbol{u}$ have the 
Jacobian $\boldsymbol{I}^{3,2}$. Any scaling of the third
column of $\boldsymbol{D}_3$ is actually irrelevant for the statement
and proof of the proposition;
it cancels out in the evaluation of \eqref{ojapde23-A} and \eqref{ojapde23-B}.
It may, however, affect the scaling of deviations
from the PDE that occur for positive structuring element radius $\varrho$.

\section{Summary and Outlook}
\label{sec-sum}

In this paper, we have reviewed selected median concepts for multivariate
data focussing on their applicability to image filtering. To this end,
the definitions of multivariate medians have been presented and traced back
to different aspects of the univariate median that they generalise,
and some important properties, especially regarding equivariance to groups
or sets of data transformations, have been discussed.
Emphasis has been put on the relation between discrete and continuous
modelling since this is a crucial issue in image and shape analysis
applications where one is concerned with discrete data arising from the
spatial sampling of continuous quantities. PDE approximation results for
multivariate image median filters were reported to the extent they are
available so far. An interesting outcome is that several affine equivariant
multivariate median concepts, despite not coinciding as discrete filters,
converge to the same process in an infinitesimal limit, which even features
a restricted projective equivariance that some of the discrete concepts
do not possess.

We hope that this condensed comparative presentation 
of multivariate median concepts
and image filters with focus on the underlying mathematical constructions
and resulting properties paves the way to efficient procedures for
shape analysis, for example shape simplification and extraction of
shape features or characteristics, and robust filtering of geometric
data. In ongoing work, we investigate the suitability of multivariate medians
to the processing of geometric and texture data on manifolds.

\end{document}